\documentclass[floats,floatfix,showpacs,amssymb,prd,twocolumn,superscriptaddress,nofootinbib,aps,10pt]{revtex4-1}

\usepackage{graphicx}   
\usepackage{epstopdf}           

\usepackage{epsfig, amssymb,graphicx} 
\usepackage{epstopdf}
\usepackage{amsmath, amsfonts}
\usepackage{bm} 

\usepackage[linktocpage]{hyperref}
\usepackage{subfigure}
\usepackage[usenames]{color}

\graphicspath{ {./plots/} }

\def\be{\begin{equation}}
\def\ee{\end{equation}}
\def\beq{\begin{eqnarray}}
\def\eeq{\end{eqnarray}}
\def\nn{\nonumber}

\begin{document}

\title{\large The nonlinear dynamical stability of infrared modifications of gravity}

\author{Richard Brito}\email{richard.brito@tecnico.ulisboa.pt}
\affiliation{CENTRA, Departamento de F\'{\i}sica, Instituto Superior T\'ecnico, Universidade de Lisboa,
Avenida Rovisco Pais 1, 1049 Lisboa, Portugal}
\affiliation{Perimeter Institute for Theoretical Physics Waterloo, Ontario N2J 2W9, Canada}

\author{Alexandra Terrana}\email{aterrana@perimeterinstitute.ca}
\affiliation{Department of Physics and Astronomy, York University
Toronto, On, M3J 1P3, Canada}
\affiliation{Perimeter Institute for Theoretical Physics Waterloo, Ontario N2J 2W9, Canada}

\author{Matthew C. Johnson}\email{mjohnson@perimeterinstitute.ca}
\affiliation{Department of Physics and Astronomy, York University
Toronto, On, M3J 1P3, Canada}
\affiliation{Perimeter Institute for Theoretical Physics Waterloo, Ontario N2J 2W9, Canada}

\author{Vitor Cardoso}\email{vitor.cardoso@tecnico.ulisboa.pt}
\affiliation{CENTRA, Departamento de F\'{\i}sica, Instituto Superior T\'ecnico, Universidade de Lisboa,
Avenida Rovisco Pais 1, 1049 Lisboa, Portugal}
\affiliation{Perimeter Institute for Theoretical Physics Waterloo, Ontario N2J 2W9, Canada}

\date{\today} 

\begin{abstract}
Scalar forces ``screened'' by the Vainshtein mechanism may hold the key to understanding the cosmological expansion of our universe,
while predicting new and exciting features in the interaction between massive bodies. Here we explore the dynamics of the Vainshtein screening 
mechanism, focusing on the decoupling limit of the DGP braneworld scenario and dRGT massive gravity. 
We show that there is a vast set of initial conditions whose evolution is well defined and which are driven to the static, screening solutions of these theories. Screening solutions are stable and behave coherently under small fluctuations: they oscillate and eventually settle to an equilibrium configuration, the timescale for the oscillations and damping being dictated by the Vainshtein radius of the screening solutions. At very late times, a power-law decay ensues, in agreement with known analytical results. 
However, we also conjecture that physically interesting processes such as gravitational collapse of compact stars may not possess a well-posed initial value problem.
Finally, we construct solutions with nontrivial multipolar structure describing the screening field of deformed, asymmetric bodies and show that higher multipoles are screened more efficiently than the monopole component.
\end{abstract}


\maketitle

\section{Introduction}

Einstein's theory of General Relativity (GR) is by far the most successful and simple theory that we have to describe the gravitational interaction. GR elegantly explains all observed phenomena on solar-system scales. To preserve the success of GR, cosmological observations (e.g. those made by the Planck satellite~\cite{Ade:2013ktc}) require dark matter and dark energy to be invoked as the drivers of clustering and accelerated expansion. This model, $\Lambda$CDM, is a compelling and plausible explanation for nearly all existing cosmological observations. However, there are various outstanding theoretical puzzles, perhaps the greatest of which is the extraordinary fine tuning that seems necessary in any theory of dark energy (e.g. the cosmological constant problem). These puzzles, along with the vigorous observational program to observe modifications of gravity in cosmology (see e.g. ~\cite{Jain:2013wgs,Huterer:2013xky}, have driven considerable interest in infrared modifications of GR~( for reviews see~\cite{Clifton:2011jh,Joyce:2014kja}).

The challenge for any such theory is sequestering the modification to large enough distance scales, so that the predictions can be reconciled with solar system tests (for a review of such tests see Ref.\cite{Will:2005va}). The class of theories we will consider in this paper introduce new scalar degrees of freedom, which potentially mediate gravitational-strength long range forces if they couple to matter. To prevent this phenomenological disaster, the scalar force must be ``screened" on sufficiently short distance scales, but active on cosmological distance scales. This feat can in principle be accomplished by making the range of the force (the Chameleon mechanism~ \cite{Khoury:2003aq}) or the coupling to matter (the Symmetron mechanism~\cite{Hinterbichler:2010es}) environmentally dependent. Theoretically consistent models displaying the Chameleon mechanism are tightly constrained~\cite{Upadhye:2012vh,2012PhRvD..85l3006D,Erickcek:2013dea,Sakstein:2013pda,2013ApJ...779...39J,Lombriser:2014dua,Vikram:2014uza}, as are versions of the Symmetron mechanism~\cite{2012PhRvD..85l3006D,Sakstein:2013pda,Taddei:2013bsk}. 

Another possibility, which we will concentrate on in this paper, is the Vainshtein mechanism~\cite{Vainshtein:1972sx}. The Vainshtein mechanism~(for a recent review see~\cite{Babichev:2013usa}) relies on non-linear derivative couplings to screen the effect of extra degrees of freedom on small scales. In effect, sources which couple to the extra degrees of freedom are dressed with a scalar profile whose gradient ensures that the magnitude of the fifth force is small.

Theories displaying the Vainshtein mechanism have a long history, beginning with its role in establishing the consistency between the massless limit of the Fierz-Pauli massive graviton~\cite{Fierz:1939ix} and GR~\cite{vanDam:1970vg,Zakharov:1970cc}. The Vainshtein mechanism is also manifest in theories with Galileon symmetry~\cite{Nicolis:2009aa} and their covariant generalizations~\cite{Deffayet:2009wt} (special cases of the most general non-linear scalar tensor theory with second order equations of motion~\footnote{The most general scalar-tensor ghost-free theory had already been discovered by Horndeski in the `70s~\cite{Horndeski:1974wa}, but didn't receive much attention at the time. This is why the generalized action for Galileons found in Ref.~\cite{Deffayet:2011gz} is now known as the Horndeski action. A generalization of the Horndeski Lagrangian to vector-tensor theories can be found in Refs.~\cite{Horndeski_vector,Jimenez:2013qsa,Tasinato:2014eka,Tasinato:2014mia,Heisenberg:2014rta}. }), as first discovered in the context of the Dvali-Gabadadze-Porrati (DGP) braneworld model~\cite{Dvali:2000hr}. More recently, the Vainshtein mechanism has played an important role in dRGT gravity~\cite{deRham:2010kj,deRham:2010ik}, the non-linear generalizations of ghost-free Fierz-Pauli massive gravity (for recent reviews on Massive gravity see Refs.~\cite{Hinterbichler:2011tt,deRham:2014zqa}), and massive bimetric gravity~\cite{Hassan:2011zd,Hassan:2011ea}. 

Within the context of these theories, considerable work has been done identifying static spherically symmetric solutions which manifest the Vainshtein mechanism~\cite{Nicolis:2004aa,Babichev:2009aa,Babichev:2009jt,Babichev:2010jd,Chkareuli:2011te,Berezhiani:2013dca,Berezhiani:2013dw,Babichev:2013pfa}. This provides an important check on the compatibility of these models with solar system tests, and insight into how the evolution of the Universe is affected on large scales. Cosmological N-body simulations which treat the matter dynamically but neglect time-derivatives of the scalar degree of freedom~\cite{2009PhRvD..80d3001S,Barreira:2013eea,Li:2013tda} (the so-called quasi-static approximation~\cite{Lue:2004rj}; see \cite{Noller:2013wca} for a critical analysis of this approximation) have also been used to study the formation of large-scale structure and make predictions for cosmological observables.

However, due to the complex structure of the non-linear equations of motion in these theories, there has been little study of fully dynamical phenomena.~\footnote{Fully dynamical phenomena have been studied in the context of K-essence, which possesses a non-linear kinetic sector (though not of the type we consider), see e.g.~\cite{Akhoury:2011hr,Akhoury:2011rc,Leonard:2011ce,Brax:2014gra}. See also Refs. \cite{Llinares:2013jua,Llinares:2013qbh} for a fully dynamical treatment of a cosmological N-body simulation coupled to a scalar field theory displaying the symmetron screening mechanism.}  Another important gap to fill is the study of solutions without spherical symmetry. Some notable exceptions are the study of scalar gravitational radiation emitted by a binary pulsar system~\cite{deRham:2012fw,deRham:2012fg}, computations of the Green's function~\cite{Chu:2012kz}, plane wave solutions~\cite{Babichev:2012qs}, studies of the characteristic matrix of the full dRGT theory~\cite{Deser:2012qx,Deser:2013eua,Deser:2013qza,Deser:2014aa}, and some results on stationary solutions~\cite{Chagoya:2014fza}. 

In this paper, we attempt to fill some of the existing gaps in the knowledge of fully dynamical and aspherical solutions in theories displaying the Vainshtein mechanism. We study two theories in the decoupling limit: DGP and dRGT gravity. In the decoupling limit, a non-gravitating scalar degree of freedom is introduced that couples to the matter sector. We consider non-relativistic sources and neglect the internal dynamics of the matter sector. Under these approximations, we consistently solve for the full dynamical evolution of the scalar on a fixed background Minkowski space with fixed sources. The two theories differ primarily in their derivative self-couplings and couplings to matter; qualitatively the dynamical behaviour is similar. 

We begin by characterizing the linear and non-linear stability of the screening solutions in the DGP model. Using numerical simulations, we find that in the fully non-linear regime a large space of initial conditions evolve to the screening solution around a fixed time-independent source at sufficiently late times. We study the linear response properties of the screening profiles, showing that the screening solutions behave much as a black hole when perturbed: there is a prompt response, followed by a universal ring-down and power-law decay of the wave form. We also examine the screening properties of static non-spherically symmetric scalar profiles and sources. 

However, we also find that for initial data sufficiently far from the screening solution, the evolution becomes ill-posed: regions form where the constant time surfaces fail to be Cauchy surfaces of the scalar equation of motion. 
Following Ref.~\cite{Leonard:2011ce}, we term this Cauchy breakdown. We study the situations in which Cauchy breakdown occurs, and show that no foliation of flat spacetime exists in which the evolution can be continued.

In addition to static sources, we consider two models of a dynamical source corresponding to spherically symmetric collapse and explosion. The collapse model includes a time-dependent pressure term in the energy momentum tensor, which changes the coupling to the screening field. In both the DGP model and dRGT massive gravity, this causes the source to loose its scalar hair, and in many cases the appropriate vacuum solution is smoothly reached. In the case of explosion, we model the source as a mass-conserving out-going null spherical shell of pressureless matter. The shell sources an outgoing pulse of the screening field, and in many cases the appropriate vacuum solution is left behind.

However, in both cases, for sufficiently dense sources or sufficiently short timescales of collapse, we find that Cauchy breakdown occurs. These examples suggest that a fully dynamical study of astrophysical phenomena for infrared modifications of gravity will require a prescription for evolving past Cauchy breakdown, or an ultraviolet completion of the theory. A complete treatment will also require stepping away from the decoupling limit, and considering realistic dynamical sources.



\section{Theories displaying the Vainshtein mechanism}\label{sec:vainshtein}

The Vainshtein mechanism relies on derivative self-interactions to screen long-range fifth forces mediated by a scalar degree of freedom. There are a variety of theories that manifest the Vainshtein mechanism including the DGP braneworld scenario, dRGT massive gravity, massive bi-metric gravity, and scalar field theories with Galileon invariance. In each case, the action contains higher-derivative interaction terms for the fluctuating degrees of freedom, but the equations of motion remain second order. 

In this paper, we restrict our focus to DGP and dRGT massive gravity. The DGP model~\cite{Dvali:2000hr} physically describes our Universe as a 3-brane embedded in a 5D bulk, introducing a brane-bending mode that from the 4D point of view corresponds to an additional scalar degree of freedom. dRGT massive gravity~\cite{deRham:2010kj,deRham:2010ik} is the non-linear generalization of the Fierz-Pauli massive graviton. The theory propagates 5 degrees of freedom, two tensor, two vector, and one scalar; we focus on the physics of the scalar sector. Attempts have been made to explain the observed accelerated expansion of the Universe in the context of both models (or close cousins), and one might expect that many of the features present in these models will be shared by other infrared modifications of gravity relying on similar non-linear derivative interactions. 

Both DGP and a sector of dRGT massive gravity are special cases of a general effective theory with action:
\begin{equation}\label{eq:action}
S = \int d^4x \  \mathcal{L}_\pi\,,
\end{equation}
where
\beq \label{Lan_general}
	\mathcal{L}_\pi &=& c_2\mathcal{L}_2+\frac{c_3}{\Lambda^3}\mathcal{L}_3+\frac{c_4}{\Lambda^6}\mathcal{L}_4+\frac{c_5}{\Lambda^9}\mathcal{L}_5\nonumber\\
	&+&\frac{\xi\pi T}{M_4} +\frac{\alpha}{M_4\Lambda^3}\partial_\mu\pi\partial_\nu\pi T^{\mu\nu}\,, 
\eeq
$\Pi_{\mu \nu} \equiv \nabla_{\mu} \partial_\nu \pi$; $c_2,\,c_3,\,c_4,\,c_5,\,\alpha,\,\xi$ are dimensionless parameters; $\Lambda$ is the strong coupling scale of the theory;
\begin{eqnarray}
\mathcal{L}_2&=&(\partial\pi)^2\,,\label{gal_lin} \\
\mathcal{L}_3&=&(\partial\pi)^2[\Pi]\,, \label{gal_cubic}\\
\mathcal{L}_4&=&(\partial\pi)^2([\Pi]^2-[\Pi^2])\,, \label{gal_quartic}\\
\mathcal{L}_5&=&(\partial\pi)^2([\Pi]^3-3[\Pi][\Pi]^2+2[\Pi^3]) \label{gal_quintic}\,,
\end{eqnarray}
with $(\partial\pi)^2 \equiv \eta^{\mu \nu} (\partial_\mu \pi) (\partial_\nu \pi)$ and square brackets are used to denote the trace, $[\Pi]=\Pi_{\mu\nu}\eta^{\mu\nu}$. The interactions in this effective theory are dictated by the Galilean symmetry $\pi \rightarrow \pi + c + b_\mu x^\mu$.

Throughout this work, we restrict our focus to the decoupling limit, in which there is no direct coupling between the scalar and gravitational sectors (note the lack of minimal coupling to gravity in Eq.~\ref{eq:action}). The decoupling limit retains the number of degrees of freedom of the theory, and is generically defined by keeping the lowest scales fixed, while sending all the higher scales to infinity,
\begin{equation} \label{eq:dgpdc}
	M_4, \ L_D, \ T_{\mu\nu} \rightarrow \infty, \hspace{0.6cm} \Lambda, \ \frac{T_{\mu\nu}}{M_4} \sim \text{constant}\,,
\end{equation}
where $M_4$ is the 4D Planck mass, $L_D$ is an additional length scale in the theory responsible for the couplings between the different modes, $T_{\mu\nu}$ is the stress-energy tensor for the matter source, and $\Lambda=(M_4/L_D^2)^{1/3}$ is the strong coupling scale of the theory~\footnote{In DGP, $L_D\equiv M_4^2/M_5^3$, where $M_5$ is the 5D Planck mass and in massive gravity, $L_D\equiv 1/m$, where $m$ is the graviton mass.}. 
In this limit we are perturbing around flat space $g_{\mu\nu} = \eta_{\mu\nu} + h_{\mu\nu}$ with $|h_{\mu\nu}| \ll 1$ so the solutions are only valid in the weak field regime, i.e., for matter sources much bigger than their Schwarzschild radius. The interesting feature of this limit is that it allows to decouple the scalar mode from gravity, but at the same time to keep the full non-linear dynamics of the scalar field and its coupling to matter. Thus within its range of validity it describes important non-linear phenomena of the theory, such as the Vainshtein mechanism. 

A viable theory of modified gravity must avoid the vDVZ discontinuity \cite{vanDam:1970vg}, meaning that it should agree with GR in the $L_D \rightarrow \infty$ limit, at least for astrophysical objects on solar system scales. This is accomplished by the Vainshtein mechanism. Thus, in the decoupling limit, we can make a direct comparison of a modified gravity theory to GR. 

The Lagrangian Eq.~\ref{Lan_general} is the most general effective theory displaying the Vainshtein mechanism with no couplings between the scalar and gravitational sectors that can be obtained from the Horndeski action~\cite{Koyama:2013paa} in the decoupling limit.  A restricted sector of the decoupling limit of dRGT massive gravity~\cite{deRham:2010ik} is recovered by setting $c_2=-3/2$, $c_3=3\alpha/2$, $c_4=-\alpha^2/2$, $c_5=0$, $\xi=1$ in the Lagrangian~\eqref{Lan_general}~\footnote{In the full decoupling limit of massive gravity the theory depends on two parameters, $\alpha$ and $\beta$ in the terminology of Refs.~\cite{Berezhiani:2013aa,Berezhiani:2013dca}. When $\beta\neq 0$ there is a nonlinear coupling term between the helicity--0 and --2 modes which cannot be removed by a local field redefinition~\cite{deRham:2010ik}. Here we will consider $\beta=0$.}. The decoupling limit of the DGP model~\cite{Nicolis:2004aa} can be obtained by setting $c_2=-3$, $c_3=-1$,  $\xi=1/2$, $c_4=c_5=\alpha=0$.

Although we mostly focus on the decoupling limit of the DGP model and dRGT massive gravity, we expect our results to generalize \emph{qualitatively} to Galileon theories which retain a direct coupling between the scalar and gravitational sectors. We work in the approximation that the radius of the matter source $T_{\mu\nu}$ is bigger than its Schwarzschild radius $R_S$. For Galileon theories, due to their Vainshtein suppression near the source, the backreaction of the scalar field on the geometry is negligible as long as $\Lambda^{-1}\gg R_S$ and if the galileon decays at infinity. As we will see below, there are also cases where the galileon field grows quadratically at infinity, which after backreacting on the metric drives it to a cosmological spacetime. Near the source, our results should also give a good qualitative approximation for this case. Therefore, most of the features we will discuss in this work are generic.

As a final caveat, there is the possibility that we do not capture important properties of the full nonlinear system beyond the decoupling limit. However, as argued in \cite{Babichev:2009aa}, the decoupling limit solutions give the local dynamics at scales within the present Hubble volume, the scale on which the infrared modifications of gravity kick in. Thus, solutions found in the decoupling limit should still appear as transients lasting for long cosmological times in the full theory \cite{de-Rham:2011ab}.

\section{The DGP model}
We will first consider the DGP model, which is the simplest setting in which the Vainshtein mechanism arises. The full action for the scalar field in the DGP decoupling limit can obtained from the Lagrangian~\eqref{Lan_general} setting $c_2=-3$, $c_3=-1$,  $\xi=1/2$, $c_4=c_5=\alpha=0$~\cite{Nicolis:2004aa}
\begin{equation} \label{eq:dgpaction}
S = \int d^4 x \left[ - 3 (\partial \pi)^2 - \frac{1}{\Lambda^3} (\partial \pi)^2 \Box \pi + \frac{1}{2 M_4} \pi T \right] \,.
\end{equation}
Notice that the scalar self interaction term $(\partial \pi)^2 \Box \pi $ is the cubic galileon~\cite{Nicolis:2009aa}, ensuring that the equations of motion for $\pi$ are second order in derivatives. If this were not the case, then by Ostrogradsky's theorem (see e.g.~\cite{Woodard:2007aa} for a clear exposition) the Hamiltonian would necessarily be unbounded from below, and stable solutions would not exist. However, ``ghost" instabilities can appear at the nonlinear level if the kinetic term for perturbations on top of a background acquire the wrong sign with reference to other fluctuating degrees of freedom. This was the main obstruction to obtaining a viable explanation for the observed cosmological accelerated expansion in the context of DGP~\cite{Luty:2003aa,Nicolis:2004aa,Koyama:2005tx,Gorbunov:2005zk,Charmousis:2006pn}.

Varying the action \eqref{eq:dgpaction} gives rise to the following equation of motion: 
\begin{align}
	& 6 \Box \pi +\frac{2}{\Lambda^3}\left(\Box\pi\right)^2-\frac{2}{\Lambda^3} \left(\nabla^{\mu}\nabla^{\nu}\pi\right)\left(\nabla_{\mu}\nabla_{\nu}\pi\right)+ \frac{T}{2 M_4} = 0\,,\label{eq:eom}
\end{align}
where the metric is included to allow for easy conversion between alternative foliations of Minkowski space. An important length scale in this theory is the Vainshtein radius $R_V$, below which the non-linearities in the equation of motion become important. These non-linearities are crucial for the Vainshtein mechanism. Continuity with GR is recovered on length scales less than $R_V$ by screening the scalar degree of freedom. This is crucial for phenomenological applications of modified gravity models. The Vainshtein radius is given by
\begin{equation}
	R_V \equiv \frac{1}{\Lambda} \left( \frac{M(r \rightarrow \infty)}{M_4} \right)^{1/3}\,,
\end{equation}
and retains its form in the decoupling limit as can be seen from~\eqref{eq:dgpdc}. Here, $M$ is the mass of the source defined by
\begin{equation}
	M(r) \equiv - 4 \pi \int_0^r dr' r'^2 T(r')\,.
\end{equation}

From now on, we will measure all quantities in terms of $\Lambda$ and $M_4$. We re-scale dimensionfull quantities by: $x^\mu\Lambda \rightarrow x^{\mu}$, $\pi/\Lambda \rightarrow \pi$ and $T / (M_4\Lambda^3) \rightarrow T$,
%
%
in which case
\begin{equation}
R_V = (M(r \rightarrow \infty))^{1/3}\,.
\end{equation}
It can be checked from the equations of motion \eqref{eq:eom} that this is equivalent to working with the dimensionless
density $\rho = \rho_d / (M_4 \Lambda^3)$ where  $\rho_d$ is the dimensionful density. For a sun-like star,
$\rho_d \sim 1.4 \times 10^3 {\rm kg/m}^3 \sim 5.9 \times 10^{-18} {\rm GeV}^4$ and the strong coupling scale is $\Lambda \sim M_5^2 / M_4 \sim (1000 {\rm km})^{-1} \sim 1.8 \times 10^{-21} {\rm GeV}$. Using $M_4 = 2.4 \times 10^{18} {\rm GeV}$ we get the dimensionless density, to be used in our adopted units, of $\rho \sim 4.2 \times 10^{26}$.

For much of this work, we consider a simple spherically symmetric non-relativistic source with central density $\rho$ and radius $R_0$ described by
\begin{equation} \label{eq:source}
T_{\mu \nu} = {\rm diag} \left( \rho \exp\left(-\frac{r^2}{R_0^2}\right), 0,0,0 \right) \ , 
\end{equation}
Note that the source is exponentially close to zero for $r\gtrsim R_0$. The Vainshtein radius for this source is given by $R_V= (M(r \rightarrow \infty))^{1/3}=\sqrt{\pi}\rho^{1/3}R_0$. For simplicity we do not treat the internal dynamics of the source, nor are we allowing it to evolve in the presence of the scalar field. This is a convenience, and not a physical choice. Nevertheless, we expect our results to be very weakly dependent on the precise radial distribution of $T(r)$ and future work will incorporate the full dynamics of a realistic source.  

One can also consider a source with non-zero pressure, which would have important effects on the solution and its stability. For an increasingly relativistic source, the trace of the energy momentum tensor shrinks, weakening the coupling between the source and the scalar (this has been shown to have important implications for the Chameleon screening mechanism~\cite{Erickcek:2013dea}). As another example, in dRGT massive gravity the disformal coupling $\partial_\mu\pi\partial_\nu\pi T^{\mu\nu}$ can give rise to instabilities whenever $p\sim \rho$~\cite{Berezhiani:2013aa}. 

\begin{figure}[hbt]
	\begin{center}
		\begin{tabular}{c}
			\epsfig{file=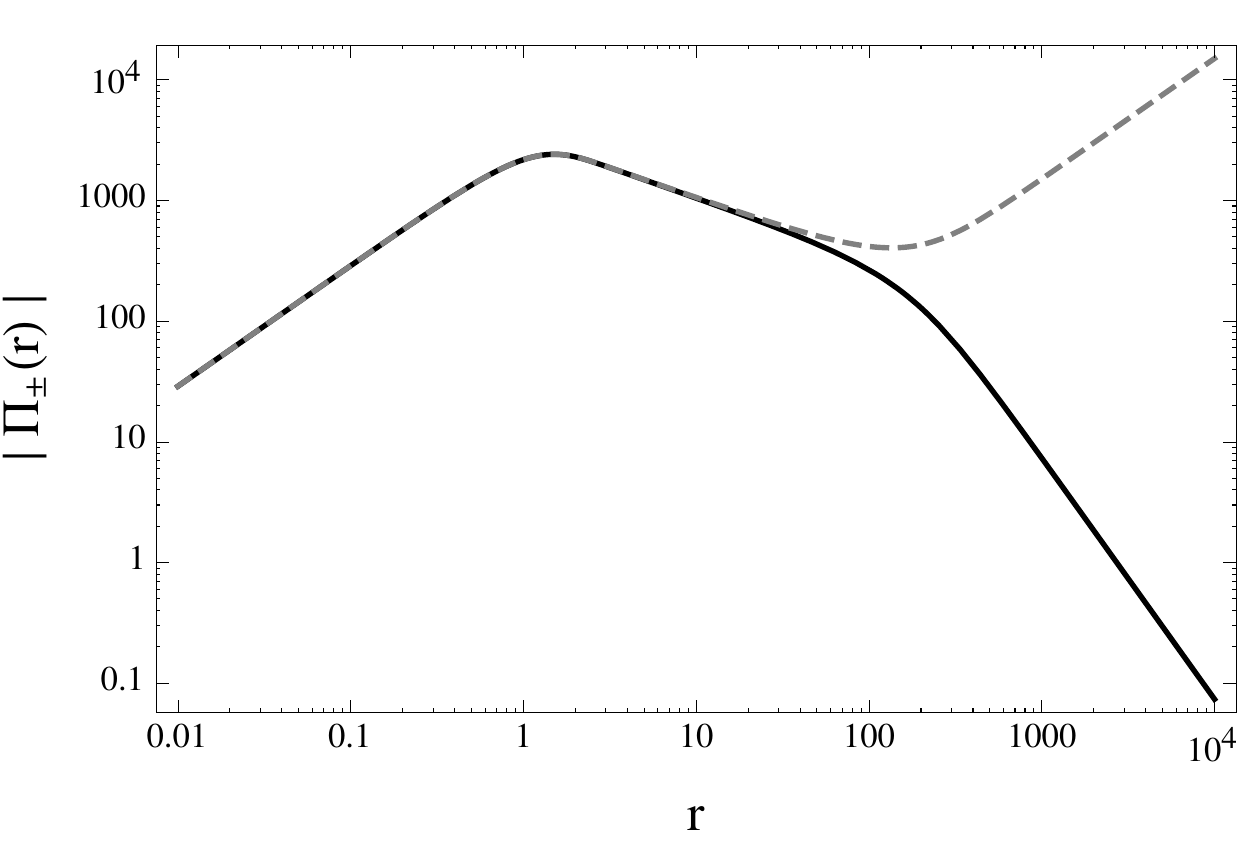,width=8.5cm,angle=0,clip=true}
		\end{tabular}
		\caption{The two static screening solutions $\Pi_+$ (solid) and $\Pi_-$ (dashed) of equation \eqref{eq:dgpeomstatic}. These solutions correspond to a source radius of $R_0=1 $ and a Vainshtein radius $R_V\simeq 1000$. Both solutions grow as $\sim r$ inside of the source and decay as as $\sim 1/\sqrt{r}$ outside the source for $0\ll r \ll R_V$. For very large distances $r \gg R_V$, $\Pi_-$ diverges as $\sim r$, whereas $\Pi_+\sim 1/r^2$.\label{fig:dgpSSS}}
	\end{center}
\end{figure}

Under the assumption of spherical symmetry, the equation of motion~\eqref{eq:eom} for $\pi(r,t)$ can be written as 
\begin{align}\label{eq:dgpeom}
	\ddot{\pi} \left( 6 + 8 \frac{\pi'}{r} + 4 \pi'' \right) & = \frac{T}{2} + 12 \frac{\pi'}{r} + 6 \pi'' + 4 (\dot{\pi}')^2 \nonumber \\
& + 8 \frac{\pi' \pi''}{r} + 4 \frac{(\pi')^2}{r^2}\,,
\end{align}
where $\dot{\pi}=\partial\pi/\partial t$ and $\pi'=\partial \pi/\partial r$. General solutions to this non-linear partial differential equation are not available, and a numerical analysis is required (see Section~\ref{sec:numerics} below). 
Nevertheless, static solutions -- which establish the viability of the Vainshtein mechanism -- can be obtained analytically.
Dropping the time-dependence in Eq.~\eqref{eq:dgpeom} allows us to integrate it to obtain a simple algebraic relation for $\Pi\equiv\pi'$
\begin{equation}\label{eq:dgpeomstatic}
4\pi r^2\left( 6\Pi(r) + 4\frac{\Pi(r)^2}{r} \right) = \frac{M(r)}{2}.
\end{equation}
This quadratic equation has two solutions given by
\begin{equation}\label{eq:screening}
	\Pi_{\pm}(r) = - \frac{3r}{4} \pm \frac{\sqrt{r M(r) + 18 \pi r^4}}{4 \sqrt{2 \pi} r}
\end{equation}
which can be integrated once more to obtain the field $\pi_{\pm}(r)$. The scalar fifth force is proportional to $\Pi_{\pm}(r) \hat{r}$, and comparing with the Newtonian gravitational force on different scales one obtains:
\begin{eqnarray} \label{eq:force}
	\biggl| \frac{F_{\Pi_+}}{F_g}  \biggl| \sim \frac{\Pi_+/M_4}{M/(M_4^2r^2)}\sim\begin{cases}
	 \left( \frac{r}{R_V} \right)^{3/2} \ \textrm{ if $r \ll R_V$} \\
       1 \ \ \ \ \ \ \ \ \ \ \  \ \textrm{ if $r \gg R_V$}\\
      \end{cases}\\
      \biggl| \frac{F_{\Pi_-}}{F_g}  \biggl| \sim \frac{\Pi_-/M_4}{M/(M_4^2r^2)}\sim\begin{cases}
	 \left( \frac{r}{R_V} \right)^{3/2} \ \textrm{ if $r \ll R_V$} \\
      \left( \frac{r}{R_V} \right)^3 \ \ \ \textrm{ if $r \gg R_V$.}\\
   \end{cases}
\end{eqnarray} 
These are both screening solutions, since the fifth force is suppressed on scales much smaller than $R_V$, but comparable to gravity at large scales. This is the simplest manifestation of the Vainshtein mechanism. For $0\ll r \ll R_V$, both solutions $\Pi_\pm$ decay as $\sim 1/\sqrt{r}$, while for $r \gg R_V$, $\Pi_+\sim 1/r^2$ and $\Pi_- \sim r$ as shown in Fig.~\ref{fig:dgpSSS}. Note that in theories for which the quartic galileon is present, such as in the decoupling limit of massive gravity studied in the next section, the suppression is even stronger: $F_\pi /F_g \sim \left( r/R_V \right)^2$ for $r \ll R_V$ \cite{deRham:2014zqa}.

The asymptotically decaying solution $\Pi_+$ gives rise to an asymptotically flat spacetime $g_{\mu\nu} \propto \eta_{\mu\nu}$, whereas $\Pi_-$ sources a spacetime with cosmological asymptotics $g_{\mu\nu} = (1 - \frac{3}{4} \frac{\Lambda^3}{M_4} r^2)\eta_{\mu\nu}$. Note that this solution is distinct from the self-accelerating solution in DGP which gives rise to a maximally symmetric de Sitter spacetime. The self-accelerating solution has no sources and a scalar field configuration of $\pi=-(1/2)\Lambda^3x_\mu x^\mu$, leading to a spacetime of the form $g_{\mu\nu} = (1 - \frac{1}{2}\frac{\Lambda^3}{M_4} x_\mu x^\mu)\eta_{\mu\nu}$ which is locally de Sitter (for times and physical distances much smaller than $L_{\text{DGP}}\equiv M_4^2/M_5^3$) \cite{Nicolis:2004aa}. 

These static solutions of the DGP theory are well known, but very little is known about how equations of the type~\eqref{eq:dgpeom} behave in the fully dynamical regime. We attempt to address this here by considering the linear and non-linear stability of the screening solutions, solving the full time-dependent equation of motion for the DGP scalar field numerically. 

\section{Linear and non-linear stability in the decoupling limit of DGP}\label{sec:screeningstability}
Stability of the screening solutions to small perturbations is a fundamental condition for their physical relevance. The analysis of fluctuations around the static screening solutions reveals that $\Pi_+$ is stable to small high-frequency fluctuations, but $\Pi_-$ is not (see also Appendix~\ref{app:IVP}). To see this, consider small perturbations $\delta\pi(r,t)$ about the screening solution, which have the action:
\begin{equation}\label{eq:Sdeltapi}
S_{\delta\pi}=\int{d^4x\left[\frac{1}{2} Z^{\mu\nu}\partial_\mu\delta\pi\partial_\nu\delta\pi+\frac{1}{2M_4}\delta\pi T\right]}\,,
\end{equation}
where the kinetic coefficients are given by the effective metric components
\begin{align}\label{eq:Zs}
	Z^{tt} = -\left( 6 + 8 \frac{\pi'}{r} + 4 \pi'' \right),\hspace{0.6cm} Z^{rr} = 6 + 8 \frac{\pi'}{r}\,, \nn\\
	r^2 Z^{\theta\theta}=r^2 \sin^2{\theta}Z^{\phi\phi}=6 + 4 \frac{\pi'}{r} + 4 \pi''.
\end{align} 
Therefore, although the field lives in flat space, it propagates in the effective metric $Z^{\mu\nu}$. For the asymptotically flat solution, the matrix $Z^{\mu\nu}$ has the signature of Minkowski spacetime. For the asymptotically non-flat solution, the kinetic terms switch sign in relation to other fluctuating degrees of freedom in Minkowski space. This ``wrong'' sign in the kinetic term indicates a ghost, first discussed in Ref.~\cite{Luty:2003aa}. Dynamically, ghosts cause problems when there is an interaction with another field whose kinetic term takes the opposite sign. In the decoupling limit with fixed sources, the scalar field evolves independently, and we therefore do not expect to see any instabilities in either branch of solutions in what follows. 

\begin{figure}[hbt]
	\begin{center}
		\begin{tabular}{c}
			\epsfig{file=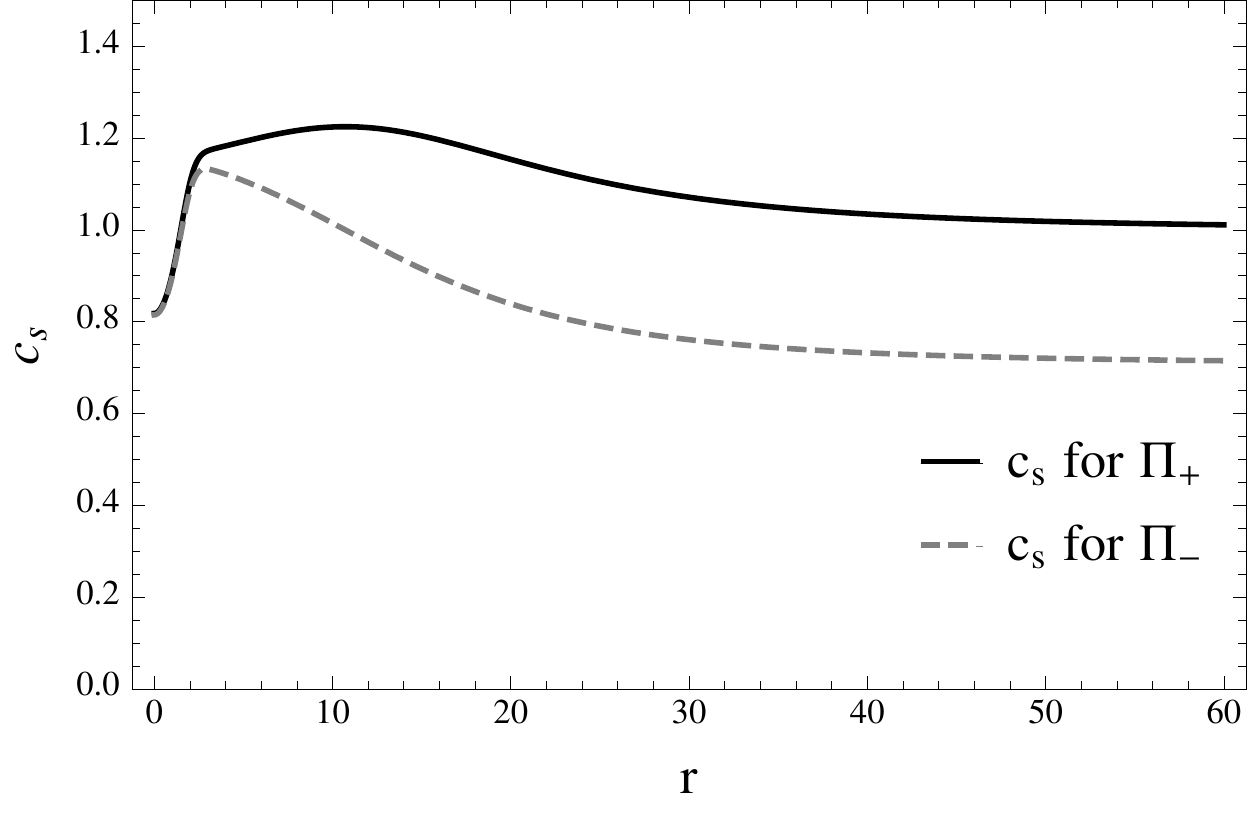,width=8.5cm,angle=0,clip=true}
		\end{tabular}
		\caption{The sound speed profiles corresponding to fluctuations on top of the screening solutions, shown here for $R_0=1$ and $\rho = 10^5$, giving $R_V \sim 80$. Superluminal propagation $c_s > 1$ can occur in both branches inside and outside of the source. For large $r$, $c_s \rightarrow 1$ for the positive branch, while $c_s \rightarrow 1/\sqrt{2}$ for the negative branch.}
		\label{fig:speed}
	\end{center}
\end{figure}

Fluctuations about the screening solutions can travel superluminally. Fig.~\ref{fig:speed} shows the sound speed profiles, calculated as $c_s = \sqrt{-Z^{rr}/Z^{tt}}$, for both branches of screening solution. In both cases, there is superluminal propagation inside and outside of the source. Note that for large $r$, fluctuations around $\Pi_+$ travel luminally, while fluctuations around $\Pi_-$ are subluminal and approach $1/\sqrt{2}$.

Although DGP has fallen out of favour due to the presence of ghost instabilities and superluminal propagation speeds~\cite{Hinterbichler:2009aa,Gregory:2008bf}, both the ghostly and asymptotically flat branches give us insight into how solutions of more complicated, and theoretically consistent, cases of the general theory Eq.~\eqref{Lan_general} behave.

\subsection{Dynamical approach to screening}\label{sec:screeningsims}
We have seen that physically appealing {\it static} profiles do exist as simple
solutions to the equations of motion. Do these equilibrium configurations actually form
from generic initial data, or is there an unstable or ill-defined evolution?
In other words, what possible initial configurations of $\pi(r,t=0)$ -- if any -- would dynamically evolve to the screening solutions? Although a complete classification of the possible initial data is not possible, we focus on two representative cases: a spherical shell collapsing on the screening solution, and evolution from vacuum.

\subsubsection{Code description\label{sec:numerics}}

The Vainshtein screening solutions arise from the non-linear self-interactions of the $\pi$ field. The solutions of interest are therefore fully in the non-linear regime of the equation of motion, Eq.~\eqref{eq:dgpeom}, where analytic solutions beyond the perturbative regime are difficult to obtain. Therefore, in order to fully explore the stability of screening solutions and investigate the general properties of time evolution, a numerical treatment is necessary.

We evolve Eq.~\eqref{eq:dgpeom} numerically using the method of lines. Spatial derivatives were discretized on a fixed grid size with typical resolutions $dr=1/25$, $dr=1/50$ and $dr=1/100$ using a second order differencing scheme. We explicitly checked that our results do not depend on the resolution. The resulting system of ordinary differential equations were then integrated using a fourth order explicit Runge-Kutta method. The stability of the evolution scheme relied on using stencils for both first and second derivatives; a fully first-order formulation of Eq.~\eqref{eq:dgpeom} (in both spatial and time derivatives) did not yield stable numerical evolution. The spatial grid was constructed on the finite interval $[r_\text{min},\ r_\text{max}]$. Our results are independent of the interval chosen as long as $r_\text{min}$ is sufficiently small and $r_\text{max}$ sufficiently large such that the outer boundary remains causally disconnected from the region under study. When dealing with the positive branch of solutions, we imposed Neumann boundary conditions at both ends $\pi'(r=r_\text{min})=\pi'(r=r_\text{max})=0$, where this boundary condition at the origin is required for the field to remain regular. However, for solutions with cosmological asymptotics, the outer boundary condition was adjusted to $\pi'(r)=\Pi_-(r)$ at $r=r_\text{max}$, where $\Pi_-(r)$ is from equation \eqref{eq:screening}. The convergence of the code with increasing resolution is as expected for a second order scheme, see Appendix~\ref{sec:convergence}. As an additional test of the code, many of our results were reproduced using the {\scshape Mathematica} software.

\subsubsection{Incoming spherical wave packet \label{sec:screen1}}
%
\begin{figure}[htb!]
		\begin{center}
		\begin{tabular}{cc}
			\subfigure[]{\includegraphics[width=4.1cm]{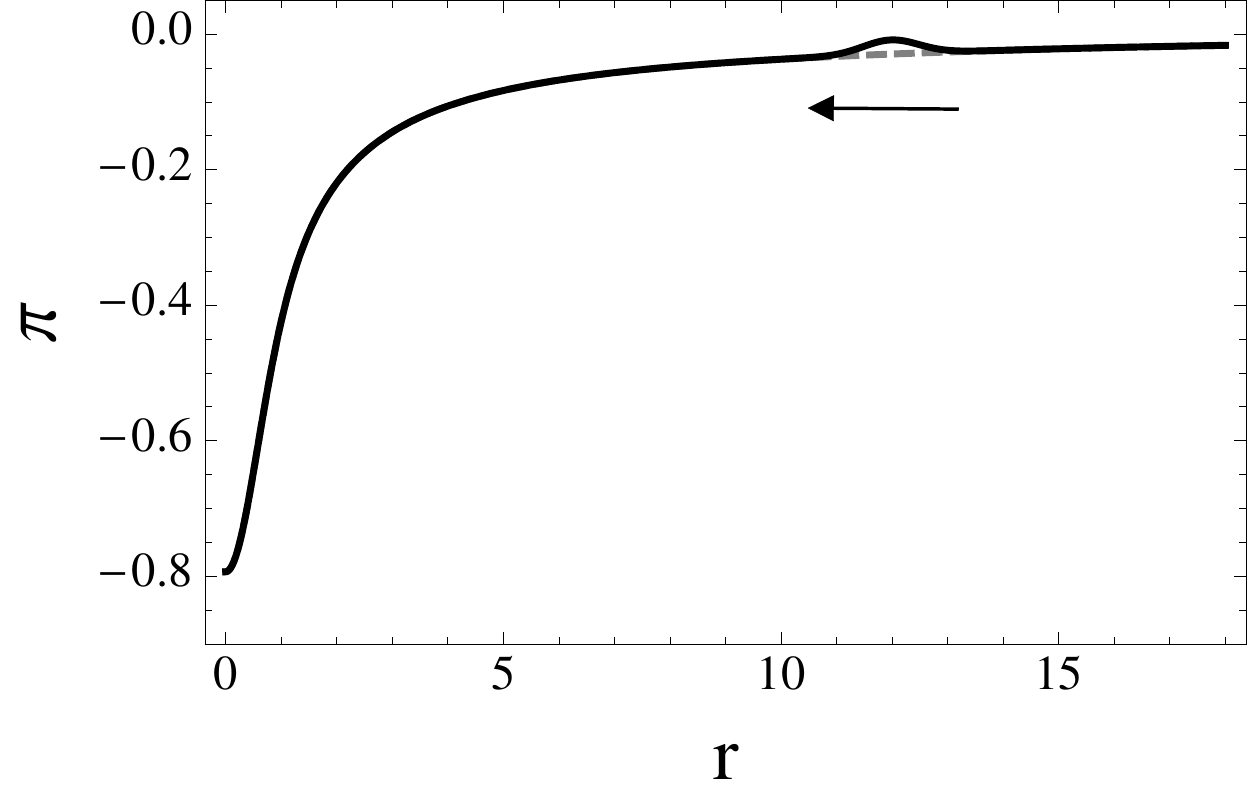}}&
			\subfigure[]{\includegraphics[width=4.1cm]{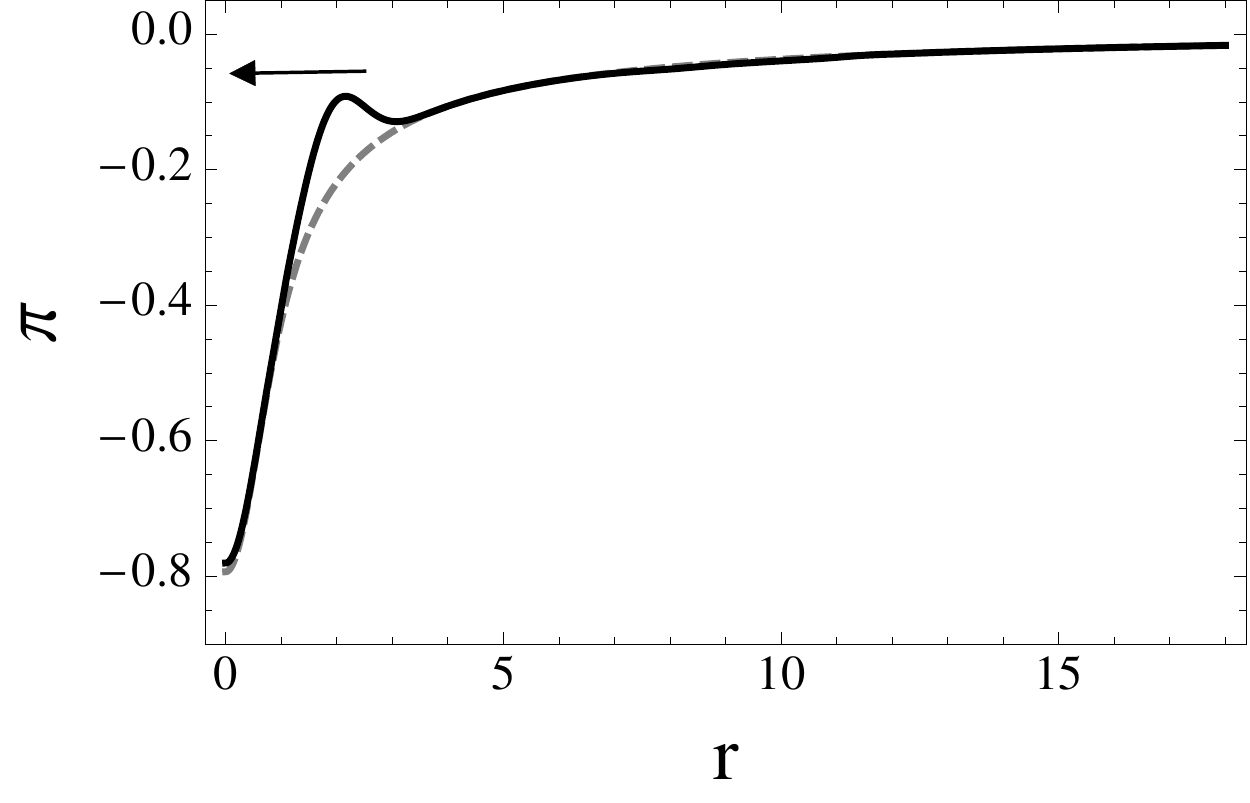}}\\
			\subfigure[]{\includegraphics[width=4.1cm]{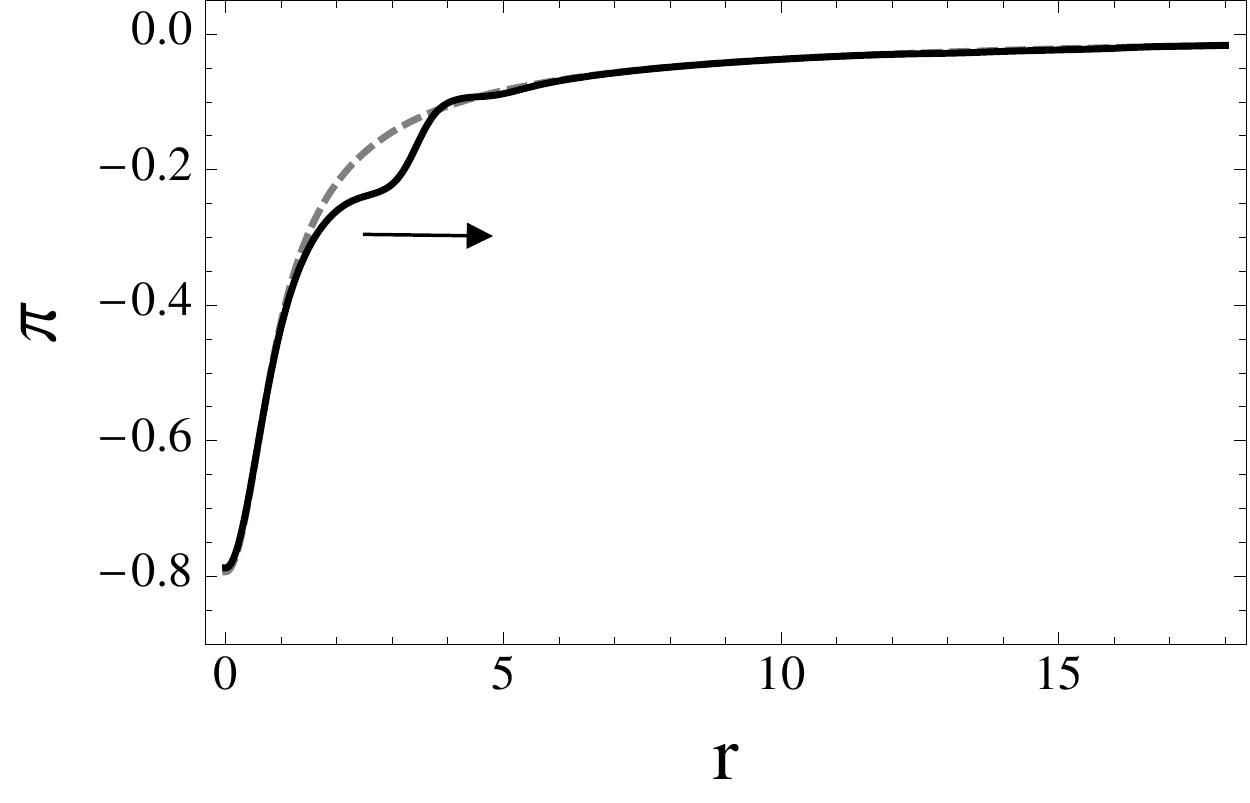}}&
			\subfigure[]{\includegraphics[width=4.1cm]{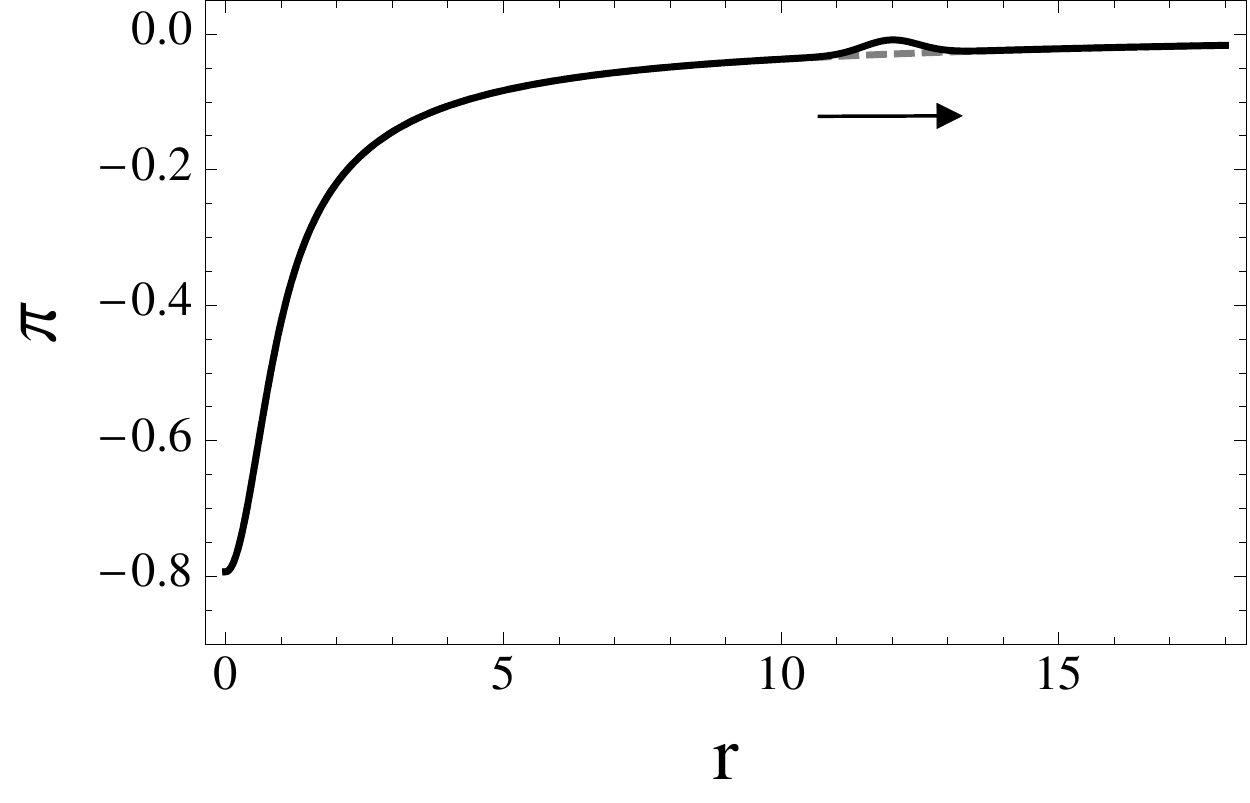}}	
			\end{tabular}
			\caption{The evolution of $\pi(r,t)$ (solid) for an initial condition of the type \eqref{eq:gaussian_initial}
			with $A=0.25,\,\sigma=0.5,\,r_w = 12$. The screening solution $\Pi_+$ is characterized by
			$\rho=100,\,R_0=0.5,\, R_V\simeq 4$. From left to right, top to bottom are snapshots taken at $t=0,\,10,\,15,\,23$.
			The initial gaussian fluctuation is seen to propagate away from the domain at close to the speed of light, leaving behind the static screening solution at very late times.
			}
			\label{fig:sol3snaps}
		\end{center}
\end{figure}
The first family of initial data we consider are spherical shells collapsing on the static screening solutions \eqref{eq:screening}. The source in the equation of motion is given by Eq.~\eqref{eq:source}, with variable $\rho$ and $R_0$. To obtain a screening solution we must ensure that there is a significant hierarchy between $R_0$ and $R_V$. For an object like the sun, with $\rho \simeq 10^{26}$, we have $R_V \sim 10^9 R_0$. Resolving such a hierarchy of scales would be computationally intractable with our fixed grid code; we must consider sources with a much more modest hierarchy. We use $R_0 = .5$ and $\rho=100$, which gives a Vainshtein radius of $R_V=\sqrt{\pi}\rho^{1/3}R_0\sim 8 R_0$. 

The initial data is given by:
%
%
\begin{eqnarray}\label{eq:gaussian_initial}
\pi(r,0) &=& \pi_\pm(r) + \frac{Ar^2}{(r^2+\epsilon^2)^{3/2}} \exp\left(-\frac{(r - r_w)^2}{2\sigma^2} \right)\,, \nn\\
\dot{\pi} (r,0) &=& -\frac{Ar^2(r-r_w)}{\sigma^2(r^2+\epsilon^2)^{3/2}}\exp\left(-\frac{(r - r_w)^2}{2\sigma^2}\right)\,.
\end{eqnarray}
where $A$ and $\sigma$ parametrize the amplitude and width of the shell, localized at $r_w$. The regulator $\epsilon \ll 1$ is to ensure that the field is well defined at $r=0$. Note that this wave packet is purely in-going.

An example of the evolution is shown in Fig.~\ref{fig:sol3snaps}, where the wavepacket is characterized by $A=0.25,\,\sigma=0.5,\,r_w = 12$. The incoming wavepacket perturbs the screening solution $\pi_+$, and then dissipates out of the computational domain leaving behind the original screening solution. Similar behavior was observed when perturbing the negative branch $\pi_-$. Exploring a wide range of $A,\,\sigma$, we find that the screening solution is stable to a range of nonlinear perturbations, although as we explain in more detail in Sec.~\eqref{sub:CB}, large perturbations are problematic.

\subsubsection{From vacuum to screening solutions\label{sec:screen2}}
%
\begin{figure*}[hbt]
	\begin{center}
		\begin{tabular}{cc}
		\epsfig{file=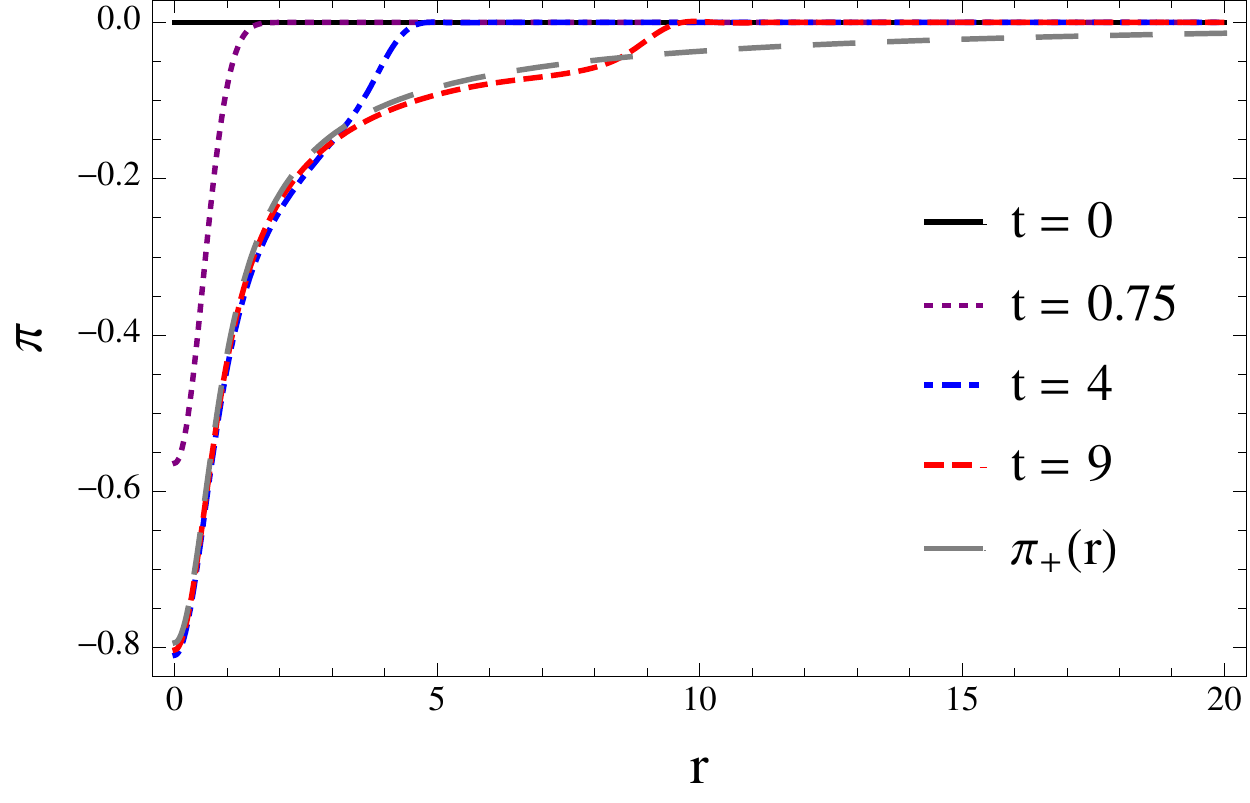,width=8.5cm,angle=0,clip=true}&\epsfig{file=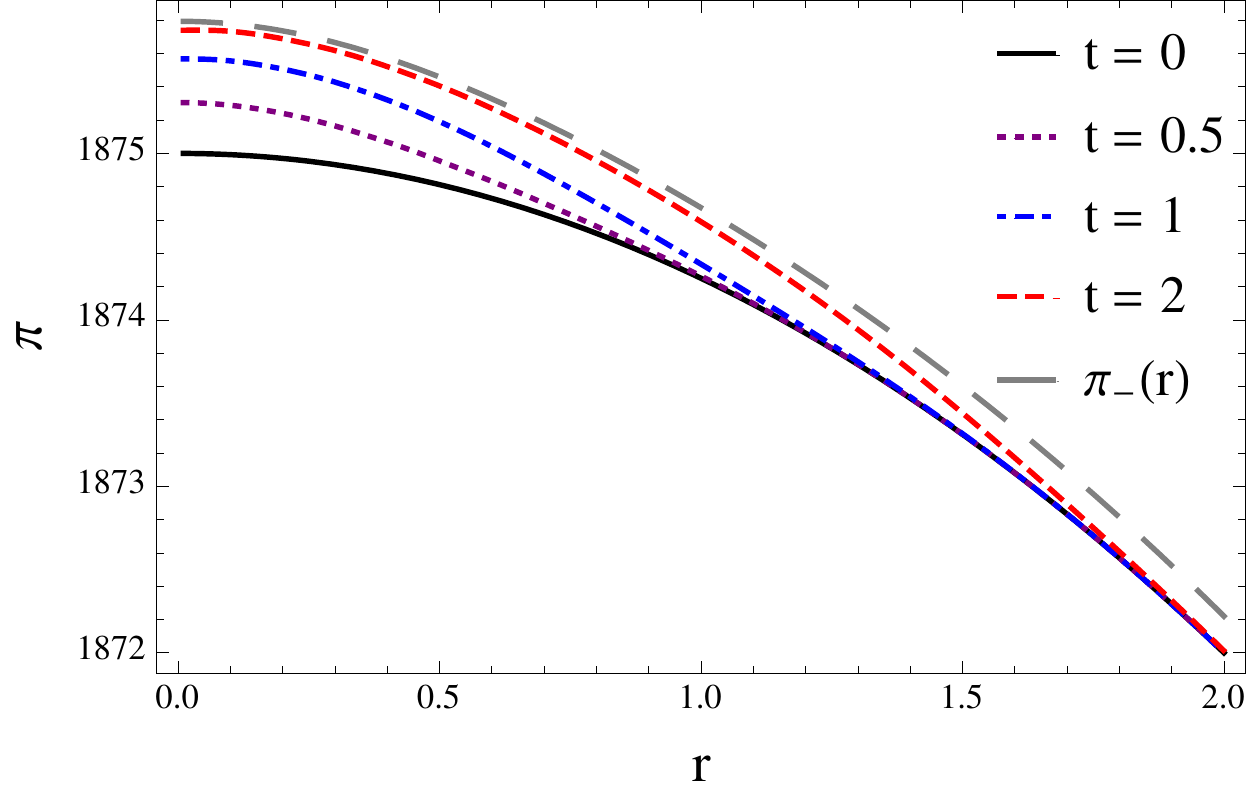,width=8.5cm,angle=0,clip=true}
		\end{tabular}
		\caption{Left panel: The evolution of $\pi(r,t)$ with $\pi(r,0)=0$ and $\rho=100, R_0=0.5$, resulting in a Vainshtein radius $R_V \simeq 4$.
		At late times the solution approaches the static, asymptotically flat screening solution $\pi_+(r)$ corresponding to the same parameters.\\
		Right: The evolution of $\pi(r,0)=-\frac{3}{4}(r^2-r_{0}^2)$ for $\rho=100,\, R_0=0.5,\, r_0=50,\, R_V \simeq 4$. The evolution drives the system towards the screening solution $\pi_-$ at late times.}
		\label{fig:sol1snapsall}
	\end{center}
\end{figure*}

The second class of initial conditions we consider are the two vacuum solutions to Eq.~\eqref{eq:dgpeom}:
\begin{eqnarray}\label{eq:vacuum_ic}
\pi(r,0)&=&0\,, \nn\\
\pi(r,0) &=& -\frac{3}{4}(r^2-r_{0}^2)\,,
\end{eqnarray}
 where $r_0$ is a normalization parameter. In the presence of a source, the vacuum initial conditions will necessarily evolve. A priori there are a number of possible endpoints to this evolution, but the expectation is that the static screening solutions $\pi_\pm$ are reached at late times. In order to determine the endpoint of evolution from both vacuum initial conditions, we perform numerical evolution in the presence of sources with varying $R_V$ and $R_0$.

Examples of time evolution from vacuum initial conditions in the presence of a source with $\rho=100,\,R_0=0.5$ are shown in Fig.~\ref{fig:sol1snapsall}. In these examples, the $\pi_+$ screening solution is reached from $\pi(r,0)=0$ and the $\pi_-$ solution is reached from $\pi(r,0) =-\frac{3}{4}(r^2-r_{0}^2)$. The initial condition was set to match the screening solution at $r=r_{0}=50$, so the evolution is most visible for small $r$ which is why we only plot up to $r=2$. The fact that the expected screening solutions are reached as the endpoint of evolution from vacuum initial conditions displays that the screening solutions are quite robust to large perturbations.

\subsection{Quasinormal modes and tails of screening solutions}\label{sec:QNM_DGP}
%
\begin{figure}
\begin{tabular}{c}
\centerline{\includegraphics[height=6 cm] {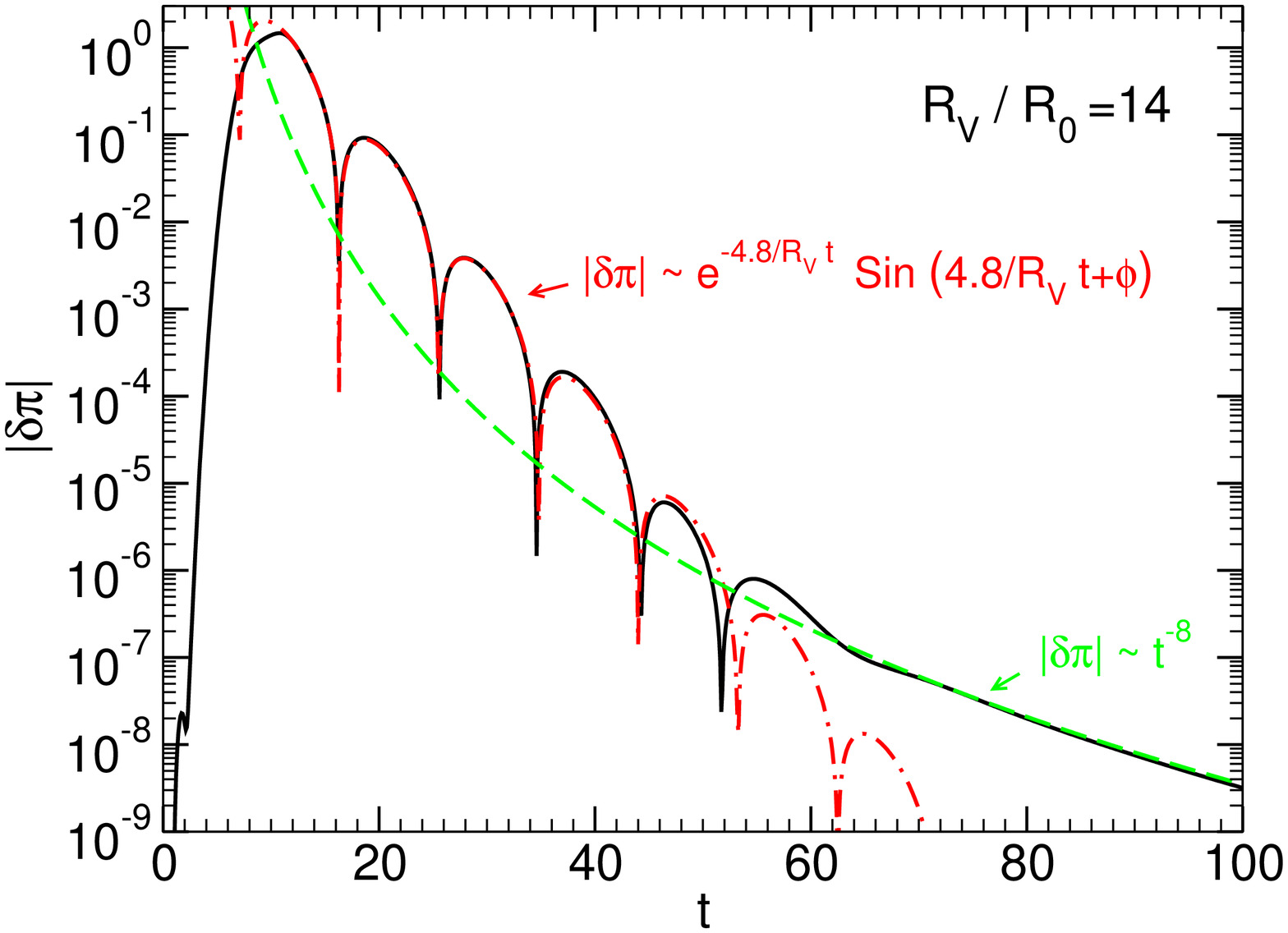}}\\
\centerline{\includegraphics[height=6 cm] {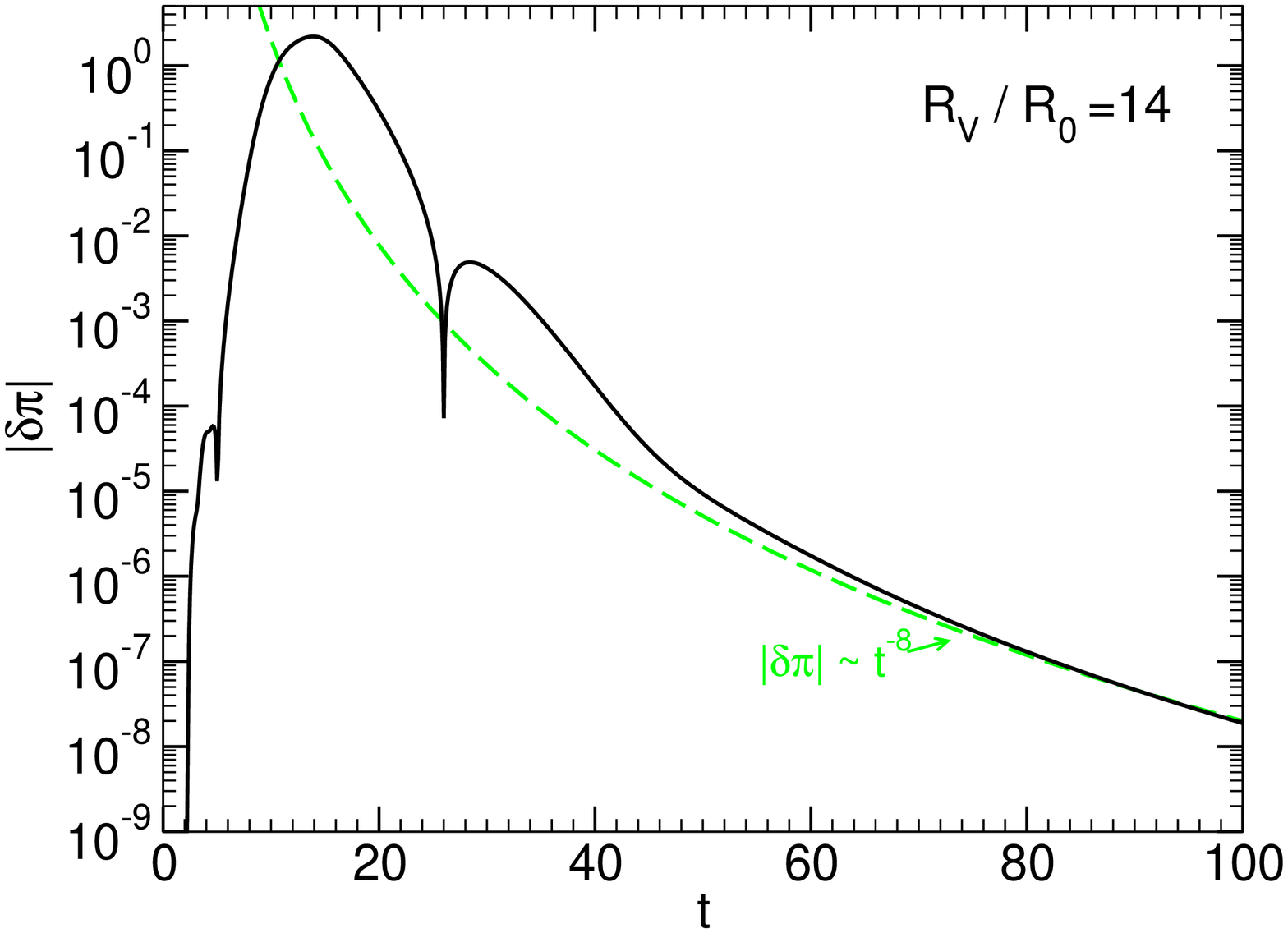}}
\end{tabular}
\caption{Top panel: Linearized time-evolution of a Gaussian wavepacket in the background of $\Pi_+$ with amplitude $A=3$,
width $\sigma=1$ and localized at $r_w=10$. The radius of the source is at $R_0=1$. The waves are
extracted at $r=2$. The intermediate-time behavior consists on an exponentially
damped sinusoid --called quasinormal mode -- and the late-time behavior is described
by a power-law falloff of the field $\delta\pi\sim t^{-8}$. The observed behavior is in agreement with a frequency-domain
numerical and analytical calculation; see text for further details.\\
%
Bottom panel: Same in the background of $\Pi_-$. The time-domain profile suggest that in this case the quality factor of the fundamental quasinormal mode $\omega_R/\omega_I$ is smaller than in the background of $\Pi_+$, i.e. $\omega_R/\omega_I<1$.
\label{fig:osci}}
\end{figure}
The previous subsections focused on completely nonlinear evolution, and suggest that when perturbed, a static screening solution
behaves as a coherent object: it vibrates and eventually settles down to the original static solution.
To understand this behavior more thoroughly, and to understand generic small fluctuations of the screening solutions~\eqref{eq:screening},
we focus now on linearized fluctuations, considering generic perturbations of the form
\be\label{pert}
\pi(t,r,\vartheta,\,\varphi)=\pi_{\pm}(r)+\epsilon \ \delta\pi(t,t)Y_{lm}(\vartheta,\,\varphi)\,,
\ee
where $\pi_\pm$ is the static solution given by~\eqref{eq:screening}, $Y_{lm}(\vartheta,\,\varphi)$ are the usual spherical harmonics and $\epsilon$ is a small bookkeeping parameter. Inserting~\eqref{pert} in equation~\eqref{eq:eom} and linearizing in $\epsilon$ we find the equation for $\delta\pi$,
\begin{equation}\label{dgp_pert}
Z^{tt}\ddot{\delta\pi} +Z^{rr}\delta\pi'' + 2 r Z^{\theta\theta}\delta\pi'-l(l+1)Z^{\theta\theta}\delta\pi=0\,,
\end{equation}
where the coefficients $Z^{tt}, Z^{rr}$ and $Z^{\theta\theta}$ are those in equation \eqref{eq:Zs}.

We evolved Eq.~\eqref{dgp_pert} in time considering an initial Gaussian wave-packet $\dot{\delta\pi}=\frac{A}{r} e^{(r-r_w)^2/2\sigma^2}$. A typical waveform is shown in Fig.~\ref{fig:osci} for $R_0=1$ and $\rho=500$. This plot shows the value of the field as a function of time at a specific position $r$, and has the same form regardless of the position. The waveform consists of three stages, familiar in the context of wave propagation in curved spacetimes~\cite{Berti:2009kk,Barausse:2014tra}: a prompt response at very early times, which depends on the details of the initial data and is the analogue of on-light-cone propagation in flat spacetime; at intermediate times the signal consists of a series of exponentially damped sinusoids, termed quasinormal modes~\cite{Berti:2009kk} which correspond to the ``characteristic modes''
of the vibrating object. In this case the vibrating object is the static screening solution, and the fluctuations are damped because the system is intrinsically dissipative: energy flows to infinity. This stage is universal and independent of the details
of the initial conditions. Finally, a power-law tail sets in at very late times caused by backscattering off the scalar profile (in complete analogy with backscattering due to spacetime curvature~\cite{Ching:1994bd,Ching:1995tj}). Comparison against the full nonlinear evolution confirmed this typical behavior.

\subsubsection{Quasinormal modes}
To quantify the three stages of evolution, it is useful to recast the evolution equation as a Schrodinger-type equation in the frequency domain.
\begin{figure}
\centerline{\includegraphics[height=5.5 cm] {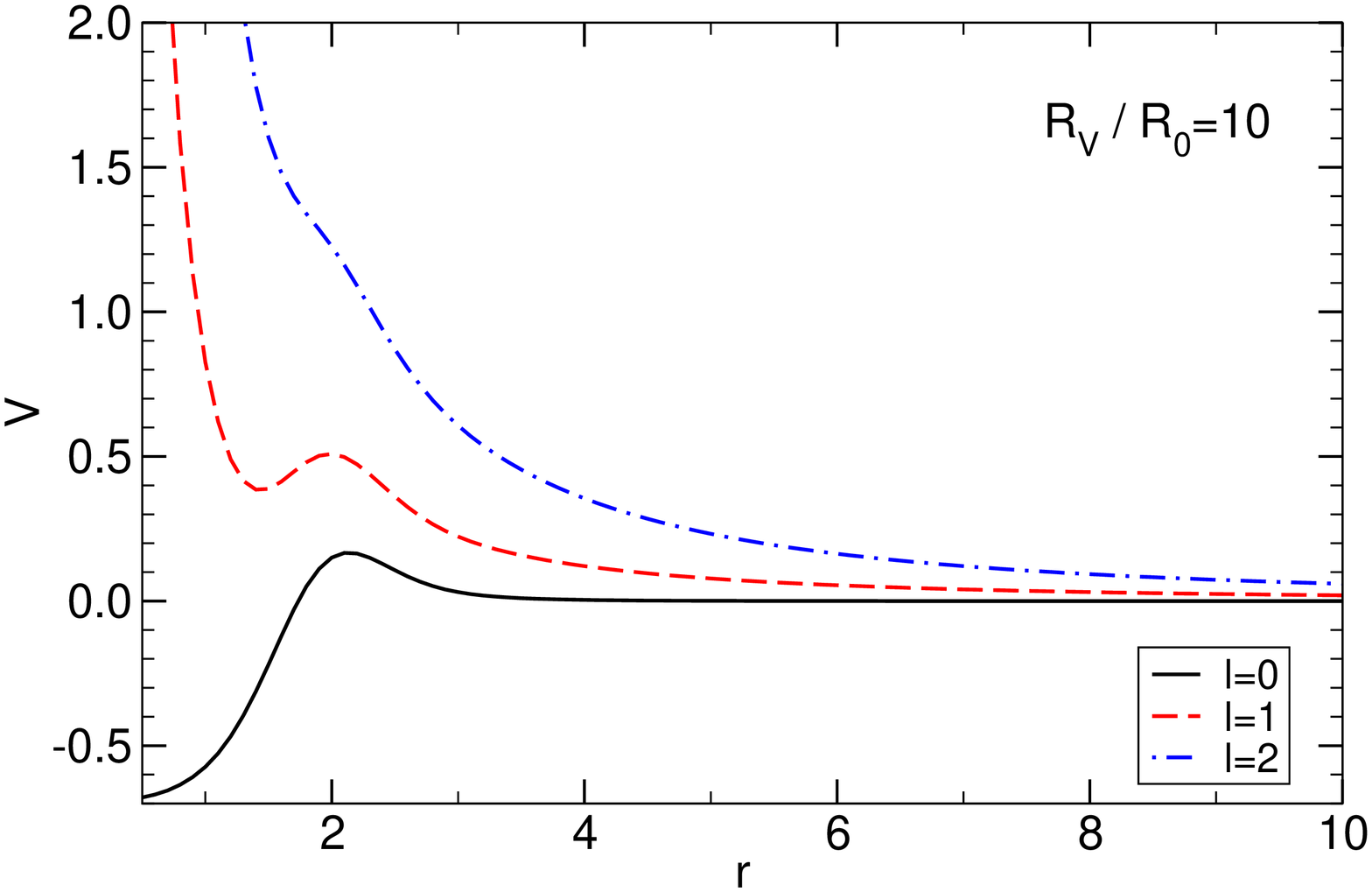}}
\caption{Effective potential in the background of $\Pi_+$ for $R_0=1,\,R_V=10$ 
and different multipoles $l$. In the background of $\Pi_-$ the potential is qualitatively similar.
\label{fig:potential}}
\end{figure}
Defining
\be
\psi(t,r)=\delta\pi(t,r)\frac{r}{\sqrt{2}}\left(-Z^{tt}Z^{rr}\right)^{1/4}\,,
\ee
and a new coordinate $r_*$ by
\be
\frac{dr}{dr_*}=f(r)\equiv \sqrt{-\frac{Z^{rr}}{Z^{tt}}}=\sqrt{\frac{6+8\Pi/r}{6+8\Pi/r+4\Pi'}}\,.
\ee
we can rewrite Eq.~\eqref{dgp_pert} as a wave equation of the form
\be\label{dgp_potential}
\left[\partial_{r_*}^2-\partial_t^2-V(r)\right]\psi(t,r_*)=0\,,
\ee
where the effective potential $V$ is given by
\beq\label{potential_linear}
&&V=-\frac{r^2Z^{\theta\theta}}{Z^{tt}}\frac{l(l+1)}{r^2}+\frac{4Z_2+3f^4(Z^{rr}-2)(Z^{rr}+6)}{4r^2f^2Z_2}\nn\\
&&+\frac{f^2Z_2(r^2(f')^2-8)-8rff'Z_2-2r^2f^3Z_2f''}{4r^2f^2Z_2}\,.
\eeq
Here, primes stand for radial derivatives and we defined $Z_2\equiv (Z^{rr})^2$. The effective potential $V(r)$ in the background of $\Pi_+$ is plotted in Fig.~\ref{fig:potential}
for the monopole, dipole and quadrupole components ($l=0,1,2$). The shape is familiar from studies of wave quasinormal modes and tails around black holes and neutron stars~\cite{Kokkotas:1999bd,Berti:2009kk,Cardoso:2008bp}. The local maximum indicates that the effective metric $Z^{\mu\nu}$ allows for unstable null circular geodesics, with the instability timescale dictating the lifetime of fluctuations. In analogy with the gravitational cases, we expect the screening solutions to support quasinormal modes. These modes can be understood as the scalar modes of vibration of a coherent object. Unlike normal modes, they decay in time due to dissipation, where in this case dissipation occurs due to the leakage of energy to infinity (see Refs.~\cite{Kokkotas:1999bd,Berti:2009kk,Konoplya:2011qq} for reviews).

To perform a quantitative analysis, it is convenient to go to fourier space:
\be
\left(\delta\pi(r,t),\psi(t,r_*)\right)=\left(\Delta\pi(r,\omega),\Psi(r,\omega)\right)e^{-i\omega t}\,.
\ee
The equation of motion \eqref{dgp_pert} and \eqref{dgp_potential} are written as
\begin{align}
&Z^{rr}\Delta\pi''+ 2 r Z^{\theta\theta}\Delta\pi'-\left(\omega^2Z^{tt}+l(l+1)Z^{\theta\theta}\right)\Delta\pi=0\,,\label{dgp_pert2}\\
&\left[\partial_{r_*}^2+\omega^2-V(r)\right]\Psi(r,\omega)=0\,.\label{dgp_potential2}
\end{align}
At the origin the equation of motion admits the behavior
\be
\label{ser_origin}
\Delta\pi(r,\omega),\Psi(r,\omega)/r\sim A_1 r^{l}+A_2 r^{-(l+1)}\,.
\ee
%
%
%
Regularity of the field and its derivatives requires that $A_2=0$. At infinity one has 
\be
\label{ser_inf}
r\Delta\pi(r,\omega),\Psi(r,\omega)\sim B_1 e^{ik_{\infty} r}+B_2 e^{-ik_{\infty} r}\,,
\ee
where $k_{\infty}=\omega$ in the background of $\Pi_+$ and $k_{\infty}=\sqrt{2}\omega$ in the background of $\Pi_-$. Requiring that the system is otherwise isolated is tantamount to demanding Sommerfeld outgoing boundary conditions, $B_2=0$.

\begin{figure}
\centerline{\includegraphics[height=6 cm] {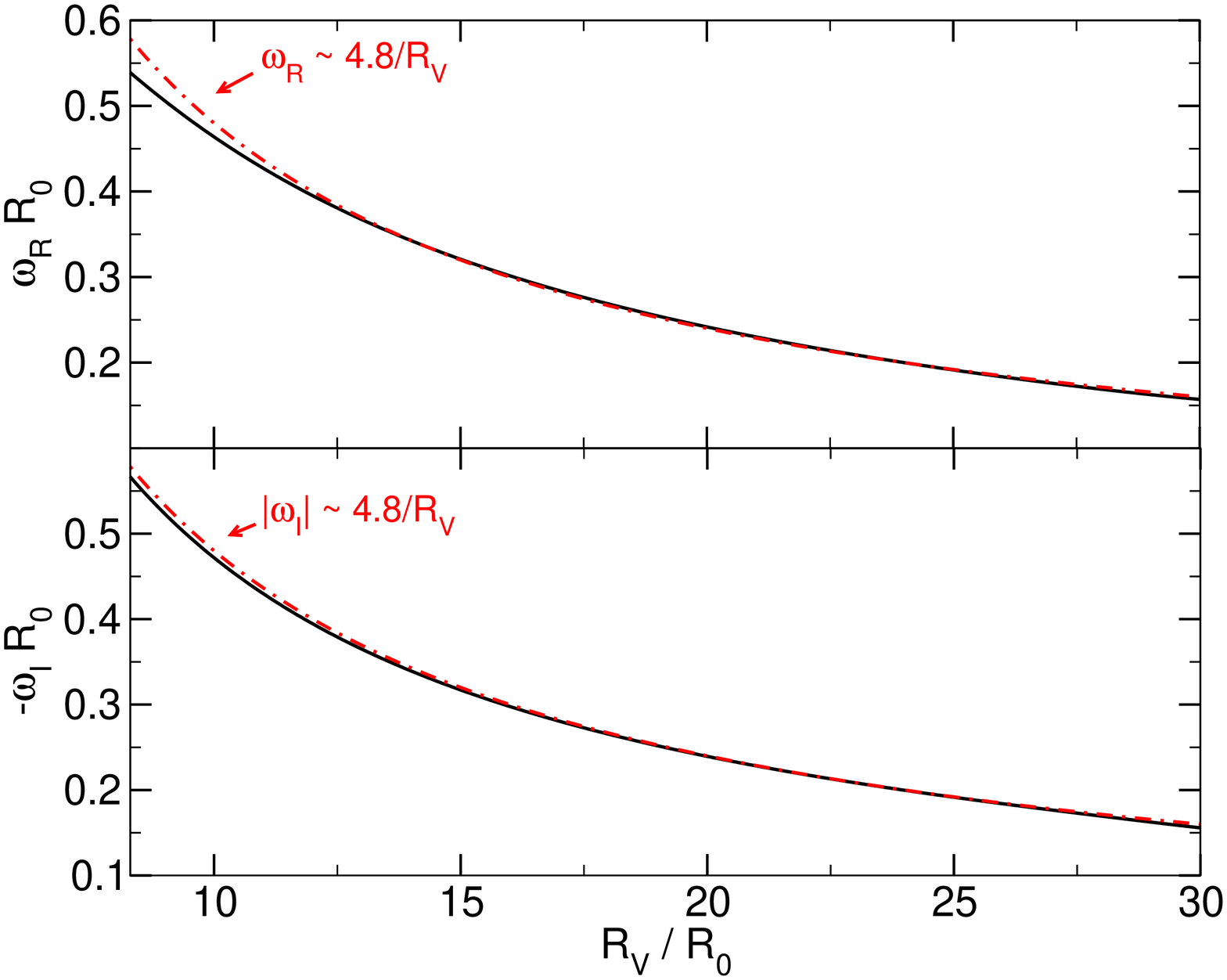}}
\caption{Fundamental quasinormal modes of the scalar field in the background $\Pi_+$. The full lines correspond to the numerical results, whereas the dashed lines show the analytic approximation at low frequencies. The top and bottom panels show
the real part, $\omega_R R_0$, and the imaginary part, $\omega_I R_0$, of the mode as a function of the Vainsthein radius $R_V/R_0$.
\label{fig:QNM}}
\end{figure}
With the above two boundary conditions, Eqs.~\eqref{dgp_pert2} or \eqref{dgp_potential2} define an eigenvalue problem for the (generically complex)
quasinormal frequency $\omega=\omega_R+i\omega_I$. To compute the eigenfrequencies we use a direct integration approach described in Refs.~\cite{Chandrasekhar:1975zza,Pani:2013pma}. Notebooks are available online~\cite{webpage}. We integrate from each of the boundaries towards a matching point where the wavefunction and its radial derivative are required to be continuous. For generic $\omega$ the continuity conditions are not satisfied, unless $\omega$ is an eigenfrequency. One can then find the eigenfrequencies using a standard shooting method. 
The eigenfrequencies are typically ordered by increasing (absolute value of) imaginary part, the fundamental mode being the largest and longest-lived. The fundamental overtone for $l=0$ is shown in Fig.~\ref{fig:QNM} as a function of $R_V/R_0$, and is well approximated by
\beq
\omega_R&\sim& \frac{4.8}{R_V}\label{wreal}\,,\\
\omega_I&\sim& -\frac{4.8}{R_V}\label{wimaginary}\,. 
\eeq
This scaling is generic and does not depend on the details of the source. The frequency domain and time domain analysis agree extremely well with each other and with the non-linear results, as summarized in Fig.~\ref{fig:osci}. In the background of $\Pi_-$ the time-domain profile suggest that the quality factor $\omega_R/\omega_I$ is smaller that in the background of $\Pi_+$, that is $\omega_R/\omega_I<1$. The method used in the frequency-domain works well when $\omega_R\gtrsim \omega_I$, thus in the background of $\Pi_-$ we have not been able to compute the quasinormal frequencies accurately. Nevertheless we expect the scaling to follow closely Eqs.~\eqref{wreal} and~\eqref{wimaginary}.  

These results are very general; perturbed screening solutions will ring, and the response will be dominated by its lowest
quasinormal modes. In the present setting, gravitational degrees of freedom of the source are frozen. Once they are allowed dynamics,
and second quasinormal mode stage will appear, corresponding to the oscillation of the source itself~\cite{Barausse:2014tra}. Source dynamics are presumably already well understood in GR, the ringdown stage we described is new and can be assigned entirely to the large-scale scalar field screening solution. The time scale of the ring-down is proportional to the Vainshtein radius and we explicitly checked that this scaling is independent on the source functional form, showing that the static solutions behave as large-scale objects localized at $R_V$.

\subsubsection{Late-time power-law tails}
A thorough study of the late-time behavior of equations of the form \eqref{dgp_potential2} was performed in Ref.~\cite{Ching:1994bd,Ching:1995tj}.
A late-time power-law tail of the form $\psi\sim t^{-\beta}$ is caused by back-scattering off the (efective) spacetime curvature {\it at large distances} (mathematically this is due to a branch cut in the Green's function) and has the form
\begin{equation}
\lim_{t\to \infty}\psi(r_*,t)=t^{-2l-\alpha}\,,\quad {\rm for}\,\,\lim_{r_*\to \infty}V=\frac{l(l+1)}{r_{*}^2}+ \frac{K}{r_{*}^{\alpha}}\,,
\end{equation}
with $l$ an integer and $K$ is a constant that depends on $l$ and $M(r\to\infty)$. In the background of $\Pi_\pm$ the effective potential~\eqref{potential_linear} has the large distance asymptotic behavior
%
%
\be
V\sim\frac{l(l+1)}{r_*^2}+\frac{K}{r_*^8}\,.
\ee
%
%
%
%
Thus, spherically symmetric fluctuations ($l=0$) are expected to decay as $t^{-8}$ at very late times for both the $\Pi_+$ and $\Pi_-$ background solutions~\footnote{For the special case of static initial data the power changes to $t^{-9}$.}. Such decay is consistent with our findings in Fig.~\ref{fig:osci}. For higher multipoles, the above analysis predicts a decay $\psi(r_*,t)\sim t^{-2l-8}$ at late times.

Note that the features we described are generic to any kind of scalar field with non-linear kinetic terms. In particular, our analysis imply that perturbations of galileons~\cite{Nicolis:2009aa} will display similar late-time behavior.

\section{Cauchy Breakdown}
\label{sub:CB}

We have shown in the previous sections that the static screening solutions are stable against a variety of fluctuations. In this section we want to give a quantitative measure of how robust the are solutions against \emph{large} perturbations. 

In theories described by Eq.~\eqref{Lan_general}, scalar fluctuations propagate in an effective metric $Z^{\mu\nu}$. As shown in  Appendix~\ref{app:IVP}, for the evolution problem to be well posed (i.e., that the solution is unique and depends continuously on the initial data), the initial data must be such that the effective metric $Z^{\mu\nu}$ has a Lorentzian signature, e.g., $\det(Z^{\mu\nu})<0$ everywhere in space, and surfaces of constant time are required to be spacelike, $Z^{tt}<0$. 




However, even within this restricted class of initial conditions, problems can still arise because of the non-linearity of the equations. Since the spacetime metric is in general different from the effective metric $Z^{\mu\nu}$, and it is possible that $Z^{tt}\to 0$ in the absence of any other pathologies like singularities or horizon formation (in fact we will mostly deal with a flat spacetime metric). For DGP the relevant components of the effective metric for a time-dependent background are given by
\begin{align}\label{eq:Zs_time}
	&Z^{tt} = -\left( 6 + 8 \frac{\pi'}{r} + 4 \pi'' \right)\,,\quad Z^{rr} = 6 + 8 \frac{\pi'}{r}-4\ddot{\pi}\,, \nn\\
	&Z^{tr} = 4\dot{\pi}'\,, \quad r^2 Z^{\theta\theta}=6 + 4 \frac{\pi'}{r} + 4 \pi''-4\ddot{\pi}\,.
\end{align} 
If at any point in spacetime $Z^{tt}\to 0$, the Cauchy problem breaks down because the surfaces of constant time become null with respect to the effective metric, i.e., $Z^{\mu\nu}\partial_{\mu}t\partial_{\nu}t\to 0$. When this happens, the numerical evolution ceases to be possible past this point, and it is possible that the theory itself ceases to be well defined. Similar issues have been reported recently in the context of $k$-essence models~\cite{Akhoury:2011hr,Akhoury:2011rc,Leonard:2011ce}.
We refer to this issue ($Z^{tt}=0$) as Cauchy breakdown, following earlier nomenclature~\cite{Leonard:2011ce}. Substituting the static screening solutions into $Z^{tt}$ gives that $Z^{tt}<-6$ for $\Pi_-$ and $Z^{tt}>12$ for $\Pi_+$, so Cauchy breakdown is not an issue initially. However, if we perturb the static screening solution, $Z^{tt}$ could pass through zero at a finite $r$ and $t$.

Besides the issue of Cauchy breakdown, we might also expect that in some situations, regions can form where $Z^{rr}\to 0$, $Z^{\theta\theta}\to 0$ or any of the eigenvalues of the matrix $Z^{\mu\nu}$ given by Eq.~\eqref{eigenvalues} cross zero. For time-dependent backgrounds we have $Z^{tr}\neq 0$, so regions where $Z^{rr}\to 0$ are not physical singularities but are rather regions where a sound horizon forms. This is a typical feature of non-linear fields with non-linear kinetic terms~\cite{Babichev:2006vx,Babichev:2010kj,Akhoury:2011hr,Akhoury:2011rc,Leonard:2011ce} and it is to be expected in regions where the propagation speed of the fluctuations is much smaller than the propagation speed of the background. However, as discussed in Appendix~\ref{app:IVP}, regions where $Z^{\theta\theta}$ (or any other eigenvalue of $Z^{\mu\nu}$) switch sign, are prone to instabilities. The timescale of these instabilities generically scales with  $\tau\sim \Lambda^{-1}$. In a dynamical setup local instabilities can arise for a small amount of time, $t_{\rm din}$, and in small regions in space. These instabilities are not necessarily catastrophic as long as the instability timescale $\tau\gtrsim t_{\rm din}$. Furthermore, looking at~\eqref{eq:Zs_time} we expect that if any unstable region forms for long times, the development of this instability will make derivatives of the fields grow and will likely be followed by Cauchy breakdown.  

Cauchy breakdown occurs in a wide variety of scenarios, which we can study using the numerical methods described in Sec.~\ref{sec:screeningstability}. For the class of initial data described in Eq.~\eqref{eq:gaussian_initial}, Cauchy breakdown occurs at fixed $\sigma$ for a sufficiently large amplitude $A$. Using this set of initial data for the $\Pi_+$ branch, neither sound horizon regions nor unstable regions form. On the other hand for $\Pi_-$, sound horizons and unstable regions can form for a finite time before Cauchy breakdown. If the fluctuations are sufficiently small these regions eventually disappear when the wave dissipates to infinity with a timescale smaller than the instability timescale. But if the fluctuations are sufficiently large, following the onset of these instabilities, Cauchy breakdown will always occur. 

\begin{figure}[hbt]
	\begin{center}
		\begin{tabular}{c}
		\epsfig{file=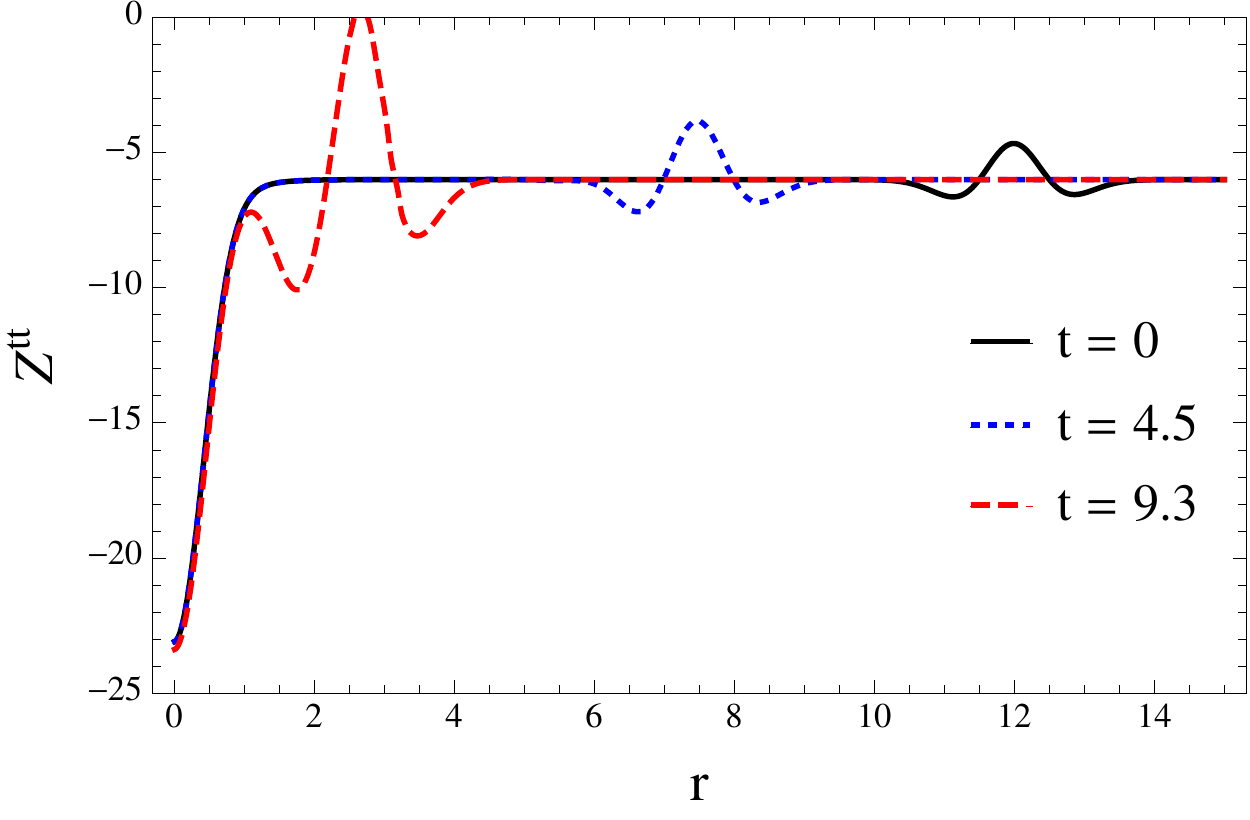,width=8.5cm,angle=0,clip=true}
		\end{tabular}
		\caption{The evolution of $Z^{tt}$ given in \eqref{eq:Zs} to Cauchy breakdown, with the initial condition \eqref{eq:gaussian_initial} with $A=1,\,\sigma=0.5,\,r_w=12$. The source is characterized by $\rho=100,\,R_0=0.5,\, R_V\approx 4$.}
		\label{fig:CB2snapsall}
	\end{center}
\end{figure}

In Fig.~\ref{fig:CB2snapsall}, we show the evolution of $Z^{tt}$ for a wave packet with $A=1,\,\sigma=0.5,\,r_w=12$; evolution cannot proceed past $t \sim 9.8$, where $Z^{tt} \sim 0$.

\begin{figure*}[htb!]
\begin{center}
\begin{tabular}{cc}
\epsfig{file=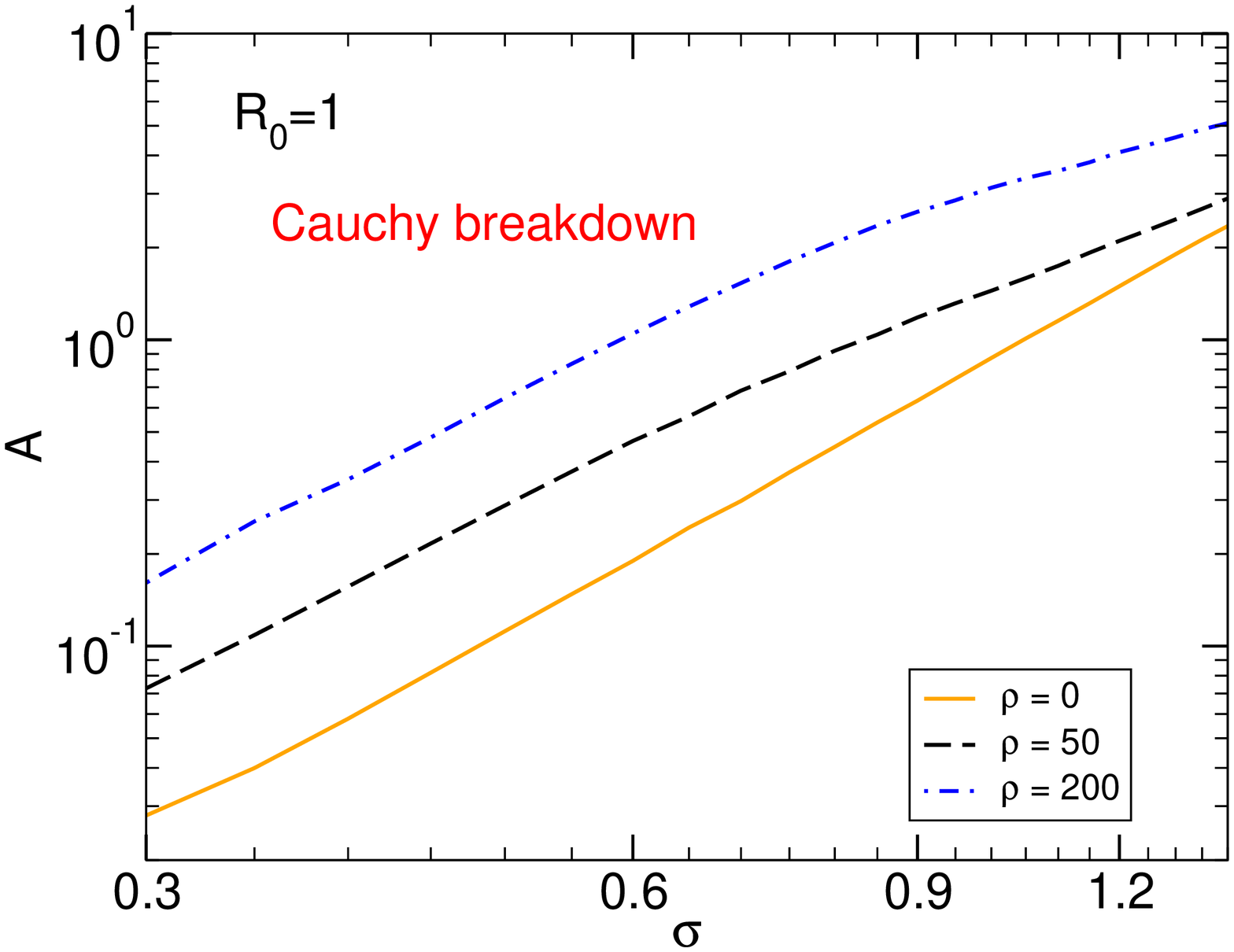,width=8.1cm,angle=0,clip=true}&
\epsfig{file=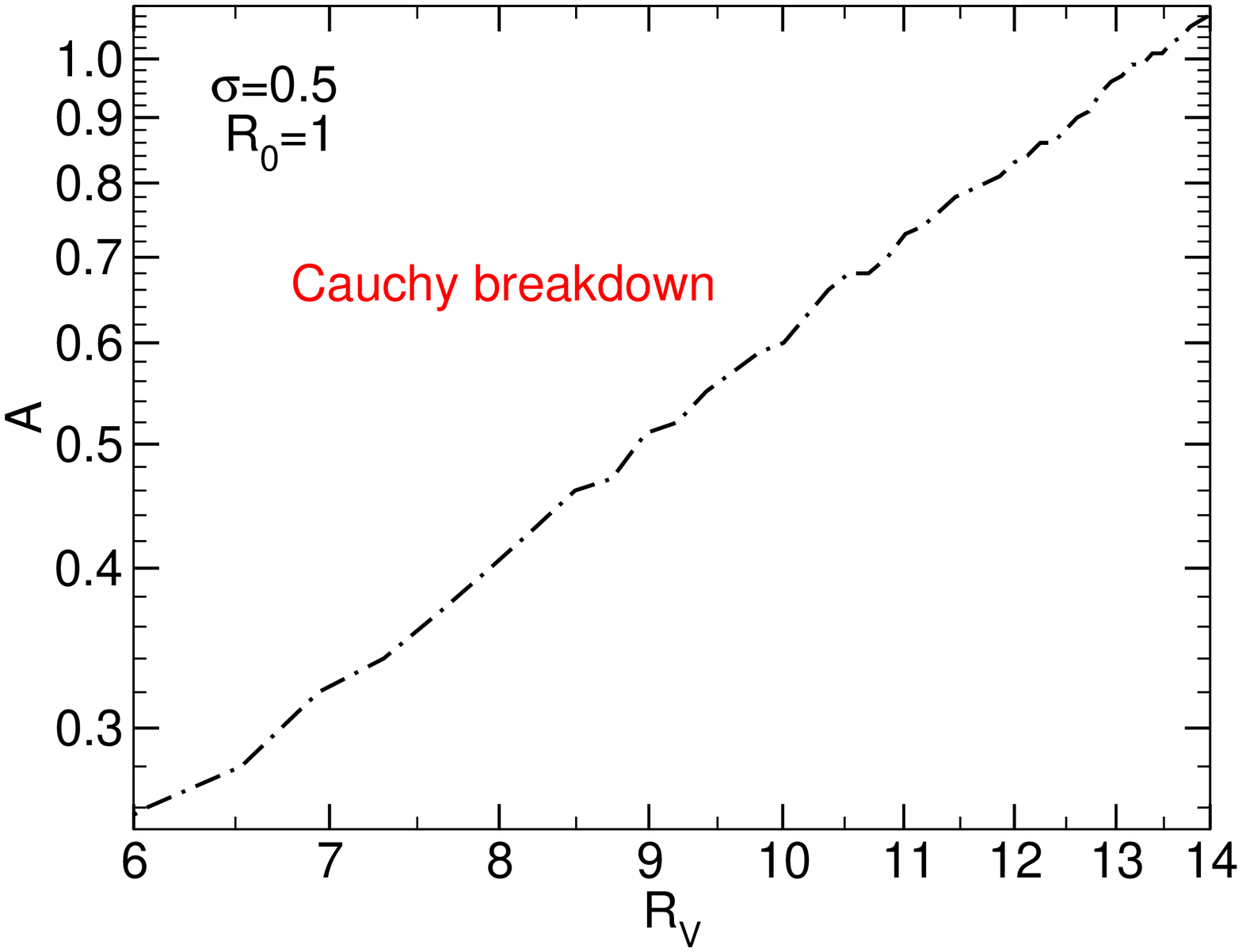,width=8.1cm,angle=0,clip=true}
\end{tabular}
\caption{Log-Log plots of the region in parameter space of initial conditions for which Cauchy breakdown occurs. Breakdown occurs above each of the lines shown, which correspond to the critical amplitude $A_{\rm crit}$.  Recall that the Vainshtein radius is defined by $R_V=\sqrt{\pi}\rho^{1/3}R_0$.}
\label{fig:cauchy}
\end{center}
\end{figure*}

To understand the set of initial conditions described by Eq.~\eqref{eq:gaussian_initial} that lead to well-defined time-evolution, we have performed an extensive search for Cauchy-breakdown in the $\Pi_+$ branch (results are qualitatively similar for the $\Pi_-$ branch). Our results are summarized in Fig.~\ref{fig:cauchy}. In the left panel, we fix the source properties to be $R_0=1$ and $\rho=50, 200$ and study wavepackets with varying width. We also compare the results to the case where there is no source. The algorithm we used to find the critical amplitudes at which Cauchy-breakdown occurs is the following: starting from  $\sigma_0=0.3$ and $A_{\rm init}=0.01$, we increase $A$ by steps of $\delta A=0.01$ until we locate the critical amplitude $A_{\rm crit}$ at which $Z^{tt}\to 0$ is reached somewhere during the evolution (the time evolution terminates when this occurs). For amplitudes above this critical threshold, Cauchy breakdown occurs. When this happens we break the loop, starting a new loop for $\sigma_1=\sigma_0+\delta\sigma$ (we used $\delta\sigma=0.05$), and using $A_{\rm crit}$ as the new $A_{\rm init}$ for this loop. We implemented this algorithm in the range $0.3 \leq \sigma \leq 1.5$. In the right panel, the width of the wave packet is fixed at $\sigma = 0.5$ and the size of the source fixed to be $R_0=1$. We then locate the critical amplitude at which Cauchy breakdown occurs for varying Vainshtein radius, or equivalently, varying central source densities $\rho$. We have explicitly checked that results for all simulations are independent of $r_w$, the initial pulse location, when $r_w\gtrsim R_V$.

Examining the left panel of Fig.~\ref{fig:cauchy}, there are several clear trends. In vacuum the critical amplitude is fairly well fit by a power law with $A_{\rm crit}\propto \sigma^{3}$. On the other hand, for both source densities, the critical amplitude is fitted by a broken power law with $A_{\rm crit}\propto \sigma^{3}$ at small widths and $A_{\rm crit}\propto\sigma^2$ at widths $\mathcal{O}(1)$ and larger. Comparing the two densities we have sampled, we can also conclude that higher density sources are more robust to Cauchy breakdown for a fixed perturbation amplitude. Since the perturbation is riding on a larger background screening solution, larger amplitudes are necessary to drive $Z_{tt} \rightarrow 0$. From the right panel of Fig.~\ref{fig:cauchy}, this increase in $A_{\rm crit}$ appears to scale roughly with $R_V^{3/2}$ (in terms of density, $A_{\rm crit}$ therefore scales like $\rho^{1/2}$).

\begin{figure}[hbt]
	\begin{center}
		\begin{tabular}{c}
		\epsfig{file=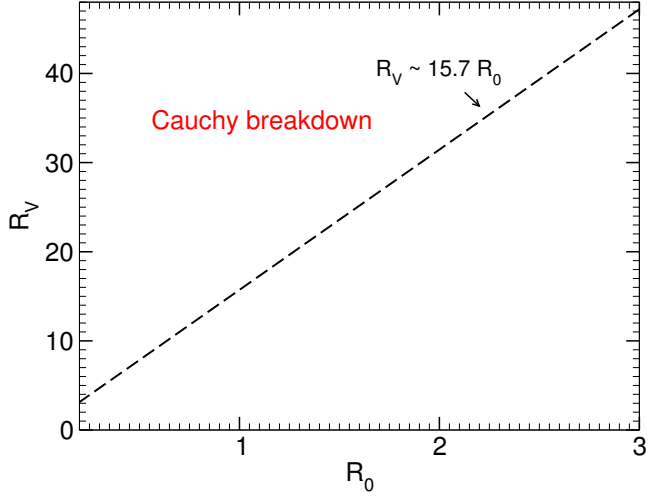,width=8.5cm,angle=0,clip=true}
		\end{tabular}
		\caption{For the initial condition $\pi(r,0)=0$, Cauchy breakdown occurs in the region above the curve. The trend shows that above $R_V/R_0\sim 15.7$ there is Cauchy breakdown.}
		\label{fig:CBtest1}
	\end{center}
\end{figure}

For vacuum initial conditions Eq.~\eqref{eq:vacuum_ic}, we also find Cauchy breakdown, which is normally preceded by the formation of a sound horizon and an unstable region for both branches. To investigate the types of sources for which this occurs, we have simulated evolution in the presence of sources with radii between $0.1 \leq R_0 \leq 3$ and Vainshtein radii between $1 \leq R_V \leq 50$. For sufficiently small central densities, a wavepacket forms, taking the initial conditions to the final screening solution. Additional fluctuations are dissipated out of the computational domain, leaving the screening solution. For $R_V$, or equivalently $\rho$, larger than a critical threshold, the wavepacket overshoots the screening solution and the $Z^{tt}$ factor to pass through zero causing Cauchy breakdown. We characterize the parameter space leading to Cauchy breakdown in Fig.~\ref{fig:CBtest1}. For $\pi(r,0)=0$ we find that for values above $R_V/R_0 \sim 15.7$ there is Cauchy breakdown. We obtain a qualitatively similar result for quadratic vacuum initial conditions.

Based on these results, we see that the Vainshtein screening solutions in DGP are dynamically stable to a wide variety of perturbations. In all cases, as long as Cauchy breakdown is avoided, the screening solution is approached at late times. Sources with large central densities (and correspondingly large hierarchies between the source size and Vainshtein radius) are more robust to perturbations. A screened object is therefore most vulnerable when it is in a low density state -- first starting to collapse. In general, the presence of sources makes the theory less susceptible to Cauchy breakdown, but there is nevertheless a restriction on initial data that leads to well-posed evolution.

\subsection{Coordinate invariance of Cauchy breakdown}

Cauchy breakdown could be either a point where the theory breaks down~\cite{ge:2Burra011cr} or an artificial problem due to the way we choose to slice the spacetime. In fact, locally, $Z^{\mu\nu}$ can always be brought to the Minkowski form by the appropriate coordinate transformation, as long as the hyperbolicity condition, $\det(Z^{\mu\nu})<0$, is met. However, since the matter fields evolve in the spacetime metric, we have to consider also the dynamics in the metric $\eta_{\mu\nu}$. To have a well-posed Cauchy problem we must have a common family of Cauchy surfaces with respect to $\eta_{\mu\nu}$ and the effective spacetime metric $Z^{-1}_{\mu\nu}$~\cite{Babichev:2007dw}, where $Z^{-1}_{\mu\nu}$ is the inverse of $Z^{\mu\nu}$, i.e., $Z^{-1}_{\mu\nu}Z^{\mu\nu}=\delta_{\mu}^{\nu}$. If $\det\left(Z^{\mu\nu}\right)\neq 0$, for spherically spacetimes we have
\beq\label{inverse}
&Z^{-1}_{tt}=\frac{Z^{rr}}{Z^{tt}Z^{rr}-(Z^{tr})^2}\,,\quad\,
&Z^{-1}_{rr}=\frac{Z^{tt}}{Z^{tt}Z^{rr}-(Z^{tr})^2}\,,\nn\\
&Z^{-1}_{tr}=\frac{Z^{tr}}{(Z^{tr})^2-Z^{tt}Z^{rr}}\,,\quad\,
&Z^{-1}_{\theta\theta}=\frac{1}{Z^{\theta\theta}}\,.
\eeq

Consider a general spacelike hypersurface, with respect to Minkowski spacetime, with unit normal $n^\mu$ such that $\eta_{\mu \nu} n^\mu n^\nu = -1$. Working in spherical symmetry, and only considering general coordinate transformations of $r$ and $t$, the most general unit normal is:
\be
n^\mu = (A(r,t), \sqrt{A(r,t)^2 - 1}, 0, 0 )\,,
\ee
for an arbitrary spacetime function $|A|>1$. We want to know if there is any choice of A which keeps the unit normal timelike with respect to the effective spacetime $Z^{-1}_{\mu\nu}$:
\be
Z^{-1}_{\mu\nu} n^\mu n^\nu < 0\,.
\ee
Expanding out we obtain,
\be
Z^{-1}_{tt}A^2 + 2 Z^{-1}_{tr} A\sqrt{A^2 -1} + Z^{-1}_{rr}(A^{2}-1) < 0\,.
\ee
In particular, when Cauchy breakdown occurs, we have $Z^{-1}_{rr}=\frac{Z^{tt}}{Z^{tt}Z^{rr}-(Z^{tr})^2}=0$, so the above inequality simplifies to:
\be
Z^{-1}_{tt} < -2 Z^{-1}_{tr} \sqrt{1 -A^{-2}}\,.
\ee
In the limit of large $A$, we have:
\be
Z^{-1}_{tt} < -2 Z^{-1}_{tr}\,.
\ee
In the limit of $A = 1 + \epsilon$ with $\epsilon \ll 1$, we have:
\be
Z^{-1}_{tt}<-\sqrt{2\epsilon}Z^{-1}_{tr}
\ee
The stronger condition is the first one. Using Eq.~\eqref{inverse} we conclude that it is possible to find a common spacelike surface in both the flat and effective spacetimes only when:
\be\label{cb_slice}
Z^{rr} < 2Z^{tr}\,.
\ee
For all cases we studied we found that this is never satisfied when Cauchy breakdown occurs. To understand this consider the DGP model with the effective metric given by \eqref{eq:Zs_time}. Using $Z^{tt}=0$ and~\eqref{eq:Zs_time}, the condition~\eqref{cb_slice} reads
\be\label{cb_slice_2}
-4(\pi''+2\dot{\pi}'+\ddot{\pi})<0\,.
\ee
Cauchy breakdown generically occurs inside regions where gradients become large and do not have a definite sign. For example, for large fluctuations of the background static solutions, they occur at the peak of the traveling wave packet (see Fig.~\ref{fig:CB2snapsall}). Approximating the wave packet by a Gaussian of the form~\eqref{eq:gaussian_initial}, we see that second derivatives are all negative at the peak of the Gaussian. Thus Eq.~\eqref{cb_slice_2} is not satisfied there. This means that, for the cases we considered, Cauchy breakdown is a real physical problem and not simply an artificial coordinate singularity.

As a final remark, notice that $Z^{tt}$ changes sign between $\Pi_+$ and $\Pi_-$, which is a result of the fact that $\Pi_+>0$ and $\Pi_-<0$. This means that there is no way for time evolution to connect one branch of solutions to the other without going through a region where Cauchy breakdown occurs. In addition, it is impossible to construct a spacetime containing local regions of each branch. For example, one cannot match the negative branch on $r<r_*$ onto the asymptotically flat solution for $r>r_*$ since $Z^{tt}$ would have to cross zero. An indication that such solutions do not exist (beyond the decoupling limit) has been presented in Ref.~\cite{Izumi:2007gs} (for similar considerations in dRGT massive gravity see Ref.~\cite{Park:2012cq}).

\section{Collapsing and Exploding Sources\label{sec:collapse}}

In this section we consider a dynamical source $T(r,t)$ in order to model astrophysical phenomena where the source undergoes gravitational collapse into a relativistic object (e.g. a neutron star) or explodes (e.g. as in a supernova). Our models for the source are not physical in the sense that there is no underlying model, but rather are intended to give qualitative information on the possible evolution of $\pi$. 

The first example we consider is a \emph{collapsing} source. For simplicity, we assume that the relative contribution from pressure and density change in time, but the source radius does not. Our model for the energy momentum tensor is:
\begin{eqnarray}\label{eq:sourcecollapse}
T_{00} &=& \rho \exp \left(-\frac{r^2}{R_0^2}\right) \,,\nn\\
T_{xx} &=& T_{yy} = T_{zz} = \frac{\rho}{3} \exp \left(-\frac{r^2}{R_0^2}\right) \left[1-e^{-t/\tau}\right]\,.
\end{eqnarray}
Thus, the source becomes relativistic, with an equation of state $p=\rho/3$ on a timescale $\tau$. We begin with the field at rest in one of the screening solutions $\Pi_{\pm}$. For adiabatic collapse $\tau \rightarrow \infty$, the field evolves as the source collapses to reach the vacuum solutions described above (either $\pi=0$ or $\pi \sim r^2$). However, a source which collapses instantaneously corresponds to a large perturbation around the vacuum solutions, which from the results of Sec.~\ref{sub:CB}, can be vulnerable to Cauchy breakdown. This suggests a critical collapse time constant, $\tau = \tau_c$, below which Cauchy breakdown occurs. 

\begin{figure*}[hbt]
	\begin{center}
		\begin{tabular}{cc}
		\epsfig{file=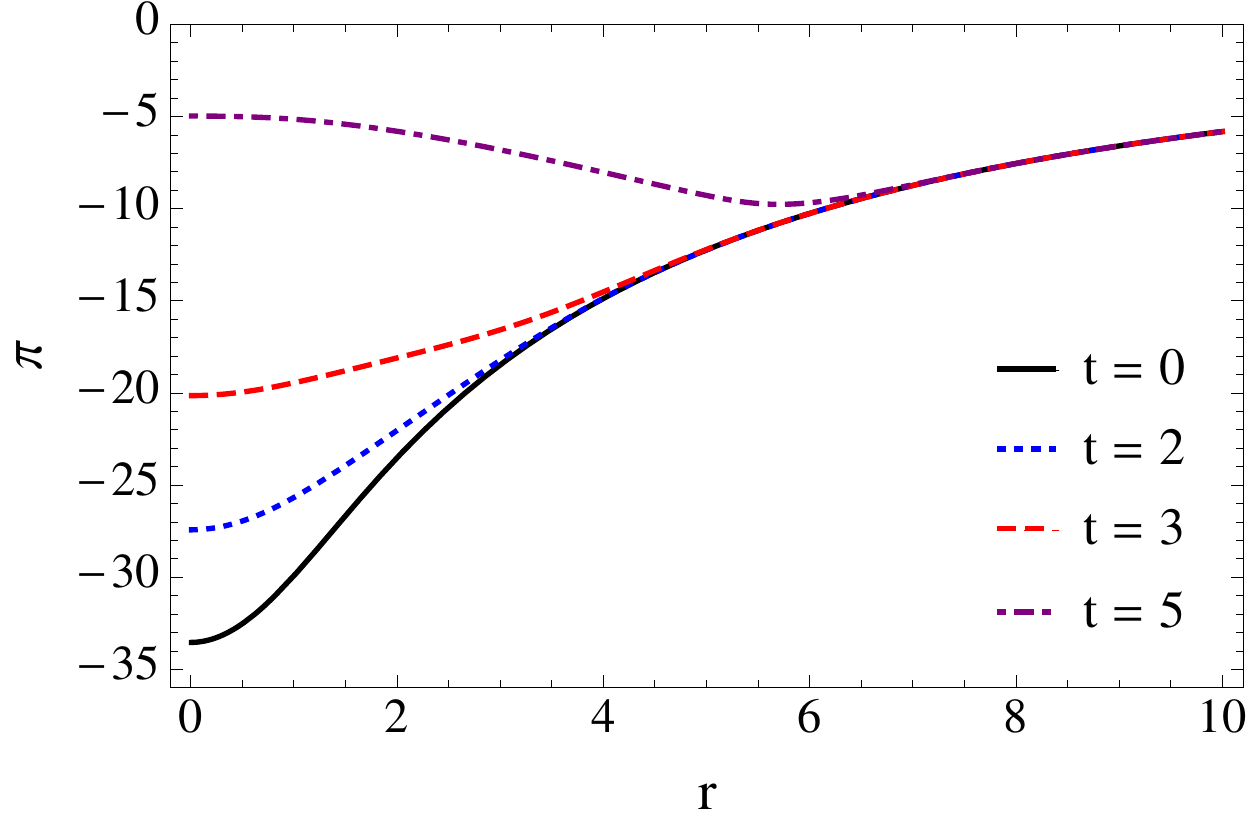,width=8.5cm,angle=0,clip=true}&\epsfig{file=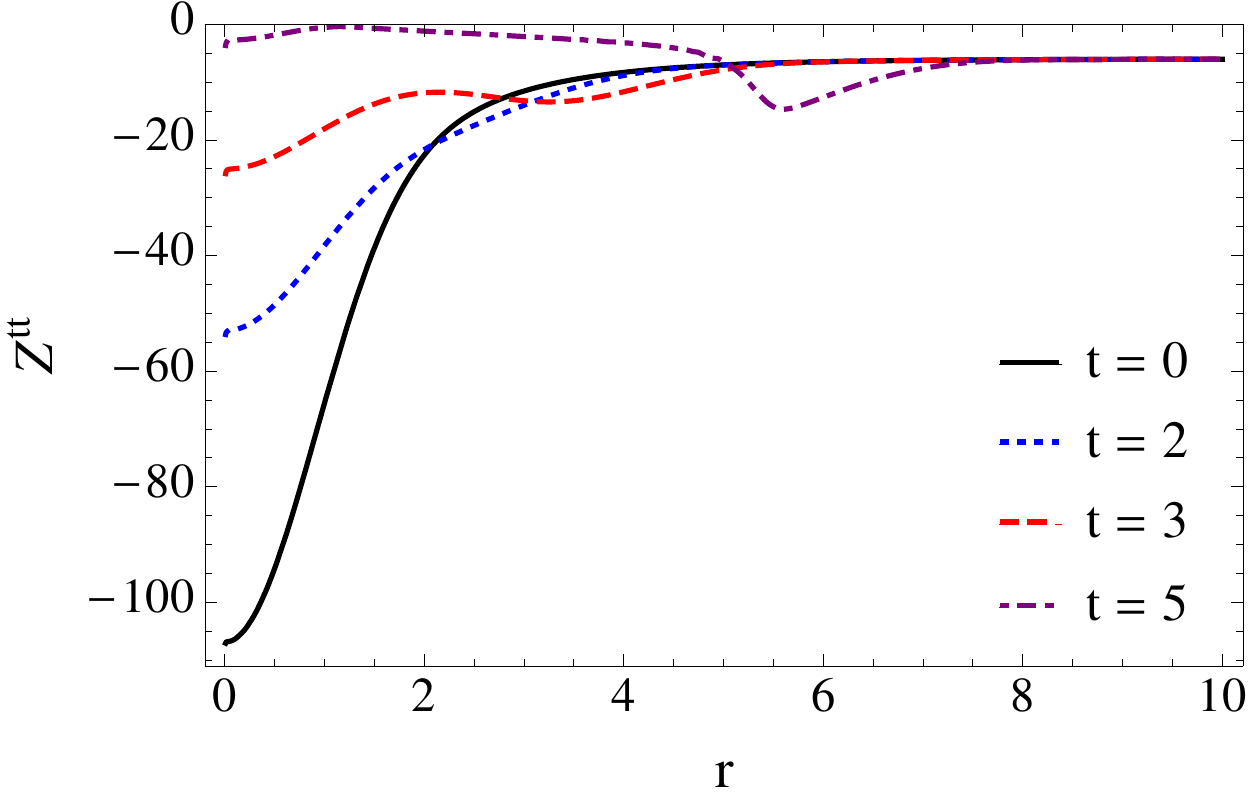,width=8.5cm,angle=0,clip=true}
		\end{tabular}
		\caption{An example of evolving towards Cauchy breakdown for a collapsing source of the form \eqref{eq:sourcecollapse} with $\rho=2000$, $R_0=1$ and $\tau=1$. Left panel: the evolution of $\pi(r,t)$ starting from $\pi_+(r)$. It is driven towards the vacuum $\pi_(r,t)\rightarrow 0$ as the source collapses (until Cauchy breakdown is reached).\\
		Right: The corresponding factor $Z^{tt}$ \eqref{eq:Zs_time} that crosses zero at $t=5$ resulting in  Cauchy breakdown}
		\label{fig:collapse}
	\end{center}
\end{figure*}

In Fig.~\ref{fig:collapse}, a sample evolution is plotted for a collapsing source defined by parameters $\rho=2000$, $R_0=1$, $\tau=1$. In this example, the perturbation created by the collapse causes Cauchy breakdown at $t=5$. Increasing the value of $\tau$ allows for a well-defined evolution. By fixing $R_0=1$ and varying $\rho$, we found that as the density increased, the corresponding time constant increased as well, in a linear fashion. Specifically, for both branches of solution, we found that $\tau_c \sim 0.0005\rho$.

Before Cauchy breakdown, the collapse seems to be generically preceded by the formation of an unstable region near the origin, where $Z^{\theta\theta}$ changes sign. Nevertheless, we always observe that the dynamical evolution is eventually stopped by $Z^{tt}\to 0$ before eventual unstable modes have time to grow.

Now let's consider the opposite effect: an \emph{exploding} source. We once again begin with an initially screened source, and then model the ``explosion" as an outgoing spherical shell of dust travelling at the speed of light:
\begin{equation}\label{eq:sourceexplode}
	T_{\mu \nu} = {\rm diag} \left(\frac{\rho}{f(t)} \exp\left(-\frac{(r-t)^2}{R_0^2}\right), 0,0,0\right).
\end{equation}
where $f(t)$ is defined so that the mass at infinity has the same constant value as previously, specifically, $M(r\rightarrow \infty)=\pi^{3/2} \rho R_0^3$ for all $t$. This implies that
\begin{equation}
	f(t)  = (1 + 2t^2/R_0^2) (1+\text{Erf}(t/R_0)) + (2t / (\sqrt{\pi} R_0)) e^{-t^2/R_0^2}.
\end{equation}
We stress that this is not a physical model for an exploding source, but rather a simple test of the response to dynamical sources.

\begin{figure*}[hbt]
	\begin{center}
		\begin{tabular}{cc}
		\epsfig{file=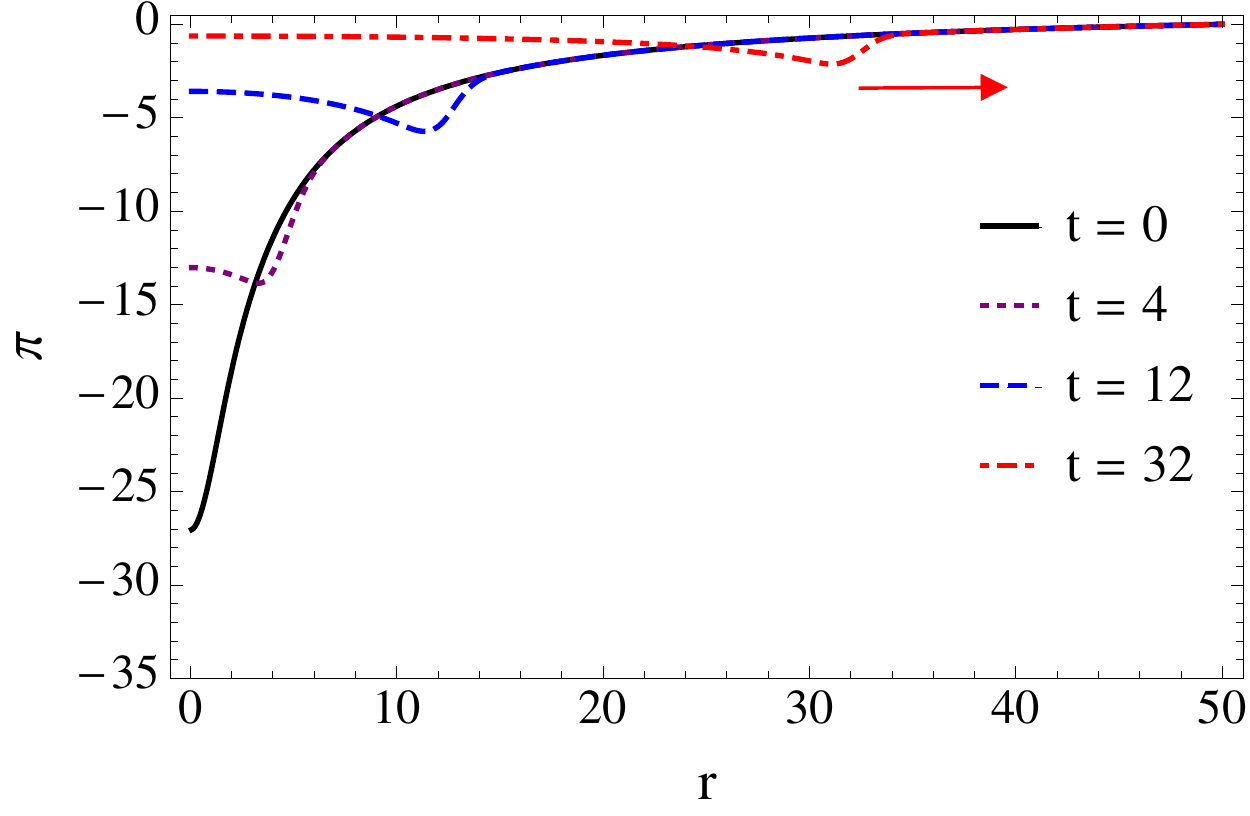,width=8.5cm,angle=0,clip=true}&\epsfig{file=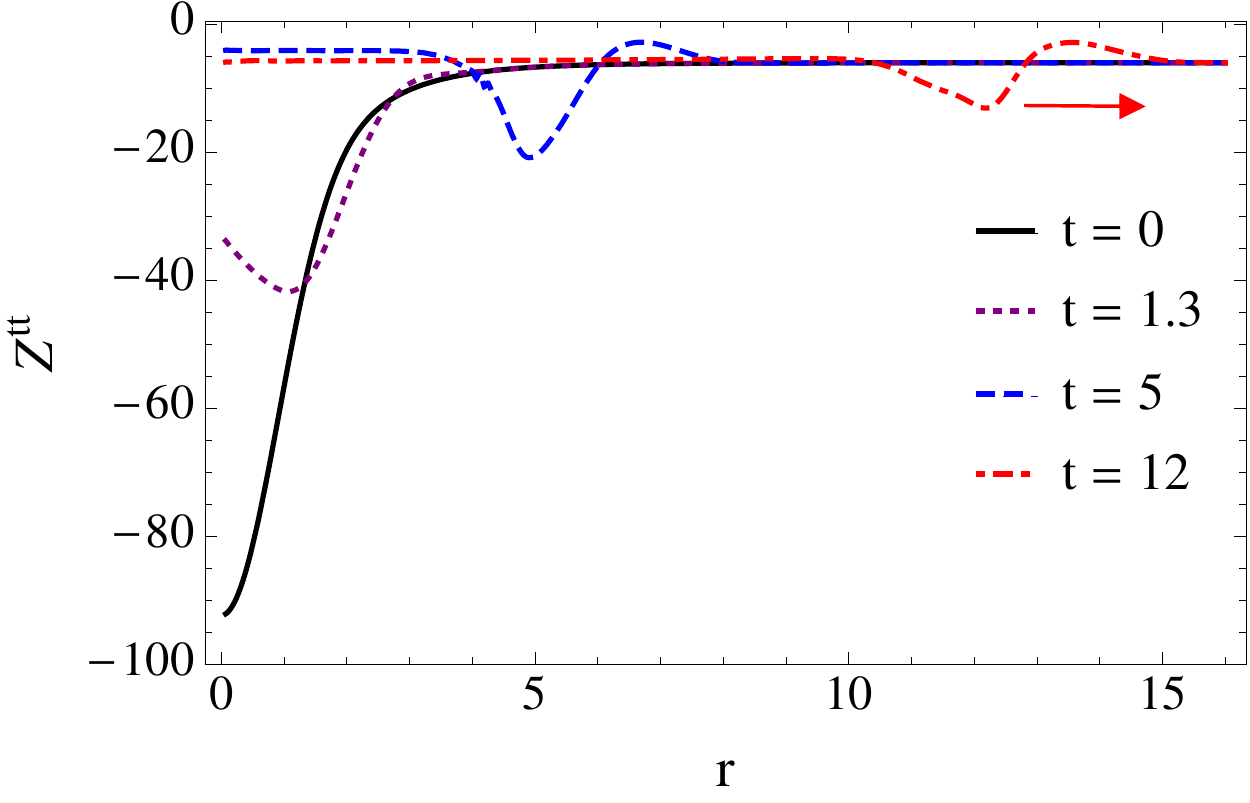,width=8.5cm,angle=0,clip=true}
		\end{tabular}
		\caption{A sample evolution for an exploding source that evades Cauchy breakdown. The source has the form \eqref{eq:sourceexplode} with $\rho=1500$ and $R_0=1$. Left panel: the evolution of $\pi(r,t)$ starting from $\pi_+(r)$. It is driven towards the vacuum $\pi_(r,t)\rightarrow 0$ as the source explodes, while a localized packet follows the travelling source off to infinity.\\
		Right: The corresponding factor $Z^{tt}$ \eqref{eq:Zs}. A perturbation is created that safely travels off to infinity without crossing zero, thus avoiding Cauchy breakdown.}
		\label{fig:explode}
	\end{center}
\end{figure*}

Starting with a screening configuration, as the source explodes, the scalar field relaxes to its vacuum state, while emitting a wave packet that travels with the source as it moves off to infinity. To illustrate this phenomena, we show the evolution of the field in Fig.~\ref{fig:explode} for the case of a source with $\rho=1500$ and $R_0=1$. Once again, one can imagine that for a source that is sufficiently dense, a sound horizon and an unstable region can form when the source explodes due to regions where fluctuations propagate subluminally. These will generically happen before $Z^{tt}\to 0$ and can eventually disappear if the source is not too dense. However if the source is too dense, the induced perturbation in the scalar field will be enough to drive $Z^{tt}$ to zero, causing Cauchy breakdown. The right panel of Fig.~\ref{fig:explode} shows the evolution of the Cauchy breakdown factor corresponding to the scalar field shown in the right panel. For this example, $Z^{tt}$ safely avoids crossing zero, so the evolution remains well-defined. 

To analyze the possibility of Cauchy breakdown in more detail we calculated the critical density $\rho_c$ for various source sizes $R_0$, so that Cauchy breakdown is inevitable for $\rho>\rho_c$. As the source size increases, the corresponding critical density increases as $\rho_c \sim 2000 R_0$ (consistent with both branches).
Once again, very dense compact objects can cause problems for the evolution. 

There are two relevant timescales for explosion or collapse: the crossing time of the source $R_0$ and the crossing time for the screening solution $R_V$. The longer timescale, $R_V$, sets the response time for the screening profile to changes in the source. In realistic scenarios, $R_0 \ll R_V$, and collapse or explosion will occur effectively instantaneously on timescales over which the screening solution can adjust. Therefore, we conjecture that Cauchy breakdown will be a problem for any realistic violent astrophysical process. However, to study breakdown in more detail, it is necessary to go beyond the decoupling limit and consider realistic dynamical sources.

\section{Asymmetric screening solutions}\label{sec:dgp_asymmetric}
In the previous sections we have been concerned about the linear and non-linear stability of the spherically symmetric screening solutions. We now wish to understand if these screening solutions can be generalized when we give up spherical symmetry. 

Our numerical search for quasinormal modes in Sec.~\ref{sec:QNM_DGP} did not yield any zero-frequency mode. In other words, we were not able to find any
regular, asymptotically flat static solution to the linearized equation of motion \eqref{dgp_pert} (apart from the trivial solution for $l=0$).
This can be considered a simple version of a ``no-hair'' result for screening solutions: no static multipoles -- other than the spherically symmetric monopole -- are allowed to anchor onto spherically symmetric sources. 

Does this result generalize for non-spherically-symmetric sources? Do the scalar multipoles anchor on higher source-multipoles?
One possibility to study this issue is to look for nonlinear, asymmetric solutions. Given the structure of the equations of motion,
such solutions are not trivial to find, although particular solutions can be built. Take for instance
\beq\label{eq:quadansatz}
\pi=\pi_{1}\cos^2\vartheta\,.
\eeq
The field equations yield powers of $\cos\vartheta$ which can be matched to $T$ order by order in powers
of $\cos\vartheta$. Given a zeroth-order source function, the zeroth-order equation can be solved for $\pi_1$, and the remaining equations will then determine the source multipoles. For example, with the ansatz Eq.~\ref{eq:quadansatz} we find the following solution for the equations of motion~\eqref{eq:eom}, yielding a quadrupolar static solution
\begin{align}
\pi &=-\frac{r^2}{24}T_0\cos^2\vartheta\,,\\
T&=T_0(r)+\frac{r}{36}\left(T_0'(108-T_0+2rT_0')\right)\cos^2\vartheta\nonumber\\
&-\frac{r}{36}\left(rT_0''(T_0-18)\right)\cos^2\vartheta \nn\\
&+\frac{r}{72}\left(-T_0'(2T_0+13rT_0')+2rT_0''(T_0-rT_0')\right)\cos^4\vartheta\,.
\end{align}
%
%
%
This nonlinear solution represents a field strongly localized close to the source. However, for most quadrupolar source distributions $T(r,\theta)$, $-T$ is not positive definite implying that there are regions where $\rho<3p$. Nonlinear solutions for higher multipoles can be found with the same scheme; they share similar properties.

\begin{figure}[t]
	\begin{center}
		\begin{tabular}{c}
		\epsfig{file=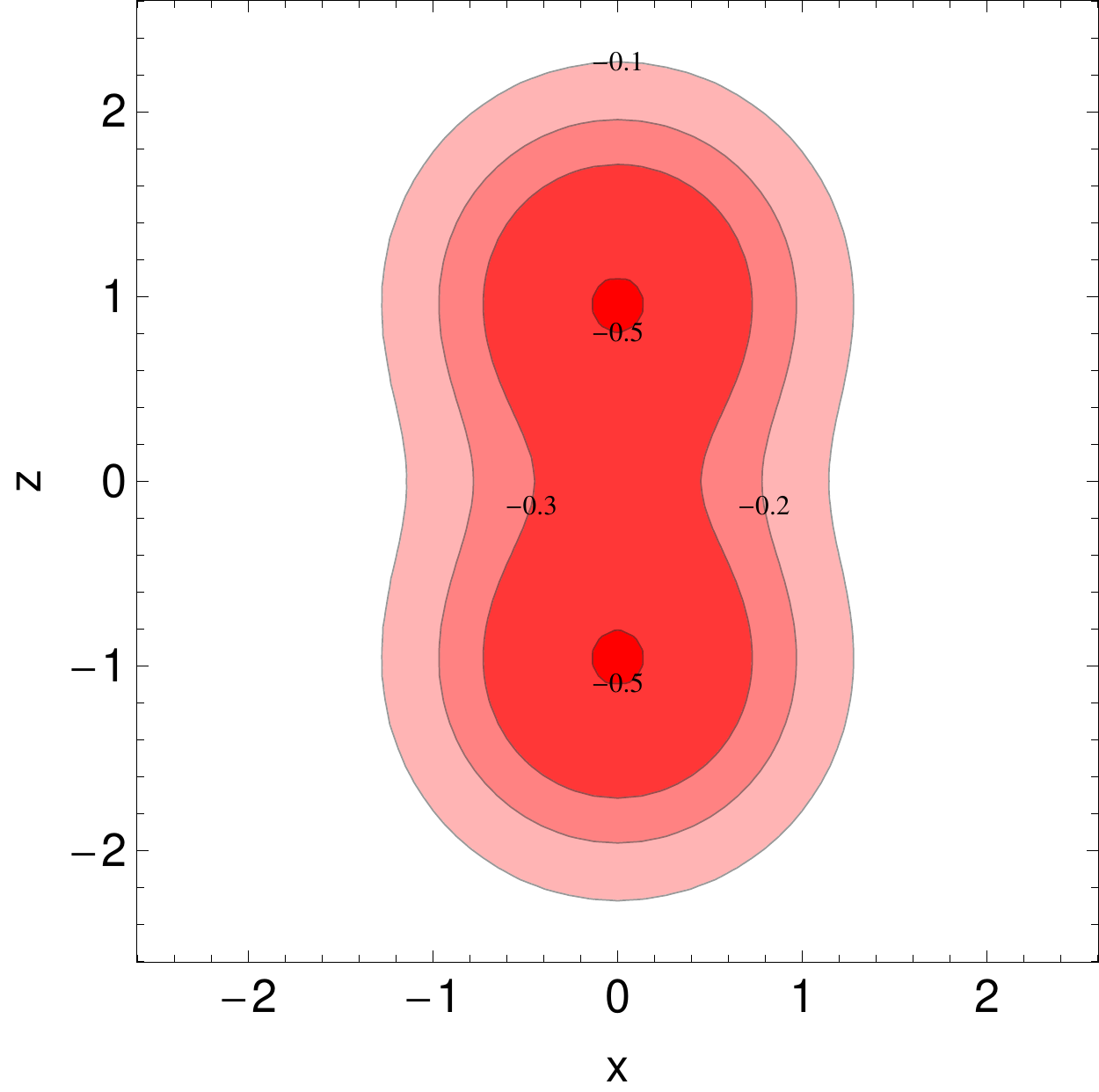,width=8.cm,angle=0,clip=true}
		\end{tabular}
		\caption{Contour plot of the $y=0$ slice of density profile \eqref{eq:sourcedensity} describing two lumps of matter, here for $\rho=1,\,z_0=0.9,\,R_0=1$. }
	\label{fig:densityplot}
	\end{center}
\end{figure}
A more robust method to look for asymmetric solutions builds on the nonlinear spherically symmetric solution for spherical sources \eqref{eq:screening}. Realistic stellar -- and other -- sources are approximately spherically symmetric, and it is therefore appropriate to search for small deviations from spherical symmetry in both the source and the field. Specifically,
we expand
\begin{eqnarray}
T&=&T_0(r)+\epsilon\sum_{l=1} t_{lm}(r)Y_{lm}(\vartheta,\,\varphi)\,,\\
\pi(r)&=&\int^r\Pi_\pm(u)du+\epsilon\sum_{l=1}\delta\pi(r)Y_{lm}(\vartheta,\,\varphi)\,.
\end{eqnarray}
We defined $T_{0}\equiv 2\sqrt{\pi}T_{00}$ and take this to be the dominant contribution.
The components $t_{lm}$ can be directly related to the more standard density moments $v_{lm}\propto t_{lm}(r)r^{l+2}dV$ where $dV$ is the volume element, in terms of which the gravitational potential multipole moments are usually expressed. 
Note that the dipole component vanishes for sources which are symmetric around the equator, while the quadrupole component is directly tied to the inertia tensor. For example, let's take two clumps of matter of the form \eqref{eq:source}, describing a deformed body,
\begin{equation} \label{eq:sourcedensity}
T(r)=-\frac{\rho}{2}\left(e^{-(x^2+y^2+(z-z_0)^2)/R_0^2}+e^{-(x^2+y^2+(z+z_0)^2)/R_0^2}\right) \,.
\end{equation}
This distribution represents two bodies localized at $\pm z_0$ on the z-axis, and is axially symmetric as shown in Fig.~\ref{fig:densityplot}. For $z_0=0$ we recover the density distribution \eqref{eq:source}, so $z_0$ can be treated as an expansion parameter. All multipoles moments $t_{lm}$ vanish for $m\neq 0$ (because the distribution is axially symmetric) and for odd $l$ (because it has equatorial symmetry). 
For small $z_0$, the lowest multipoles are

\beq\label{source_multipole_l0}
T_0&=&2\sqrt{\pi}t_{00}=-\rho\,e^{-r^2/R_0^2}\,,\\
t_{20}&=&\frac{8}{3}\sqrt{\frac{\pi}{5}}\frac{r^2z_0^2}{R_0^4}\,T_0\,,\label{source_multipole_l2}\\
t_{40}&=&\frac{32\sqrt{\pi}}{315}\frac{r^4z_0^4}{R_0^8}\,T_0\,.\label{source_multipole_l4}
\eeq

For a general source with multipoles $t_{lm}(r)$, the equation of motion for $\delta \pi$ is given by:
\begin{widetext}
\be
4r^2(3r+4\Pi_\pm)^2\delta\pi''+2\left(-12r\Pi_\pm-8\Pi_\pm^2+r^2(-18+T_0)\right)(l(l+1)\delta\pi-2r\delta\pi')=-r^3(3r+4\Pi_\pm)t_{lm}
\ee
\end{widetext}
This inhomogeneous ordinary differential equation can easily be integrated to yield solutions for $\delta\pi$. Solutions
exist for any source and decay at large distances as $r^{-l-1}$.     

\begin{figure}[htb!]
\centerline{\includegraphics[height=6 cm] {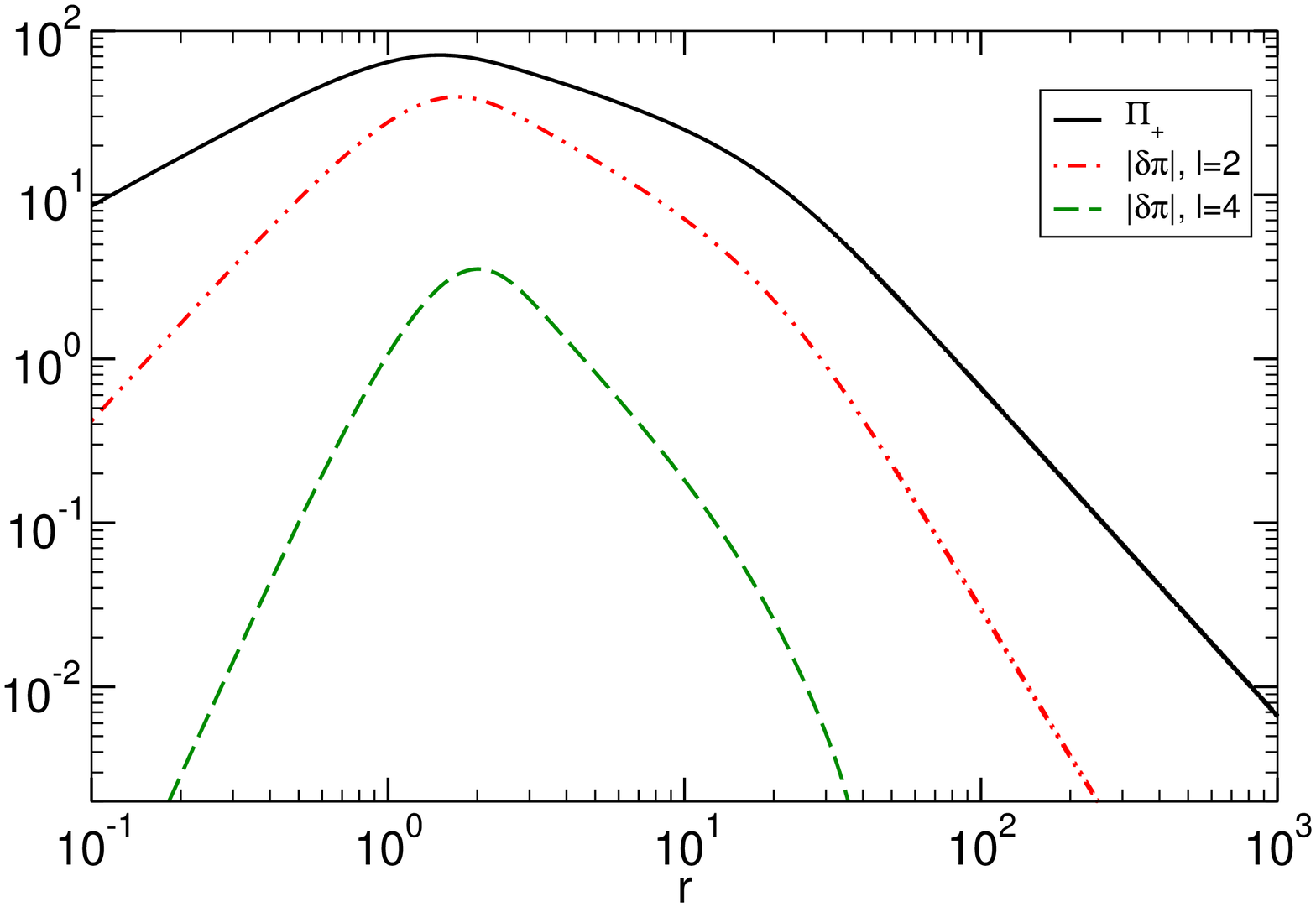}}
\caption{Hairy solutions for different multipoles $l$, here shown for $R_V=100$, $z_0=0.9$, and compared to $\Pi_+$. For $R_0\ll r\ll R_V$ the field decays as $\sqrt{\rho} z_0^lr^{-l/2}$ while for $r\gg R_V$ they decay as $\rho z_0^lr^{-(l+1)}$.
\label{fig:hair}}
\end{figure}

Computing the components $t_{lm}(r)$ of the source Eq.~\eqref{eq:sourcedensity} around the $\Pi_+$ background, we integrate the equation above requiring regularity at the origin and vanishing field at infinity. This can be done using a standard shooting method using the constant $A_1$ of the expansion at the origin~\eqref{ser_origin} as a shooting parameter. Some solutions are shown in Fig.~\ref{fig:hair} (results for $\Pi_-$ are qualitatively similar). For very large densities the screening behavior of the different multipoles is apparent. For $R_0\ll r\ll R_V$ the field decays as $r^{-l/2}$ indicating that in the Vainshtein regime higher multipoles have a stronger suppression than the monopole. The suppression of their contribution to the fifth force compared to the \emph{multipoles} of the Newtonian gravitational force $F_g\sim r^{-(2+l)}$ are given by
\beq \label{eq:force_multi}
\biggl| \frac{F_{l>0}}{F_g}  \biggl| \ \sim \left( \frac{r}{R_V} \right)^{1+l/2} \ \textrm{if $R_0\ll r \ll R_V$}\,.
\eeq 
Finally, our results show that for $R_0\ll r\ll R_V$ the field is proportional to $\sqrt{\rho}z_0^l$, whereas for very large distances it is proportional to $\rho z_0^l$, for any multipole $l$. We conclude that tidal forces due to the scalar are subdominant to gravitational tidal forces inside the Vainshtein radius, and so the spherically symmetric approximation is, in general, a very good approximation to compute the fifth force around non-spherical stars.

\section{Comparing with Massive Gravity in the Decoupling Limit}

Let us now make a comparison with ghost-free dRGT massive gravity \cite{deRham:2010kj},\cite{deRham:2010ik} which shares many features of the DGP model described above in the decoupling limit. Here, we consider a specific class of massive gravity in which one of its two free parameters is set to zero, making it possible to completely decouple the scalar and tensor modes. We'll see that a few features of massive gravity make the analysis different than DGP: a quartic galileon term, an extra free parameter $\alpha$ in the theory, and an extra non-minimal coupling to matter in the equations of motion of the form $\nabla_\mu\nabla_\nu\pi T^{\mu\nu}$. Strong constraints on this coupling can be found in Refs.~\cite{Sakstein:2014isa,Brax:2014vva}. Also note that for a time-dependent field this coupling will change the Vainstein radius~\cite{Sakstein:2014isa}. However for small fluctuations around the screening solutions, this change is in general highly suppressed so we will keep working with the static quantity $R_V$. 

The scalar mode that arises in massive gravity is essentially the longitudinal mode of the graviton, and in the decoupling limit, described in Sec.~\ref{sec:vainshtein}, its dynamics can be solved for, independently of the other degrees of freedom. Following the derivation in the Appendix~\ref{app:dec}, under the assumption of spherical symmetry $\pi=\pi(t,r)$, the equation of motion~\eqref{eq:mgeom} for the longitudinal mode of the massive graviton in the decoupling limit for a non-relativistic source ($T_{0i}=T_{ij}=0$) is
\begin{align}
	& -T + 2\alpha T^{00}\ddot{\pi} = -3\ddot{\pi}+3\pi'' + \frac{6}{r}\pi' + \nonumber \\
	& 6\alpha\left( 
	\frac{2}{r}\ddot{\pi}\pi'-\frac{1}{r^2}(\pi')^2-(\dot{\pi'})^2+\ddot{\pi}\pi''-\frac{2}{r}\pi'\pi'' \right) + \nonumber \\
	 & 6\alpha^2\left( \frac{2}{r}\pi'(\dot{\pi'})^2 - \frac{1}{r^2}\ddot{\pi}(\pi')^2 -\frac{2}{r}\ddot{\pi}\pi'\pi'' + \frac{1}{r^2}\pi''(\pi')^2\right).
	 \label{eq:mgeom2}
\end{align}
Recall that we are imposing our choice of units: $M_4 = \Lambda = 1$.
Further, we assume the same static source of density $\rho$, radius $R_0$ and mass $M$ as given in \eqref{eq:source}. For the metric, the spherically symmetric ansatz is $h_{00}=a(r,t)$ and $h_{ij} = f(r,t)\delta_{ij}$. Once a solution for $\pi(r,t)$ is found, one can then find the metric functions using equations that result from variation of \eqref{eq:mgLdc} with respect to $h_{\mu\nu}$:
\begin{align} \label{eq:af1}
	f' &= -\frac{M}{r^2} + \pi'(1-\alpha\frac{\pi'}{ r}), \\ \label{eq:af2}
	a' & = -\frac{M}{r^2} +r \ddot{\pi}-r\ddot{f}-\pi'-2\alpha\ddot{\pi}\pi'\,.
\end{align}

Analysis of \emph{static} screening solutions can be found in Refs.\cite{Chkareuli:2011te,Berezhiani:2013aa}. Let us highlight some of their results. In this case, we can write \eqref{eq:mgeom2} as a cubic (rather than quadratic as in DGP) polynomial in $\lambda=\pi'/r$:
\begin{equation}\label{eq:lambdapoly}
	3\lambda-6\alpha\lambda^2+2\alpha^2\lambda^3=\frac{M(r)}{4\pi r^3}\,.
\end{equation}

There are three solutions to \eqref{eq:lambdapoly}, that we denote as $\lambda_1$, $\lambda_2$, $\lambda_3$. When the solutions are evaluated for small values of $r$ outside of the source ($R_0<r\ll R_V$), it is clear that only $\lambda_1$ is real, and $\lambda_2,\ \lambda_3$ are imaginary in this regime. Therefore, we can take $\lambda_1$ to be our static solution and disregard $\lambda_{2,3}$. The expression for $\lambda_1$ in this limit is valid for positive or negative values of $\alpha$ and can be written
\begin{equation}
	\lambda_1 (R_0<r\ll R_V) \sim \frac{1}{\alpha}+\frac{R_V}{|\alpha|^{2/3}r}+\frac{r}{2|\alpha|^{4/3}R_V}\,.
\end{equation}
For the sake of completeness, we give the full expression here 
{\small
\begin{align}\label{static_mg}
	& \lambda_1 =   \frac{1}{\alpha} + \nonumber \\
	& \frac{(2\pi)^{1/3}r}{\alpha\left( 4\pi r^3 + \alpha M(r) + \sqrt{-16\pi^2r^6+\alpha M(r)\left(8\pi r^3 + \alpha M(r)\right)} \right)^{1/3} }\nonumber \\
	& +\frac{\left( 4\pi r^3 + \alpha M(r) + \sqrt{-16\pi^2r^6+\alpha M(r)\left(8\pi r^3 + \alpha M(r)\right)} \right)^{1/3} }{2 \alpha  (2\pi)^{1/3} r } \,.
\end{align}
}
The interesting fact is that the asymptotic behavior of $\lambda_1$ is very different depending on the sign of $\alpha$:
\begin{eqnarray} \label{eq:lambdainfty}
	\lambda_1(r\rightarrow \infty) = \begin{cases} 
      \frac{3+\sqrt{3}}{2\alpha}& \textrm{ if $\alpha>0$} \\
      0 & \textrm{ if $\alpha<0$}\,.\\
   \end{cases}
\end{eqnarray}
Plugging the asymptotics \eqref{eq:lambdainfty} into \eqref{eq:af1}, \eqref{eq:af2}, we see that when $\alpha<0$ it is possible to have an asymptotically flat spacetime: $a\simeq f \simeq M/r$. However, when $\alpha>0$, we obtain a non-trivial background with cosmological asymptotic behaviour: $a\simeq -r^2\lambda_1/2$, $f\simeq r^2\lambda_1(1-\alpha\lambda_1)/2$. This begs the question: what value can this free parameter $\alpha$ take? It was shown in \cite{Berezhiani:2013aa} that $\alpha>0$ is required to avoid a ghost instability, so let us examine this in more detail.

\begin{figure}[hbt]
	\begin{center}
		\begin{tabular}{c}
		\epsfig{file=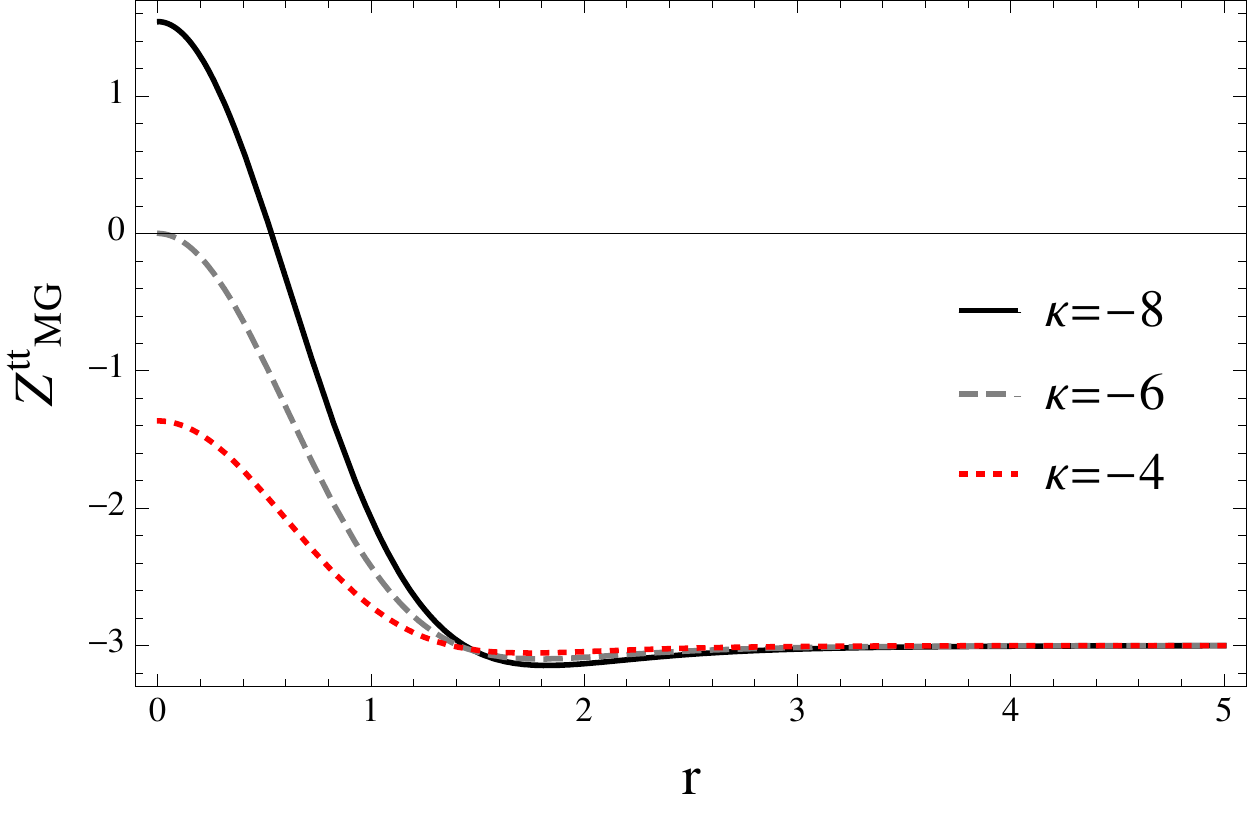,width=8.5cm,angle=0,clip=true}
		\end{tabular}
		\caption{The factor $Z^{tt}_\text{MG}$ for various values of $\kappa \equiv \alpha\rho$. The solution is unstable for $\kappa<-6$. }
	\label{fig:mgGplots}
	\end{center}
\end{figure}

As introduced in section~\ref{sec:screeningstability} and further discussed in Appendix~\ref{app:IVP} the stability of the solution can be inferred from the term multiplying $\ddot{\pi}$ in Eq.~\eqref{eq:mgeom2}. For this theory, the factor we are concerned with is
	\begin{align}\label{eq:g00mg}
		Z^{tt}_\text{MG} = & -3 - \alpha\left( 2 T^{00} - 6\pi'' - 12\frac{\pi'}{r} \right)	\nonumber \\
		& - \alpha^2\left( \frac{6(\pi')^2}{r^2}+\frac{12\pi''\pi'}{r} \right)\,. 
	\end{align}
The requirement for the stability of the static solution~\eqref{static_mg} against high-frequency modes can be shown to be equivalent to require $Z^{tt}_\text{MG}<0$ at all points in spacetime, which can be used to put constraints on the theory's free parameter $\alpha$. Analyzing the $Z^{tt}_\text{MG}$ factor leads to the realization that it is possible to have $Z^{tt}_\text{MG}>0$ inside the source for negative values of $\alpha$. The novel coupling of $\pi$ to the energy-momentum tensor plays a key role for this to happen. Note that in the background of a static solution, $Z^{tr}_\text{MG}=0$, so the condition for the Cauchy breakdown $Z^{tt}_\text{MG}=0$ is also the condition for the solution to be marginally stable. In general $Z^{tt}_\text{MG}<0$ for all positive $\alpha$ and the spatial components of the effective metric $Z^{rr}_\text{MG}$ and $Z^{\theta\theta}_\text{MG}$ are positive for all values of $\alpha$ (as long as we neglect pressure), so our only concern is that $Z^{tt}_\text{MG}$ becomes positive when $\alpha$ is negative. Setting $R_0=1$ in Eq.~\eqref{eq:source} and using the relation $\pi' = \lambda_1 r$, the term $Z^{tt}_\text{MG}$ can be written in terms of a single parameter $\kappa \equiv \alpha\rho$. One finds that $Z^{tt}_\text{MG}>0$ inside the source for $\kappa<-6$. Therefore, as long as $\alpha>-6/\rho$, the solution is stable, as shown in Fig.~\ref{fig:mgGplots}.

For physically realistic values ($\rho\sim 10^{26}$ for a sun-like source), the window of stability $-6/\rho<\alpha<0$ is quite small so we conclude that the only valid static screening solution in the massive gravity decoupling limit is the one with cosmological asymptotics. As shown in \cite{Berezhiani:2013aa}, this solution is stable against linear perturbations. In addition, fluctuations remarkably propagate with sub-luminal speeds.

We have thus seen how the extra free parameter $\alpha$ and the new coupling $\nabla_\mu\nabla_\nu\pi T^{\mu\nu}$ give some qualitative differences to the static spherically symmetric solutions in massive gravity as compared to DGP. But at this point, the study of dynamical solutions is quite similar to the DGP case, except we only have one branch of static solution to analyze. We can use the same numerical method to solve the time-dependent equation~\eqref{eq:mgeom2} and perform the same tests as we did in the DGP model. 

\subsection{Linear and non-linear stability of dRGT gravity in the decoupling limit}

An analysis of the non-linear stability of the static screening solution~\eqref{static_mg} revealed what we expected from our detailed study of the DGP model: the static solution is reached as the endpoint of the evolution for the gaussian wavepacket considered in Section~\eqref{sec:screen1}, as long as the fluctuation is small enough compared to the background solution; the static solution is also reached considering the quadratic vacuum initial conditions described in Section~\eqref{sec:screen2}, as long as $R_V/R_0$ is sufficiently small.

However, for large perturbations we found some different qualitative features which are due mainly to two reasons: the additional coupling to matter $\nabla_\mu\nabla_\nu\pi T^{\mu\nu}$; and the big hierarchy between $Z^{\theta\theta}$ and $Z^{tt}$ of the effective metric, $Z^{tt}\gg Z^{\theta\theta}$~\cite{Nicolis:2009aa}. The components of the effective metric can be found in the Appendix~\ref{app:dec} (see Eq.~\eqref{eq:Zmg_time}).

Due to the extra coupling to matter the theory is less prone to suffer Cauchy breakdown near the source than in DGP. This can be traced back to the fact that inside the source, fluctuations are highly suppressed due to this term~\cite{Berezhiani:2013aa}. In fact, when pressure is neglected, $Z_\text{MG}^{rr}$, $Z_\text{MG}^{\theta\theta}$ will always change sign before Cauchy breakdown occurs due to the big hierarchy between the different components of the metric $Z^{\mu\nu}$. The conditions for this to occur are qualitatively similar to the ones we found in Section~\eqref{sub:CB}. For  very large fluctuations, we were not always able to evolve past these points, most likely due to the excitation of high-frequency unstable modes. However, we expect that the formation of these unstable regions causes an enhancement of gradients of the field, making Cauchy breakdown inevitable. Cauchy breakdown was more easily observed outside the source where we checked that the condition~\eqref{cb_slice} was satisfied, showing once again that Cauchy breakdown is a coordinate independent phenomenon.

In conclusion, even though the extra coupling to matter renders the theory less prone to Cauchy breakdown inside the source, problems can arise for sufficiently large perturbations of the screening solutions~\eqref{static_mg}. On the other hand, for the class of well-behaved initial conditions the screening solution is stable and behaves as a coherent object with radius $R_V$, as we discuss below.

\subsubsection{Quasinormal modes and tails}

As discussed before in the DGP case, when perturbed, the static solutions~\eqref{static_mg} vibrate and eventually relax to an equilibrium state again. Performing the same analysis as in Section~\ref{sec:QNM_DGP}, we find that the waveform consists of the expected three stages, a prompt response at very early times, quasinormal modes at intermediate times and a power-law tail at very late times.

We saw in the previous section that the new coupling $\nabla_\mu\nabla_\nu\pi T^{\mu\nu}$ is important for the stability of the solutions. If this coupling is absent the behavior of perturbations on top of the static solution~\eqref{static_mg}, for $\alpha<0$ and $\alpha>0$ is very similar to the DGP case (see Fig.~\ref{fig:osci}). The introduction of this new coupling makes the solution for $\alpha<0$ unstable, which can be clearly seen in a time-domain analysis of the linear equation around the this background. On the other hand perturbations on top of the asymptotically growing solution with $\alpha>0$, are stable and have a clear quasinormal ringdown similar to the one shown in the top panel of Fig.~\ref{fig:osci}.
 
\begin{figure}
\centerline{\includegraphics[height=6 cm] {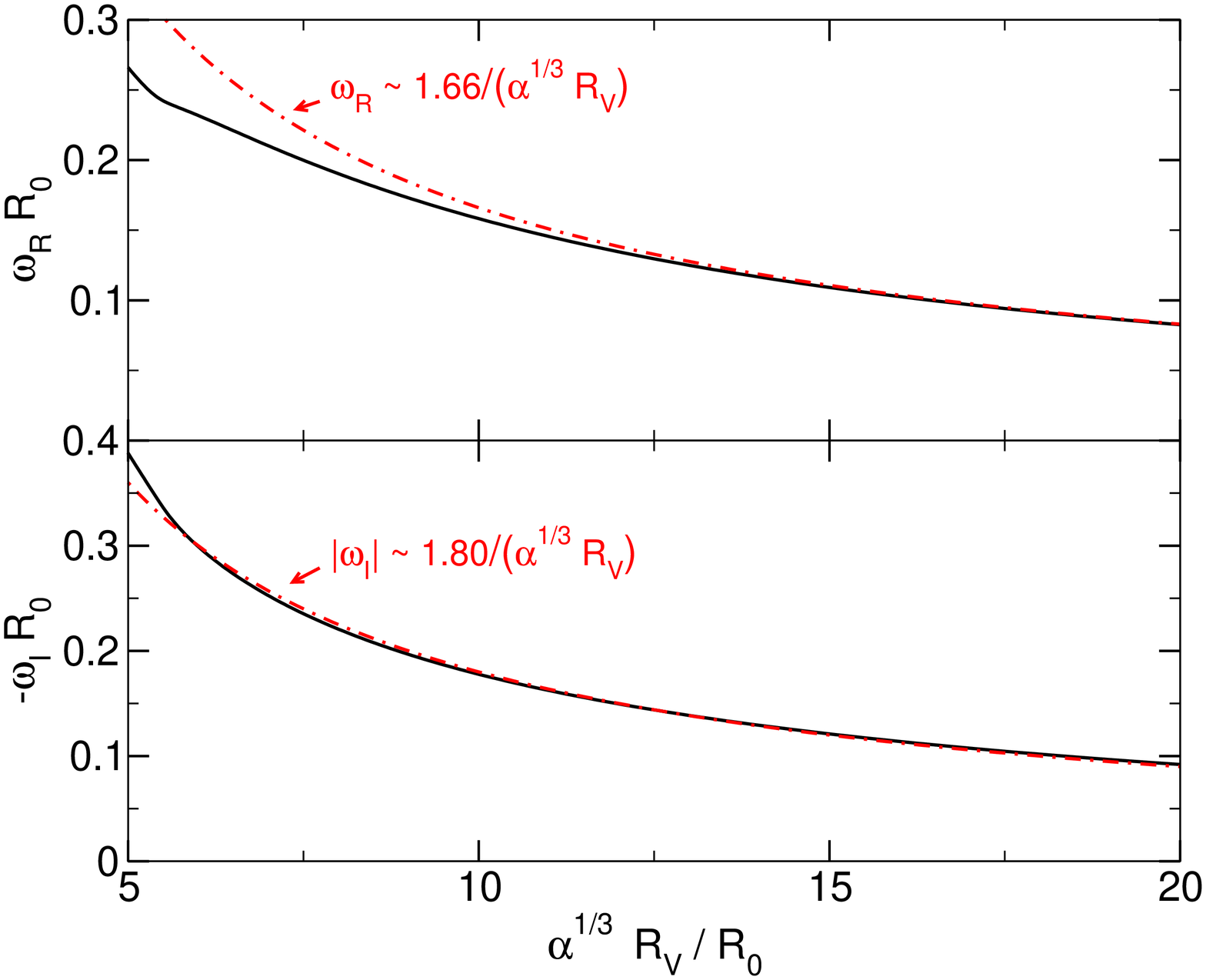}}
\caption{Fundamental quasinormal modes of the scalar field in the decoupling limit of massive gravity. The full lines correspond to the numerical results, whereas the dashed lines show the analytic approximation at low frequencies. The top and bottom panels show
the real part, $\omega_R R_0$, and the imaginary part, $\omega_I R_0$, of the mode as a function of $\alpha^{1/3} R_V/R_0$. 
\label{fig:QNM_quartic}}
\end{figure}

A frequency domain analysis also shows that for $R_V\gg R_0$ the quasinormal frequencies follow the same trend as in DGP and are given by  
\beq
\omega_R&\sim& \frac{1}{\alpha^{1/3}R_V} \label{wreal_quartic}\,,\\
\omega_I&\sim& -\frac{1}{\alpha^{1/3}R_V} \label{wimaginary_quartic}\,.
\eeq
This is shown in Fig.~\eqref{fig:QNM_quartic}, where we plot the fundamental quasinormal modes. One can understand this scaling from the fact that the coupling constant $\alpha$ can be reabsorbed into $\Lambda$ and so the effective Vainsthein radius is given by $\tilde{R}_V\equiv \alpha^{1/3}R_V$.

In the background of~\eqref{static_mg}, we can compute a wave equation of the form~\eqref{dgp_potential2}, where the effective potential has the large distance asymptotic behavior
\be
V\sim\frac{l(l+1)}{r_*^2}+\frac{K}{r_*^8}\,,
\ee
where $K$ is once more a constant that depends on $l$ and $M(r\to\infty)$. This behaviour is independent of $\alpha$ and does not depend on the new coupling $\nabla_\mu\nabla_\nu\pi T^{\mu\nu}$ or the specific form of $T^{\mu\nu}$. Thus, the above analysis suggests that scalar perturbations of the static solution generically decays as $\psi(r_*,t)\sim t^{-2l-8}$ at late times, just as in DGP. 

%

\subsection{Collapsing and Exploding Sources\label{sec:collapse_mg}}

Compared to DGP, the extra coupling to matter in massive gravity makes the scalar evolution less prone to Cauchy breakdown, although it can not be avoided for sufficiently large fluctuations. Does this hold for the dynamical sources that we considered in Section~\ref{sec:collapse}? In general yes, although some additional subtle issues are worth pointing out. 

For the case of the collapsing source~\eqref{eq:sourcecollapse}, when pressure becomes important, unstable regions near the source are inevitable, as was pointed out in~\cite{Berezhiani:2013dw}. Instabilities as well as sound horizons form during collapse as the pressure becomes significant $p \sim \rho$, as can be inferred by looking at Eq.~\eqref{eq:Zmg_time}. The spatial components, and in particular $Z^{\theta\theta}$, will change sign at some point in space, in a finite time. Although this behavior was also found in DGP, due to the extra coupling to matter, this effect is enhanced here. However,  this is in general followed by Cauchy breakdown which makes it impossible to follow the development of the instability. This is not only dependent on the density of the source but also on the decay time--scale, with a relation similar to the one found in DGP. For large time--scales (or very small densities) the field will evolve without instability or breakdown to a different solution with the same asymptotics (not to the vacuum solution, due to the extra coupling to matter). But for realistic source densities, Cauchy breakdown seems to be inevitable.

Surprisingly, the exploding source~\eqref{eq:sourceexplode} seems to avoid Cauchy breakdown for moderately high densities $\rho\sim 10^6$ (and $\alpha\sim 1$, recall that $\alpha$ can be reabsorbed). Once more, this is mainly due to the hierarchy $Z^{tt}\gg Z^{ii}$ in this background. However, for sufficiently large $\alpha$ and $\rho$, regions where the eigenvalues of $Z^{rr}$ and $Z^{\theta\theta}$ change sign can form. In these regions, the field fluctuations propagate at extreme subluminal velocities, thus leaving time for instabilities to grow. Similar to the above cases, these instabilities can eventually cause Cauchy breakdown.

\subsection{Asymmetric screening solutions}

%
\begin{figure}[htb!]
\centerline{\includegraphics[height=6 cm] {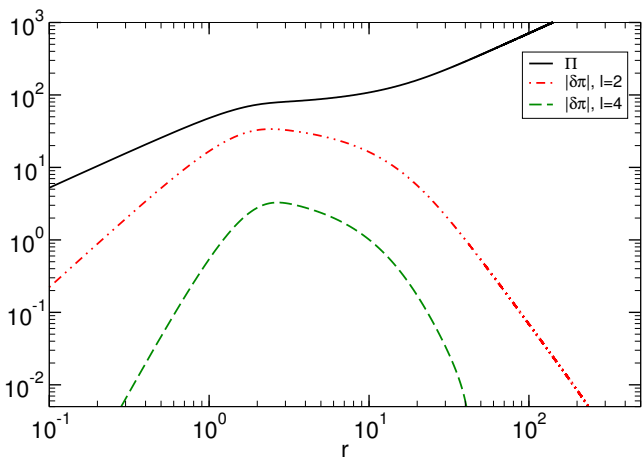}}
\caption{Hairy solutions for different multipoles $l$, here shown for $R_V=100$, $z_0=0.9$, $\alpha=1/3$, and compared to $\Pi\equiv \pi'$. For $R_0\ll r\ll R_V$ the field decays as $\rho^{1/3}z_0^lr^{-l/4}$ while for $r\gg R_V$ they decay as $\rho z_0^lr^{-(l+1)}$
\label{fig:hair_quartic}}
\end{figure}

Let us close our discussion on massive gravity by following Section~\eqref{sec:dgp_asymmetric} to compute asymmetric screening solutions for the source~\eqref{eq:sourcedensity}. These turn out to be very similar to the ones found in the DGP model. The asymptotic form of the  scalar multipole components are the same as in DGP, namely $\delta\pi\sim A_1 r^l$ at the origin and $\delta\pi\sim r^{l+1}$ at infinity. Some solutions are shown in Fig.~\ref{fig:hair_quartic}. For $R_0\ll r\ll R_V$ the field decays as $r^{-l/4}$ indicating that in the Vainshtein regime higher multipoles have a stronger suppression.  The suppression of their contribution to the fifth force compared to the \emph{multipoles} of the Newtonian gravitational force $F_g\sim r^{-(2+l)}$ is given by
\beq \label{eq:force_multi_mg}
\biggl| \frac{F_{l>0}}{F_g}  \biggl| \ \sim \left( \frac{r}{R_V} \right)^{1+3l/4} \ \textrm{if $R_0\ll r \ll R_V$}\,.
\eeq 
This shows that in this case, in the Vainshtein regime solutions are generically more suppressed in massive gravity than in DGP.

\section{Conclusions}

The theoretical challenge posed by explaining the observed accelerated expansion of the Universe has put theoretical physics at a crossroads. One can postulate dark energy  as the cause, possibly in the form of a cosmological constant, and be content with an anthropic explanation for the accelerated expansion. On the other hand, one can question the validity of GR on large distance scales, and be left with the need to explain why local departures have not been observed. In this paper, we have explored the latter possibility, studying the linear and non-linear stability of the screening solutions that restore the predictions of GR on short distance scales. 

The two theories we have studied, the DGP braneworld scenario and dRGT massive gravity, are examples where non-linear derivative interactions give rise to modifications of gravity only in the infrared through the Vainshtein screening mechanism. We have focused on the decoupling limit of these theories, in which a non-gravitating scalar degree of freedom is introduced that couples to the matter sector. Using analytic and numerical methods, we have taken the first steps towards establishing the fully non-linear dynamical stability of the Vainshtein screening solutions in spherical symmetry. We have also derived some properties of screening solutions beyond spherical symmetry. Our main results are as follows:

\begin{itemize}
\item Using numerical simulations we have shown for the first time that in the decoupling limit of both DGP and dRGT massive gravity, the Vainshtein screening solutions are dynamically accessed from a wide variety of initial conditions beyond the linear regime.
\item We have shown that the screening solutions behave as a coherent object much like a star or black hole under linear perturbations: a prompt response due to the primary scattering is followed by a universal series of damped oscillations known as quasinormal modes, which is then followed by a universal power-law decay. This analysis also shows that spherical sources can only support a monopole configuration of the scalar $\pi$; any multipolar ``hair" on spherically symmetric screening solutions is radiated away. 
\item However, for sufficiently large perturbations, regions of spacetime form in which there is no longer a well-defined Cauchy problem, a phenomenon which (following previous nomenclature) we term Cauchy breakdown. We have shown that in general this is not a coordinate singularity, but a real physical problem. In the absence of a new physical principle for what occurs in such regions, the future evolution is undetermined. This phenomenon is a general feature of theories with derivative self-interactions.
\item For sources which undergo collapse into a relativistic object or explode, we have shown that Cauchy breakdown generically occurs when there is a large hierarchy between the radius of the source and the Vainshtein radius. This is the case for realistic astrophysical objects, and hence there is the danger that Cauchy breakdown will occur in a complete  description of violent astrophysical phenomena such as supernovae or the formation of neutron stars and black holes. 
\item Finally, by considering non-spherically symmetric sources, we have shown that for both DGP and dRGT massive gravity the tidal components of the scalar fifth force are subdominant to the gravitational tidal field, and that tidal forces are screened more effectively than the monopole. 
\end{itemize}

Our results represent a first step towards establishing the nonlinear dynamical stability of infrared modifications of gravity. However, Cauchy breakdown is likely to be an important obstruction to determining the stability of a variety of cases of potential physical interest. What might be necessary to determine evolution past these points? In this work, we have neglected quantum corrections, which by a naive analysis become relevant whenever derivatives of the $\pi$ field become sufficiently large compared to the scale $\Lambda$. 



However, the Vainshtein mechanism itself changes the scale at which fluctuations become strongly coupled. On top of a background configuration, the strong coupling scale gets ``redressed'' by the effective metric and is given by $\tilde{\Lambda}\equiv \Lambda Z^{1/2}$ \cite{deRham:2014zqa}, where $Z$ schematically represents the relative strength of the eigenvalues of a slowly varying effective background $Z^{\mu\nu}$. For DGP and dRGT massive gravity, $Z$ can symbolically be written as $Z\sim 1 + \partial^2\pi_0/\Lambda^3$. In the non-linear regime $r \ll R_V$, small fluctuations around the static screening solutions see an effective metric with $Z \gg 1$, implying that $\tilde{\Lambda} \gg \Lambda$. Thus, for the static backgrounds that we considered, quantum corrections are suppressed. However, for big fluctuations around the static solutions the condition $Z\gg 1$ does not hold in general, leaving the possibility for Cauchy breakdown to occur, in which case $\tilde{\Lambda}\to 0$. This means that at this point fluctuations become infinitely strongly coupled~\cite{ge:2Burra011cr}, signaling that the classical theory can no longer be trusted. 
The scale $\Lambda$ (and the redressed scale $\tilde{\Lambda}$) is the strong coupling scale of the theory, but not necessarily the theory's cutoff. 
Hitting the scale $\Lambda$ or $\tilde{\Lambda}$ does not automatically imply a breakdown of the physical theory, but rather a breakdown of perturbativity. This means that quantum loops should be taken into account, but it does not necessarily mean that new physics is required. On the other hand, understanding how to evolve past Cauchy breakdown may require an understanding of how infrared modifications of gravity can be UV completed~\footnote{It is curious to note that our results clearly show that the screening solutions behave as coherent objects extended up to $R_V$, which could be closely related to some proposals for the UV completion of these theories~\cite{Dvali:2010jz,Dvali:2011th,Dvali:2010ns}.}.

Outside the regime of Cauchy breakdown, it is possible to study a variety of situations of physical interest. In particular, the formation of cosmological large scale structure, and perhaps some solutions in the strong field regime. In future work, we will tackle these problems, extending our analysis beyond the decoupling limit, and considering evolution that includes the internal dynamics of realistic sources. We hope that this work will produce new predictions for observables, aiding in the search for the cause of the observed accelerated expansion of the Universe.

\begin{acknowledgments}
We wish to thank N. Afshordi, K. Hinterbichler, J.L. Lehners, L. Lehner for important discussions and collaboration at early stages of  this work. We are indebted to Claudia de Rham and Jeremy Sakstein for a critical reading of the manuscript and for useful suggestions.
R.B. acknowledges financial support from the FCT-IDPASC program through the grant SFRH/BD/52047/2012, and from the Funda\c c\~ao Calouste Gulbenkian through the Programa Gulbenkian de Est\' imulo \`a Investiga\c c\~ao Cient\'ifica.
We acknowledge financial support provided under the European Union's FP7 ERC Starting Grant ``The dynamics of black holes:
testing the limits of Einstein's theory'' grant agreement no. DyBHo--256667.
This research was supported in part by Perimeter Institute for Theoretical Physics. 
Research at Perimeter Institute is supported by the Government of Canada through 
Industry Canada and by the Province of Ontario through the Ministry of Economic Development 
$\&$ Innovation.
MCJ is supported by the National Science and Engineering Research Council through a Discovery grant.
This work was also supported by the NRHEP 295189 FP7-PEOPLE-2011-IRSES Grant, and by FCT-Portugal through projects
CERN/FP/123593/2011 and IF/00293/2013.
Computations were performed on the ``Baltasar Sete-Sois'' cluster at IST.
\end{acknowledgments}

\appendix

\subsection{Convergence properties}\label{sec:convergence}
\begin{figure*}[htb!]
\begin{center}
\begin{tabular}{cc}
\epsfig{file=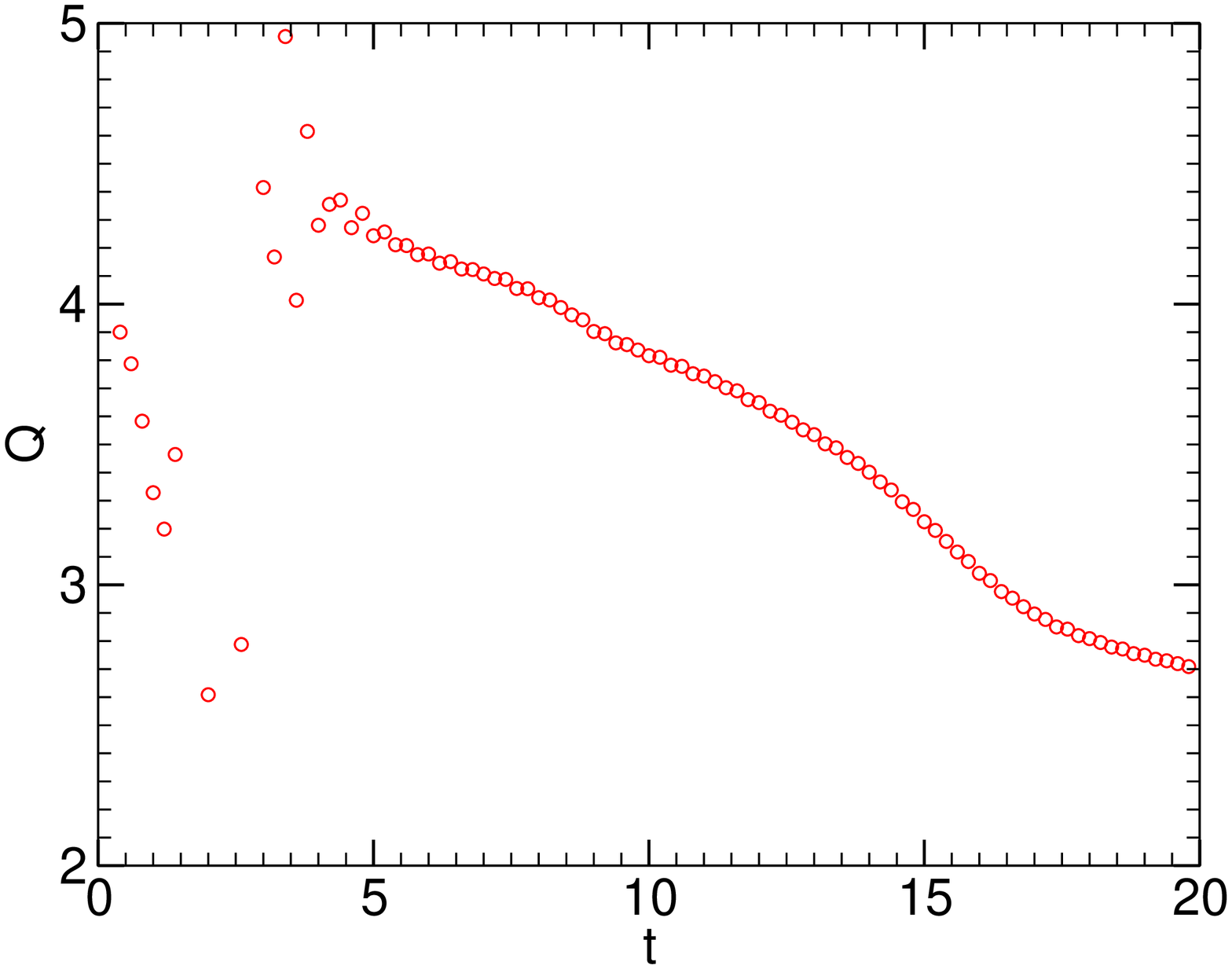,width=8.1cm,angle=0,clip=true}&
\epsfig{file=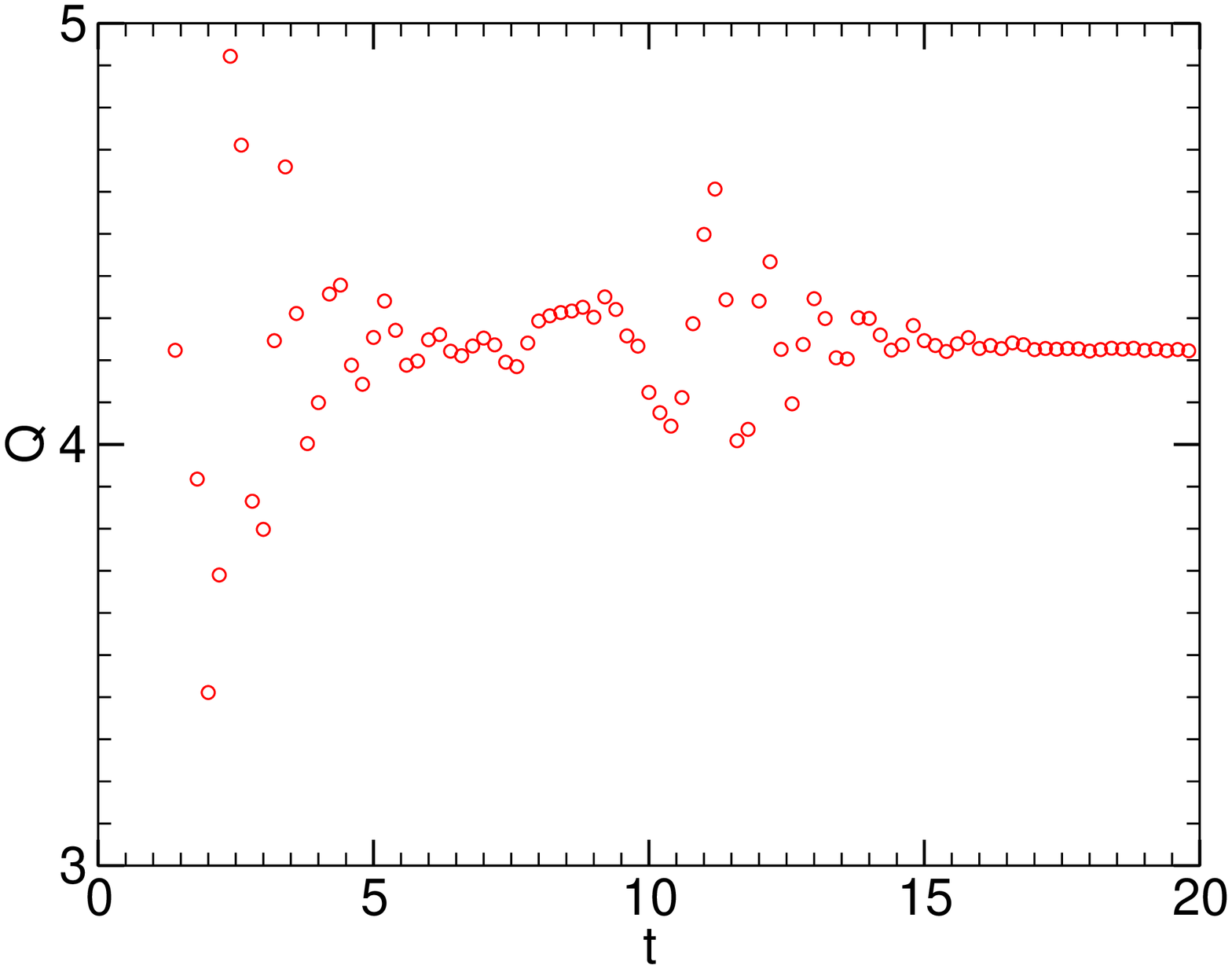,width=8.1cm,angle=0,clip=true}
\end{tabular}
\caption{The convergence factor $Q(t)$ defined by Eq.~\eqref{eq:Q} as a function of time.
Left panel refers to initial data of the form $\pi(r,0)=0,\,\rho=50,\,R_0=1$,
while the right panel refers to Eq.~\eqref{eq:gaussian_initial} for $\rho=200,\,R_0=1,\,A=0.002,\,\sigma=1,\,r_w=10,\,\epsilon=0.001$.
}
\label{fig:Qplots_2}
\end{center}
\end{figure*}

The convergence properties of our numerical results can be understood by varying the grid size. For a $p^{th}$ order scheme, the convergence ratio defined by
\begin{equation} \label{eq:Q}
	Q  = \frac{||\pi^{4h} - \pi^{2h}||_2}{||\pi^{2h} - \pi^{h}||_2}
\end{equation}
yields $Q=2^p$ in the continuum limit, $h \rightarrow 0$. Here, the superscript on the numerical solution $\pi$ refers to the size of the spacing of the grid used, and $|| \cdot ||_2$ is the $\ell_2$-norm. For example, for the second-order order scheme we use, $Q\simeq 4$. 
We have evolved both sets of initial conditions introduced in Sections~\ref{sec:screen1}--\ref{sec:screen2}, in particular Eq.~\eqref{eq:gaussian_initial} with $\rho=200,\,R_0=1,\,A=0.002,\,\sigma=1,\,r_w=10,\,\epsilon=0.001$ and $\pi(r,0)=0$, respectively.
The results are summarized in Fig.~\ref{fig:Qplots_2}, and are compatible
with second-order convergence.
%

\subsection{Cauchy problem and stability}\label{app:IVP}

Let us first consider the Lagrangian \eqref{Lan_general} for a spherically symmetric field $\pi(t,r)$. The equations of motion coming from this Lagrangian have the form of a Monge-Amp\'ere equation
\be\label{monge}
A\ddot{\pi}+B\dot{\pi}'+C\pi''+D+E\left[\ddot{\pi}\pi''-(\dot{\pi}')^2\right]\,,
\ee
where $A$, $B$, $C$ and $D$ are at most functions of $\pi$ and its first derivatives, and we assume $(A+E\pi'')\neq 0$. We wish to understand when this type of equation plus its initial conditions describe a well-posed initial value problem, commonly known as Cauchy problem (see e.g. Chapter V of~\cite{Courant:book} and Appendix 1 where the Cauchy problem for the Monge-Amp\'ere equation is considered). A family of curves $\varphi(t,r)=0$ are characteristics of this equation if
\be\label{charac}
(A+E \pi'')\varphi_t^2+(B-2E \dot{\pi}')\varphi_t\varphi_r+(C+E \ddot{\pi})\varphi_r^2=0\,,
\ee
where $\varphi_r\equiv \partial_r\varphi$ and $\varphi_t\equiv \partial_t \varphi$. Along a characteristic curve solutions are constant, so we can write
\be
\frac{d\varphi}{d\lambda}=\varphi_r \frac{dr}{d\lambda}+\varphi_t \frac{dt}{d\lambda}=0\,,
\ee
where $\lambda$ denotes the parameter along which the characteristic curve is constant. Substituting in Eq. \eqref{charac}, we have
\be
(A+E \pi'')\left(\frac{dr/d\lambda}{dt/d\lambda}\right)^2-(B-2E \dot{\pi}')\frac{dr/d\lambda}{dt/d\lambda}+(C+E \ddot{\pi})=0\,.
\ee
Solving with respect to $\frac{dr/d\lambda}{dt/d\lambda}\equiv dr/dt$ we find two roots
\beq\label{drdt}
u_{1}&\equiv&\frac{dr}{dt}=\frac{B-2E \tilde{s}+ \Delta}{2(A+E \tilde{r})}\,,\\
u_{2}&\equiv&\frac{dr}{dt}=\frac{B-2E \tilde{s}- \Delta}{2(A+E \tilde{r})}\,,\\
\eeq
where $\Delta^2=B^2-4AC+4DE$ and we used Eq.~\eqref{monge} in the form
\be\label{monge_det}
(A+E \pi'')(C+E \ddot{\pi})-\frac{1}{4}(B-2E \dot{\pi}')^2+\frac{1}{4}\Delta^2=0\,.
\ee

The type of this equation is determined by the discriminant $\Delta^2$:
\begin{itemize}
\item If $\Delta^2>0$, the equation is hyperbolic (two roots);
\item If $\Delta^2=0$, the equation is parabolic (one root);
\item If $\Delta^2<0$, the equation is elliptic (imaginary roots).
\end{itemize}
For galileon-like models, the discriminant $\Delta$ depends on the first derivatives of $\pi$. A well-posed initial value problem is equivalent to requiring that the equation be hyperbolic, i.e. $\Delta^2>0$. Note that we also must require the initial conditions to satisfy $A+E \pi''< 0$, which is equivalent to requiring the constant time $t=t_0$ hypersurface to be spacelike everywhere.

Let us see how this can be understood in terms of scalar perturbations $\delta\pi$ about a background $\pi_0$. The action for $\delta\pi$ is given by
\begin{equation}
S_{\delta\pi}=\int{d^4x\left[\frac{1}{2} Z^{\mu\nu}\partial_\mu\delta\pi\partial_\nu\delta\pi\right]}\,,
\end{equation}
where $Z^{\mu\nu}$ is the effective metric, dependent on $\pi_0$ and its derivatives, on which $\delta\pi$ propagates. Requiring the equations of motion for $\delta\pi$ to be hyperbolic is equivalent to requiring the effective metric $Z^{\mu\nu}$ to have a Lorentzian signature, i.e., $\det(Z^{\mu\nu})<0$, which in  a spherically symmetric spacetime can be written as
\be
\det(Z^{\mu\nu})=[Z^{tt}Z^{rr}-(Z^{tr})^2](Z^{\theta\theta})^2\sin^2\theta<0\,.
\ee
From Eq.~\eqref{monge_det} this can be shown to be equivalent to the condition $\Delta^2>0$ for the background $\pi_0$. 

Furthermore, the initial value problem is well posed only if the initial data are set up on a hypersurface $\rm \Sigma$ which is spacelike with respect to the effective metric $Z^{-1}_{\mu\nu}$, where $Z^{-1}_{\mu\nu}$ is defined such that $Z^{\mu\lambda}Z^{-1}_{\lambda\nu}=\delta^{\mu}_{\nu}$, i.e., we require the 1-form $\partial_{\mu}t$ to be timelike with respect to $Z^{\mu\nu}$, $Z^{\mu\nu}\partial_{\mu}t\partial_{\mu}t<0$. Thus the initial data must be such that $Z^{tt}< 0$, which is equivalent to requiring $A+E \pi''< 0$ for the background $\pi_0$. However, note that in general even for an initially well-posed Cauchy problem, due to the non-linearity of the field equations, the \emph{global} existence and uniqueness of solutions cannot be guaranteed. 

As a final remark let us consider the stability under high-frequency perturbations of the background $\pi_0$. Local stability at a given point in spacetime $p_0$ requires the metric to have a Lorentzian signature at $p_0$ which may appear either in the form of ghost or gradient instabilities~\footnote{Ghost instabilities are characterized either by a wrong-sign of the time-component of the field equations, which is characterized by $\det(Z^{\mu\nu})>0$ when all the other components have the ``correct'' signature, or by a relative overall difference of sign of the effective metric with respect to the gravitational metric, when coupling to other fields is considered. On the other hand gradient instabilities arise when $\det(Z^{ij})<0$, where the indices $i,\,j$ take values on the 3-dimensional spacelike hypersurfaces $t=\rm{const}$.}. Through this work we have not considered coupling to other dynamical fields, but if this is taken into account we must also require the matrix $Z^{\mu\nu}$ to have the same signature of the gravitational metric.  This requirement is met as long as the matrix $Z^{\mu\nu}$ is non-singular and when diagonalized has the signature $(-,+,+,+)$. In spherical coordinates this is equivalent to requiring
\beq\label{eigenvalues}
&&\frac{Z^{tt}+Z^{rr}-\sqrt{4(Z^{rt})^2+(Z^{tt}-Z^{rr})^2}}{2}<0\,,\\
&&\frac{Z^{tt}+Z^{rr}+\sqrt{4(Z^{rt})^2+(Z^{tt}-Z^{rr})^2}}{2}>0\,,\\
&&Z^{\theta\theta}>0\,.
\eeq
Note that this requirement is not met in the background of the $\Pi_-$ branch (see Section~\ref{sec:screeningstability}), so this solution would be unstable if it were coupled to another field or if the dynamical degrees of freedom of the source were taken into account, as discussed in the main text.

\subsection{Decoupling limit of Massive gravity}\label{app:dec}

The Lagrangian for the decoupling limit of dRGT massive gravity \cite{deRham:2010kj} is
\begin{align} \label{eq:mgLdc}
	\mathcal{L}=-&\frac{1}{2}h^{\mu\nu}\mathcal{E}_{\mu\nu}^{\alpha\beta}h_{\alpha\beta}+h^{\mu\nu}X^{(1)}_{\mu\nu}+\frac{\alpha}{\Lambda^3}h^{\mu\nu}X^{(2)}_{\mu\nu}\nonumber \\
	+&\frac{\beta}{\Lambda^6}h^{\mu\nu}X^{(3)}_{\mu\nu}+T_{\mu\nu}h^{\mu\nu}\,,
\end{align}
where $\mathcal{E}_{\mu\nu}=\mathcal{E}_{\mu\nu}^{\alpha\beta}h_{\alpha\beta}$ is the linearization around $\eta_{\mu\nu}$ of the Einstein tensor $G_{\mu\nu}$ ($g_{\mu\nu}=\eta_{\mu\nu}+h_{\mu\nu}$). The tensors $X^{(n)}_{\mu\nu}$ are special conserved combinations of $\Pi \equiv \nabla_\mu \partial_\nu \pi$ such that $\partial^\mu X^{(n)}_{\mu\nu} = 0$. They are defined as
\begin{eqnarray} \label{eq:Xmunu}
	X^{(1)}_{\mu\nu}&=&[\Pi]\eta_{\mu\nu}-\Pi_{\mu\nu}\,,\\
	X^{(2)}_{\mu\nu}&=&\frac{1}{2}([\Pi]^2-[\Pi^2])\eta_{\mu\nu}-[\Pi]\Pi_{\mu\nu}+\Pi_{\mu\nu}^2\,, \\
	X^{(3)}_{\mu\nu}&=&([\Pi]^3-3[\Pi][\Pi^2]+2[\Pi^3])\eta_{\mu\nu}\,, \nonumber\\
	&-&3([\Pi]^2-[\Pi^2])\Pi_{\mu\nu}+6[\Pi]\Pi_{\mu\nu}^2-6\Pi^3_{\mu\nu}\,.
\end{eqnarray}
Square brackets are used to denote the trace: $[\Pi]=\Pi_{\mu\nu}\eta^{\mu\nu}$. The parameters $\alpha \ \text{and} \ \beta$ are the two free parameters of the theory. The scale $\Lambda$ is the strong coupling scale of this theory and is given by $\Lambda=(m^2M_4)^{1/3}$ where $m$ is the mass of the graviton. The derivatives in the above expression are to be evaluated on flat space.

For general values of $\alpha \ \text{and} \ \beta$, it is impossible to completely decouple the tensor and scalar modes, and their solutions must be found simultaneously. However, we'll focus on the special case where $\beta=0$ in which it is possible to completely decouple $\pi$ and $h_{\mu\nu}$ through a field redefinition. Then we can write $\mathcal{L}=\mathcal{L}_{h_{\mu\nu}}+\mathcal{L}_\pi$ and solve the equation of motion for $\pi$ independently of the metric.

After performing the field redefinition $h_{\mu\nu}\rightarrow h_{\mu\nu}+\eta_{\mu\nu}\pi+\alpha\partial_\mu\pi\partial_\nu\pi/\Lambda^3_3$, the scalar and tensor modes can be decoupled, and the resulting Lagrangian for the scalar sector is of form~\eqref{Lan_general}:
\begin{eqnarray} \label{mgL}
	\mathcal{L}_\pi &=& -\frac{3}{2}\mathcal{L}_2 + \frac{3}{2}\frac{\alpha}{\Lambda^3}\mathcal{L}_3 - \frac{1}{2}\frac{\alpha^2}{\Lambda^6}\mathcal{L}_4 +\frac{\pi T}{M_4} \nonumber\\
	&+&  \frac{\alpha}{M_4\Lambda^3}\partial_\mu\pi\partial_\nu\pi T^{\mu\nu}\,,
\end{eqnarray}
where $\mathcal{L}_2,\ \mathcal{L}_3,\ \mathcal{L}_4$ are the galileon Lagrangians given in Eqs.~\eqref{gal_lin},~\eqref{gal_quartic} and~\eqref{gal_quintic}, respectively.
%
%
Varying these Lagrangians with respect to $\pi$ can be done using $\delta\mathcal{L}_n/\delta\pi = -2\mathcal{L}_{n+1}/(\partial\pi)^2$ so that the resulting scalar equation of motion is
%
%
%
\begin{align}\label{eq:mgeom}
&	3\square\pi - 3\frac{\alpha}{\Lambda^3}\left[(\square\pi)^2-\left(\nabla_\mu\nabla_\nu\pi\right)\left(\nabla^\mu\nabla^\nu\pi\right)\right] + \frac{\alpha^2}{\Lambda^6}\left[(\square\pi)^3\right. \nonumber \\
	&\left. -3\square\pi\left(\nabla^\mu\nabla^\nu\pi\right)\left(\nabla_\mu\nabla_\nu\pi\right)+2\left(\nabla_\mu\nabla_\nu\pi\right)\left(\nabla^\nu\nabla_\gamma\pi\right)\left(\nabla^\gamma\nabla^\mu\pi\right)\right] \nonumber \\
	& + \frac{T}{M_4}-2\frac{\alpha}{M_4\Lambda^3}\nabla_\mu\nabla_\nu\pi T^{\mu\nu} = 0\,.
\end{align}

Small perturbations around a time-dependent background propagate on the effective metric $Z^{\mu\nu}$, which can be computed perturbing Eq.~\eqref{eq:mgeom}. For a spherically symmetric background the components of this metric are given by:  
\begin{align}\label{eq:Zmg_time}
Z_\text{MG}^{tt} = & -3 - \alpha\left( 2 T^{00} - 6\pi'' - 12\frac{\pi'}{r} \right)	\nonumber \\
		& - \alpha^2\left( \frac{6(\pi')^2}{r^2}+\frac{12\pi''\pi'}{r} \right)\,,\nn\\
Z_\text{MG}^{rr} = & 3 + \alpha\left( -2 T^{11} + 6\ddot{\pi} - 12\frac{\pi'}{r} \right)	\nonumber \\
		& + \alpha^2\left( \frac{6(\pi')^2}{r^2}-\frac{12\ddot{\pi}\pi'}{r} \right)\,,\nn\\
Z_\text{MG}^{tr} = & -6\alpha\dot{\pi}'+12\alpha^2\frac{\dot{\pi}'\pi'}{r}\,,	\nonumber \\
r^2Z_\text{MG}^{\theta\theta} = & 3 + \alpha\left( -\tilde{T}^{22}-\tilde{T}^{33} + 6\ddot{\pi}- 6\pi'' - 6\frac{\pi'}{r} \right)	\nonumber \\
		& + 6\alpha^2\left(\frac{\pi''\pi'}{r}-\frac{\ddot{\pi}\pi'}{r}-\pi''\ddot{\pi}+(\dot{\pi}')^2\right)\,,\nn\\	
\end{align} 
where we defined $\tilde{T}^{22}\equiv T^{22}r^2$ and $\tilde{T}^{33}\equiv T^{33}r^2\sin^2\theta$ such that $\tilde{T}^{ii}$ are functions of $r$ and $t$ only.

\bibliography{ref}  

\end{document}